\documentclass[]{article}
\usepackage{color}
\usepackage{amsmath,amsfonts,amssymb}
\usepackage{graphicx}
\usepackage{jabbrv}
\usepackage{authblk}
\usepackage{url}
\usepackage{geometry}
\title{Hybrid integrated semiconductor lasers with silicon nitride feedback circuits}

\author[1,3,*] {Klaus-J. Boller} 
\author[1]{Albert van Rees}
\author[1,2]{Youwen Fan}
\author[1]{Jesse Mak}
\author[1]{Rob E.M. Lammerink}
\author[1]{Cornelis A.A. Franken}
\author[1]{Peter J.M. van der Slot}
\author[1]{David A.I. Marpaung}
\author[3,1]{Carsten Fallnich}
\author[2]{J\"{o}rn P. Epping}
\author[2]{Ruud M. Oldenbeuving}
\author[2]{Dimitri Geskus}
\author[2]{Ronald Dekker}
\author[2]{Ilka Visscher}
\author[2]{Robert Grootjans}
\author[2]{Chris G.H. Roeloffzen}
\author[2]{Marcel Hoekman}
\author[2]{Edwin J. Klein}
\author[2]{Arne Leinse}
\author[2]{Ren\'e G. Heideman}

\affil[1]{Laser Physics and Nonlinear Optics, Mesa$^{+}$ Institute for Nanotechnology, Department for Science and Technology, Applied Nanophotonics, University of Twente, Enschede, The Netherlands}
\affil[2]{LioniX International BV, Enschede, The Netherlands}
\affil[3]{University of M\"{u}nster, Institute of Applied Physics, Germany}
\affil[*]{Corresponding author: k.j.boller@utwente.nl}

\begin{document}
	
\maketitle
	 
\abstract{Hybrid integrated semiconductor laser sources offering extremely narrow spectral linewidth as well as compatibility for embedding into integrated photonic circuits are of high importance for a wide range of applications. We present an overview on our recently developed hybrid-integrated diode lasers with feedback from low-loss silicon nitride (Si$_3$N$_4$ in SiO$_2$) circuits, to provide sub-100-Hz-level intrinsic linewidths, up to 120~nm spectral coverage around 1.55~$\mu$m wavelength, and an output power above 100~mW. We show dual-wavelength operation, dual-gain operation, laser frequency comb generation, and present work towards realizing a visible-light hybrid integrated diode laser.}

\section{Introduction}
The extreme coherence of light generated with lasers has been the key to great progress in science, for instance in testing nature’s fundamental symmetries~\cite{biesheuvel_2016NCom, gabrielse_2002PRL}, properties of matter~\cite{hansch_1972NPS, bagdonaite_2014AJ}, or for the detection of gravitational waves~\cite{abbott_2016PRL}. While fundamental research has been, and still is, based on very diverse types of lasers, the development has been different with applications. Here, with billions of pieces fabricated per year, the diode laser (semiconductor laser) is by far most prevalent, due to a unique set of advantages. Lithographic fabrication and integration on a chip reduces mass, size and cost per piece, and the laser lifetimes can reach the 100,000-hour level. Generating light in a semiconductor junction enables ease of operation directly with an electric current, with up to 85\% power efficiency~\cite{crump_2006CLEO}. The wavelength coverage and tunability of diode lasers reaches from the near-UV into the mid-infrared, while optical integration provides excellent intrinsic stability vs. mechanical and acoustic perturbations.

With these advantages, diode lasers are essential for photonics as key enabling technology. Narrow linewidth and wavelength tunable diode lasers can serve high-end and upcoming applications. Prominent examples are monitoring and sensing in fabrication~\cite{leong_2005SMS, Jac14, technobis_2017}, bio-sensing~\cite{he_2011NT}, monitoring the integrity of civil structures~\cite{alalusi_2009SPIE, rothberg_2017OLE}, laser ranging (LIDAR) for autonomous traffic~\cite{hecht_2018OPN} or sensing of rotation with optical gyros~\cite{tran_2017OE, srinivasan_2014OE, gundavarapu_2019NPhot}.

With sufficient coherence, diode lasers can play a great role in precision metrology and timing, such as in portable atomic clocks~\cite{lezius_2016O,newman_2019O,jiang_2011NP}, including satellite-based GPS systems~\cite{ESA16,gill_2005Met}. When integrating narrow-linewidth semiconductor lasers into functional photonic circuits, they may serve as on-chip light engines, for instance, to drive Raman and Brillouin lasers~\cite{spillane_2002N,li_2017O,gundavarapu_2019NPhot}. A most recent development is driving Kerr frequency combs with narrowband diode lasers~\cite{stern_2018N, raja_2019NatCom, pavlov_2018NPhot} which complements the frequency combs provided by mode-locked diode lasers~\cite{wang_2017LSA}. Specifically, if the combs comprise narrowband comb lines, dual-comb metrology~\cite{coddington_2009NPhot, suh_2018Science}, spectroscopic detection~\cite{rieker_2014O,coddington_2016O} or dual-comb imaging~\cite{bao_2019O} can move towards chip-based formats~\cite{spencer_2018N}. Narrow-linewidth diode lasers may also be beneficial for fully integrated, chip-based quantum frequency combs that can generate highly complex entangled optical states~\cite{reimer_2018FoO}.

Of widest relevance is the role of diode lasers in communication and information technology, for instance as key component of the global fiber network~\cite{kikuchi_2016JLT} or within data centers~\cite{yue_2019PJ}. By lowering the phase noise of diode lasers, coherent optical communications based on phase-encoding~\cite{winzer_2006ProcIEEE, zhang_2009PTL} is expected to increase the transmission rates noticeably~\cite{beppu_2014OFC}. Following the relation between the bit rate B and symbol rate S (baudrate), B=$log_2$S, quadrature amplitude modulation with 4096 symbols (QAM 4096) promises a 12-times higher transmission rate. For further increased data rates, diode laser driven Kerr frequency combs can increase the number of wavelength channels available for coherent transmission~\cite{pfeifle_2014NPhot}. Similarly, low-noise diode lasers are foreseen as information carriers for processing of information with optical methods~\cite{seeds_2006JLT, capmany_2007NPhot}. This can be seen from recent progress in integrated microwave photonics~\cite{zhuang_2008Leos, zhuang_2011OE, burla_2013OE, marpaung_2015O,marpaung_2019NP}, photonic analog-to-digital conversion~\cite{khilo_2012OE} and generation of low-noise and widely tunable microwave to terahertz signals with integrated diode lasers~\cite{carpintero_2012OL, seeds_2015JLT}.

The absolutely central property in these applications is the laser’s spectral linewidth, which is a measure for the degree of spectral purity, also called coherence. Narrowing the linewidth increases the amount of information and precision to be gained in sensing and metrology, and it increases the data rate through optical interconnects and in optical processing.

As the frequency fluctuations of lasers are caused by a variety of different processes~\cite{lang_1985JQE} involving very different time scales, determining the coherence properties of laser light requires comprehensive measurements~\cite{ruthman_1978ProcIEEE, thomson_1982ProcIEEE, didomenico_2010AO, barnes_1971IEEETransIM}. A key coherence property and signature of spectral quality is the Schawlow-Townes limit, also called quantum limit, fundamental linewidth, intrinsic linewidth or fast linewidth~\cite{schawlow_1958PR, fleming_1981APL, henry_1982JQE, wiseman_1999PRA}. At a given output power, this fundamental bandwidth can only be reduced by increasing the lifetime of photons in the laser resonator, which is the main approach towards the various diode laser designs that we present here.

Depending on the application, also the slow linewidth can be of major importance, i.e., the linewidth obtained after longer averaging, often named full-width at half-maximum (FWHM) linewidth. This measure comprises also technical noise such as from thermal drift, or from noise in the pump current. The FWHM bandwidth can partly be reduced with optimizing the laser design for highest passive stability, such as provided by photonic integration. Long-term frequency stability requires that the laser is frequency tunable, such that an electronic servo control can minimize the detuning from a stable reference used as frequency discriminator~\cite{drever_1983APB}. However, to avoid that such active stabilization adds too much noise on its own, e.g., quantum noise from photo detection in the frequency discriminator, and to suppress noise also at higher noise frequencies, it remains essential to reduce the intrinsic laser linewidth~\cite{day_1992JQE, lin_2012OL}.

The remainder of this manuscript is organized as follows. In section 2, we describe the state of the art with respect to narrowing of the intrinsic linewidth of diode lasers. In section 3, we briefly discuss the physics of linewidth narrowing. In section 4, we describe a hybrid InP-Si$_3$N$_4$ laser based on two intracavity micro-ring resonators and a single gain section. In section 5, we show that be adding a second gain section that both the output power can be significantly increased and the intrinsic linewidth reduced. In section 6, we show that, by adding a third intracavity microring resonator, a record-low intrinsic linewidth can be realized. In section 7 we describe hybrid lasers producing a frequency comb or providing widely tunable dual wavelength operation, and discuss the possibility to extend the oscillating wavelength down into the visible. We conclude with a summary and outlook in section 8.

\section{State of the art}

The typical FWHM bandwidth of commercially available, integrated diode lasers has remained for long at relatively high levels around a MHz~\cite{ward_2005JSTQE, lavery_2013JLT, akulova_2002JSTQE}, with lowest values of 170 and 20 kHz achieved so far (at 1.5~$\mu$m wavelength~\cite{okai_1990PTL} and at 850~nm~\cite{price_2006PTL}, respectively). The lowest intrinsic linewidth achieved with a monolithic diode laser is about 2~kHz (FWHM 180~kHz)~\cite{spiessberger_2011APB}. Much smaller bandwidths have been obtained with non-integrated lasers that use bulk optical gratings~\cite{toptica_2019}. Miniaturized bulk components have been very effective as well \cite{luvsandamdin_2014OE}, particularly high-Q whispering gallery mode resonators~\cite{liang_2015NC} or Bragg fibers~\cite{lin_2012OL, wei_2016OE, morton_2018JLT}. In connection with extensive electronic servo stabilization, for research purposes, even diode lasers with bulk optical feedback can reach the sub-Hz-range~\cite{zhao_2010OC,alnis_2008PRA,stoehr_2006OL}. But due to the large size, mass and acoustic perturbation sensitivity, this route remains unattractive for mobile, handheld and space applications, and in all applications that are to serve big volumes. Similarly, due to size restrictions, lack of long-term stability or diffraction loss in coupling from free space to tightly guiding waveguides, even miniaturized bulk optical sources are less suitable to feed integrated photonic circuitry, e.g., in integrated microwave photonics~\cite{roeloffzen_2013OE, marpaung_2013LPR, marpaung_2019NP} or for integrating optical beam steering~\cite{doylend_2012OL}.   
\begin{figure}[tbp]
	\centering
	\includegraphics[width=0.75\linewidth]{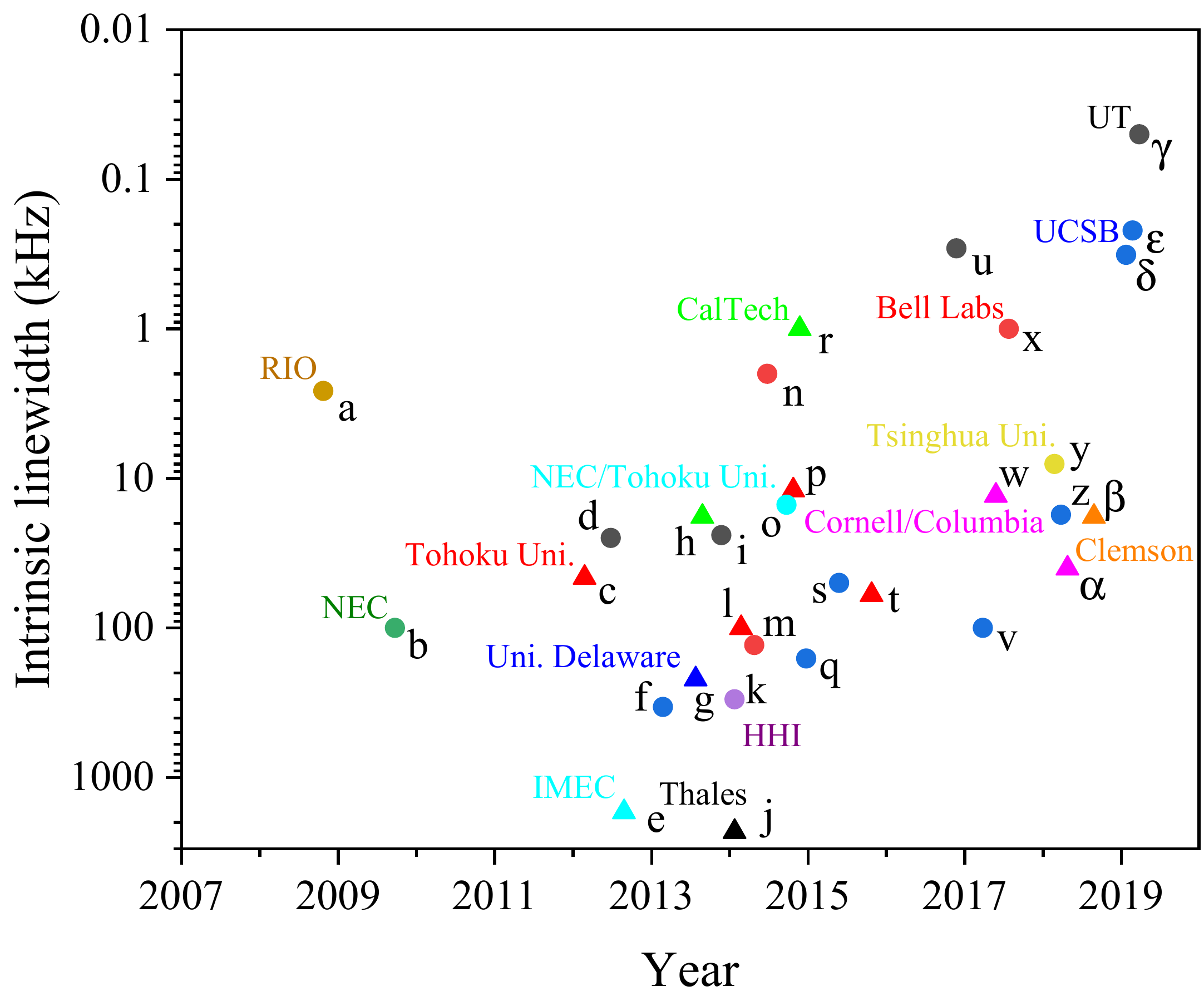}
	\caption{\label{fig:overview_linewidth_ref}Overview of intrinsic linewidth reported for hybrid or heterogeneously integrated diode lasers. a-\cite{alalusi_2009SPIE}, b-\cite{matsumoto_2010OFCC}, c-\cite{nemoto_2012APE}, d-\cite{oldenbeuving_2013LPL}, e-\cite{keyvaninia_2013OE}, f-\cite{hulme_2013OE}, g-\cite{yang_2014OE}, h-\cite{santis_2014PNAS}, i-\cite{fan_2014SPIE}, j-\cite{duan_2014JSTQE}, k-\cite{defelipe_2014PTL}, l-\cite{kita_2014STQE}, m-\cite{dong_2014OE}, n-\cite{debregeas_2014ISLC}, o-\cite{kobayashi_2015JLT}, p-\cite{tang_2015OL}, q-\cite{srinivasan_2015PJ}, r-\cite{santis_2015CLEO}, s-\cite{komljenovic_2015JSTQE}, t-\cite{kita_2016STQE}, u-\cite{fan_2017CLEO}, v-\cite{komljenovic_2017JLT}, w-\cite{stern_2017OL}, x-\cite{verdier_2018JLT}, y-\cite{li_2018JLT}, z-\cite{tran_2018ECOC}, $\alpha$-\cite{stern_2018N},  $\beta$-\cite{zhu_2018CLEO}, $\delta$-\cite{huang_2019O}, $\varepsilon$-\cite{xiang_2019OL}, $\gamma$-\cite{fan_2019ARXIV}.}
\end{figure}
Many orders of magnitude smaller linewidths than with monolithic diode lasers have been achieved with hybrid and heterogeneously integrated diode lasers, ultimately reaching into the sub-kHz-range (see Fig.~\ref{fig:overview_linewidth_ref}). The highest degree of intrinsic coherence so far is generated with an InP-Si$_3$N$_4$ hybrid integrated diode laser~\cite{fan_2019ARXIV}. There, we employed a low-loss Si$_3$N$_4$ waveguide circuit comprising microring resonators for extending the photon lifetime, imposing single-frequency oscillation, and wavelength tuning.

\begin{figure}[tbp] \label{hybrid_laser_general}
	\centering
	\includegraphics[width=0.35\linewidth]{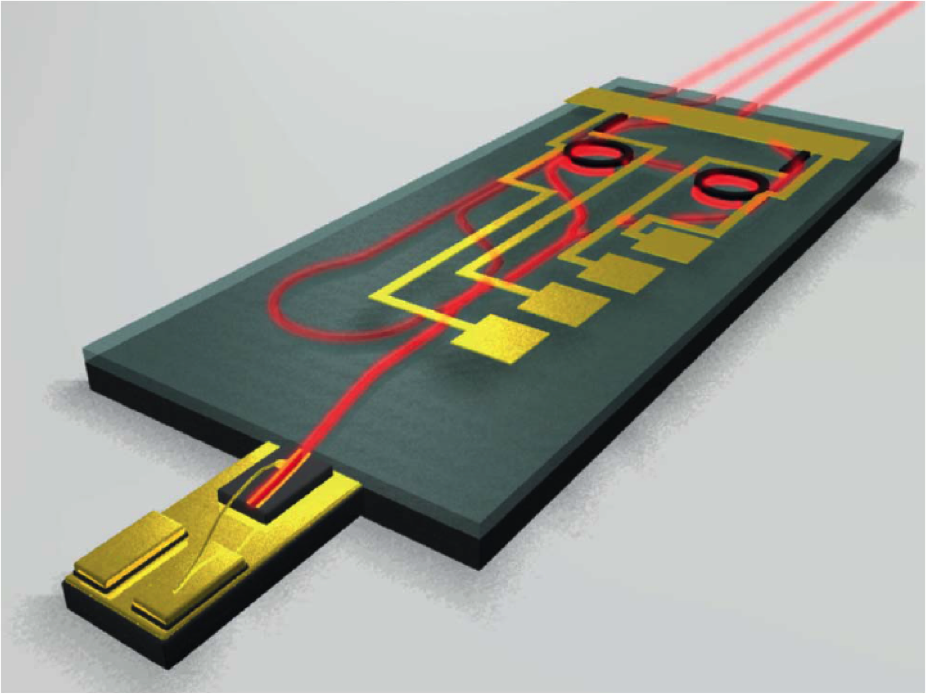}
	\includegraphics[width=0.55\linewidth]{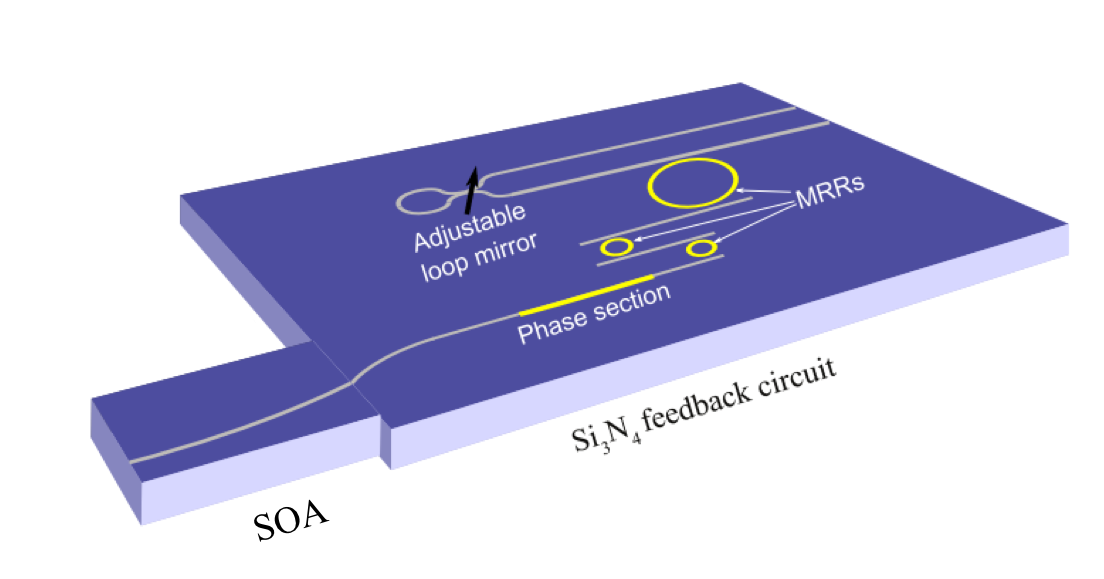}
	\includegraphics[width=0.7\linewidth]{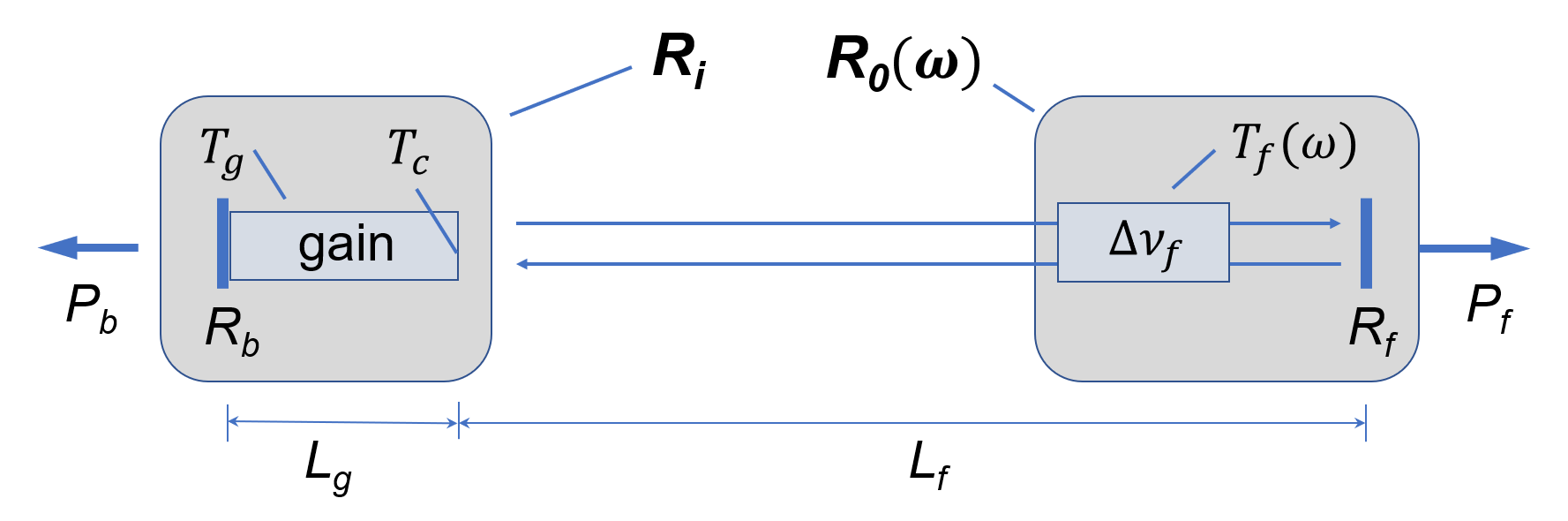}
	\caption{\label{fig:hybrid_laser_general} (\textbf{left}) Schematic view of a hybrid integrated diode laser with two microring resonators (MRRs) in Vernier configuration for spectrally selective feedback. The lower left part is a semiconductor double-pass amplifier forming the gain section (red) with electrodes for pumping (gold) and one of the laser cavity end mirrors. The upper right part is the waveguide feedback circuit showing Si$_3$N$_4$ waveguides (red), electrode pads and leads (gold) and heaters (black) for tuning the laser. (\textbf{right}) InP-Si$_3$N$_4$ with three MRRs and a Sagnac loop mirror. (\textbf{lower}) General scheme: Extending the laser cavity by adding to the gain section of length $L_g$ and double-pass reflectance $R_i$ a long and low-loss feedback arm of length $L_f$ ($L_f \gg L_{\textrm{g}}$ and $R_0 \gg R_i$) increases the photon lifetime and narrows the laser linewidth. The overall feedback reflectance is $R_0(\omega)$=$R_f \cdot T_f^2(\omega)$, where $R_f$ is the end mirror reflectance and $T_f(\omega)$ is the single-pass transmittance of the Vernier filter having a spectral bandwidth $\Delta \nu_f$.}
\end{figure}
All hybrid and heterogeneously integrated diode lasers make use of additional waveguide circuits fabricated in a different, low-loss material platform, while light is generated and amplified in a semiconductor material gain section. A schematic view of hybrid integrated lasers based on frequency selection with two or three microring resonators is shown in Fig.~\ref{fig:hybrid_laser_general}. The low-loss dielectric part of the circuit increases the photon lifetime of the laser resonator in order to reduce the laser linewidth. At the same time, the narrowband transmission of resonators imposes single-frequency oscillation via intracavity spectral filtering. Although the cavity extension is aiming on increasing the photon lifetime, it should be noted that all integration, whether hybrid or heterogeneous, inevitably causes extra roundtrip loss due to imperfect optical coupling at the interface between the distinct platforms and materials, and due to losses in the feedback circuit, both of which decreases the photon lifetime. It is thus important to reduce coupling as well as propagation loss in the feedback arm.

Besides using Bragg waveguides from Si~\cite{santis_2014PNAS, santis_2015CLEO, santis_2018PNAS, huang_2019O}, polymer~\cite{klein_2012IEEECInPRM, defelipe_2014PTL}, or doped silica (SiO$_2$) \cite{numata_2010OE, numata_2012IOPConf, lin_2012OL}, spectral filtering and extending the cavity length has mostly been based on microring resonators, employing Si waveguides~\cite{kita_2015APL,hulme_2013OE,kobayashi_2015JLT, tran_2020JSTQE}, SiON~\cite{matsumoto_2010OFCC}, SiO$_2$~\cite{debregeas_2014ISLC} and Si$_3$N$_4$~\cite{oldenbeuving_2013LPL, fan_2014SPIE, fan_2017CLEO}. While initially the linewidth was in the order of hundreds of kilohertz~\cite{hulme_2013OE, fan_2017CLEO} the lowest value obtained with silicon is now 220~Hz~\cite{tran_2020JSTQE}.

Using silicon waveguides as feedback circuits is beneficial for several reasons. The relatively high index contrast, $\Delta n \approx 2$ between the Si core and a SiO$_2$ cladding, allows tight guiding that enables using sharply bent waveguides without much radiation loss. Furthermore, techniques have been developed that allow wafer-scale heterogeneous integration with InP gain elements based on molecular or adhesive bonding ~\cite{roelkens_2010LPR, santis_2018PNAS}. There, optical coupling to the gain section is achieved with tapered vertical transitions~\cite{yariv_2007OE, huang_2019O}.

However, silicon also introduces a fundamental limitation. The lowest achievable linewidth becomes limited through nonlinear loss~\cite{vilenchik_2015PROC} beyond certain intracavity intensities and laser powers, specifically via two-photon absorption~\cite{kuyken_2017NANOPHOT}. This limits the linewidth to values above a few hundred Hertz~\cite{santis_2018PNAS}. The reason is that the photon energy for telecom wavelengths around 1.55 $\mu$m ($\approx$~0.8 eV) is close to the relatively small electronic bandgap of silicon ($\approx$~1.14 eV, corresponding to 1.1 $\mu$m) while the laser intracavity intensities can become high. Specifically, high intensities easily occur when selecting, within a wide semiconductor gain spectrum with a laser intracavity spectral filter, single longitudinal mode oscillation in an optically long laser resonator. The reason is that small-sized (integrated) spectral filters, in order to resolve single modes of the laser resonator, need to have a high finesse, i.e., they need to exhibit low loss per filter rountrip. Accordingly, there will be a significant power enhancement in such filters and, due to tight guiding, also high intensities. Avoiding nonlinear loss by reducing the laser power with weaker pumping (and subsequent amplification) is not a solution because lowering the power of a laser oscillator increases the intrinsic linewidth as well~\cite{schawlow_1958PR}. These considerations indicate that, after transition loss between platforms and other linear loss is minimized with advanced fabrication, it is ultimately the electronic bandgap of the materials chosen for the passive part of the circuit which sets the fundamental linewidth limits.

Dielectric materials, such as Si$_3$N$_4$ and silica (SiO$_2$), provide much larger bandgaps than silicon ($\approx$ 5~eV and 8~eV, respectively) which safely excludes two-photon absorption. The silica platform, having weakly doped silica as core material, offers extremely low loss and thus narrow linewidth, such as shown with feedback from a straight Bragg waveguide grating at 1.064~$\mu$m \cite{numata_2012IOPConf}. The drawback of silica waveguides is its low index contrast, $\Delta n = $ $10^{-2}$ to $10^{-4}$, which leads to weak optical guiding. Weak guiding restricts silica to circuits with low curvature radius, i.e., to large circuits with relatively low functionality, making sharp spectral filtering for single-mode selection in long laser resonators difficult.

Ultimate linewidth narrowing of integrated semiconductor lasers is thus most promising with the Si$_3$N$_4$ platform~\cite{taballione_2019OE} or other high-contrast and low-loss dielectrics, such as LiNbO$_3$ bonded on insulator~\cite{poberaj_2012LPR}, LiNbO$_3$ bonded on silicon nitride~\cite{chang_2017OL},  Ta$_2$O$_5$ in SiO$_2$~\cite{belt_2017O} or AlN in SiO$_2$~\cite{jung_2013OL}. A further advantage of high-contrast platforms is that the mode field diameter can be matched to the relatively small mode field diameter found in semiconductor amplifier waveguides. With Si$_3$N$_4$, this currently promises coupling loss as low as 0.2 dB~\cite{roeloffzen_2018JSTQE}.

\section{Intrinsic linewidth of extended cavity hybrid integrated diode lasers}
\label{section3}

Single-frequency oscillation of extended cavity diode lasers is readily obtained by narrowband spectral filtering within the cavity. This was first demonstrated with a free-space cavity extension and feedback from a bulk diffraction grating~\cite{mooradian_1981JQE}. With an integrated waveguide circuit, much finer narrowband spectral feedback filtering can be achieved with microring resonators in Vernier configuration. A variety of arrangements for the ring resonators and the semiconductor gain section is described in~\cite{komljenovic_2017AS} for heterogeneously integrated lasers. Determining appropriate design values for microring radii and power coupling coefficients for a given gain bandwidth is described in~\cite{oldenbeuving_2013LPL} for the example of a hybrid integrated InP-Si$_3$N$_4$ laser with feedback from two waveguide microring resonators as shown in the upper left panel of Fig.~\ref{fig:hybrid_laser_general}. A generic scheme for determining the laser linewidth of such laser, or also with three or more resonators, is shown in the lower panel.

Precisely predicting the intrinsic linewidth of the laser linewidth of hybrid and heterogeneously integrated diode lasers is difficult for several reasons. The first is the relatively high complexity of the laser cavity with its feedback circuitry, as compared to simple Fabry-Perot lasers. Embedding microring resonators inside a laser cavity means that the temporal response of the cavity cannot be described with a simple exponential decay law. Another aspect is that the intensity in the gain section, and thus also the spatial distribution of the inversion density, varies notably with the propagation coordinate, which is due to a relatively high roundtrip loss. This means that standard simplifications, for instance, the mean field approximation for the gain section, are not well justified. Furthermore, the linewidth depends on a larger set of experimental parameters, many of which are not well known, such as the intrinsic losses in the amplifier waveguide, or the coupling loss between the different platforms realized after integration. Other parameters are difficult to determine because they depend on the laser's operating conditions. Examples are pump current induced temperature changes in the waveguide of the semiconductor amplifier causing thermally induced phase shifts, or the exact relation between heater currents and the optical roundtrip length of the microring resonator, both depending on details of the heat sink design and fabrication.

The most realistic calculation of all laser properties, including the intrinsic laser linewidth is likely to require numerical methods, such as based on transmission line models~\cite{javaloyes_2019, vpi_2019}. We have previously used numerical methods to calculate the intrinsic linewidth for a laser as in Fig.~\ref{fig:hybrid_laser_general}, in order to reveal the detailed influence of coupling losses at the interface of platforms on the linewidth~\cite{fan_2017OE}. The closest approximations using analytic expressions are still given in the early work of Henry~\cite{henry_1982JQE,henry_1986JLT}, Patzak et al.~\cite{patzak_1983EL}, Kazarinov and Henry~\cite{kazarinov_1987JQE}, Koch and Koren~\cite{koch_1990JLT}, Ujuhara~\cite{ujihara_1984JQE} and Bjork and Nilsson~\cite{bjork_1987JQE}. Summarizing all expressions~\cite{fan_2017OE} predicts the intrinsic linewidth as

\begin{equation} \label{eq1}
\Delta\nu_{\textrm{ST}}=\frac{1}{4\pi} \, \cdot \,  \frac{v_g^2 h \nu n_{\textrm{sp}} \gamma_{\textrm{tot}} \gamma_m (1+\alpha_H^2)}{P_b  \left(1+\frac{r_b}{r_0(\omega)}\frac{1-R_0(\omega)}{1-R_b}\right)} \; \cdot   \;  \frac{\alpha_P}{F^2}.
\end{equation}

In Eq.~\ref{eq1}, $v_g=c n_g$ is the group velocity in the gain section, $h\nu$ is the photon energy. $n_{\textrm{sp}}$, assuming typical values around 2, is the spontaneous emission enhancement factor that takes into account the reduction in inversion due to reabsorption by valence band electrons. $\alpha_H>0$ is Henry's linewidth enhancement factor. The factor describes the strength of gain-index coupling in the gain section~\cite{henry_1982JQE}, a coupling that is caused by the strongly asymmetric gain spectrum provided by semiconductor junctions~\cite{vahala_1983APL}. The linewidth increasing effect associated with $\alpha_H>0$ is that spontaneous emission events not only add randomly phased contributions to the laser field. These events, via a reduction of laser inversion, also increase the refractive index, which increases the phase noise further.

The spatially averaged roundtrip loss coefficient,  $\gamma_m=-1/(2L_g)\ln[(R_b R_0(\omega))]$, is determined by the output coupling, where $R_b=|r_b|^2$ denotes an approximately frequency independent (broadband) power reflectance of the gain section back facet, and $L_g$ is the length of the gain section. All optical properties of the feedback arm are lumped into a complex-valued reflectivity spectrum for the electric field, $r_0(\omega)$. This spectrum contains the optical length of the feedback arm as a frequency-dependent phase shift,  $r_0(\omega)=|r_0(\omega)|e^{i \phi(\omega)}$, and also the overall power reflectance, $R_0(\omega)=|r_0(\omega)|^2$, to include highly frequency selective filtering or output coupling. The feedback arm reflectance, $R_0(\omega)=T_f(\omega)^2 R_f$, is given by the end mirror reflectance, $R_f$, and the transmission spectrum of the intracavity spectral filter, $T_f(\omega)$. The loss coefficient $\gamma_{\textrm{tot}}= -1/(2L_g)\ln[(R_i R_0(\omega))]$ is the spatial average of all loss per roundtrip. Here $R_i=R_b T_{g}^2 T_c^2$ lumps all loss of the remaining roundtrip, i.e., all imperfect transmission and reflection, into an intrinsic reflectance.  The power transmission in a single pass through the gain section is $T_{g}=e^{(-\gamma_g L_{g})} <1$, with $\gamma_g$ the intrinsic passive loss constant of the gain waveguide. $T_c <1$ specifies the mode coupling loss per transmission through the interface between platforms. $P_b$ is the output power from the back diode facet. The factor in brackets next to $P_b$ is bigger than 1 and accounts for additional output power emitted at other ports of the laser cavity, for instance $P_f$ in Fig. \ref{fig:hybrid_laser_general}. The longitudinal Petermann factor, $\alpha_P$, is usually very close to 1, except if spontaneous emission becomes strongly amplified in a single-pass due to extremely small feedback (if $R_b$, $R_0 \ll1$)~\cite{ujihara_1984JQE, henry_1986JLT}.

Linewidth narrowing via cavity length extension is expressed in Eq.~\ref{eq1} by the factor $F$ as
\begin{equation} \label{eq2}
F=1+A+B,
\end{equation}
where
\begin{equation} \label{eq3}
A=\frac{1}{\tau_g} \cdot \left( \frac{d\phi_0(\omega)}{d\omega}    \right)
\end{equation}
and
\begin{equation} \label{eq4}
B=\frac{\alpha_H}{\tau_g} \cdot \left( \frac{d\ln{ | r_0(\omega) |}}{d\omega}    \right).
\end{equation}
In Eqs.~\ref{eq3} and \ref{eq4}, $\tau_g=2n_g L_g/c$ denotes the roundtrip time in the gain section, and $\phi_0(\omega)$ the additional optical phase accumulated by light when travelling forth and back through the feedback arm.  

Term $A$ can be interpreted as the ratio between the optical length of the laser cavity extension  and the optical length of the gain section. Physically, the term describes the factor by which the photon lifetime of the laser is increased by the additional travel time through the extended cavity, with regard to the roundtrip time through the solitary diode gain element. It should be noted that the presence of resonators in a Vernier filter increases the optical length of the feedback arm by a factor that grows linearly with the number of roundtrips through each resonator. To give an example, we consider a resonator of geometrical length $L_r$ and effective group index $n_{eff}$. For simplicity we assume that the ad and drop ports are separated by half a roundtrip, $L_r/2$, and that losses are much smaller than the power coupling coefficient at the add and drop ports, $\kappa^2$. Then, at resonance, the optical length of the resonator becomes multiplied with a roundtrip factor of $M=1/2+(1-\kappa^2)/\kappa^2$, i.e., the  effective optical length of the resonator becomes $n_{eff}L_r \cdot M$. As a consequence, $A$ is biggest, and the length-related linewidth reduction via $F$ in Eq.~\ref{eq1} is strongest, if the laser frequency is resonant with the Vernier filter frequency. 

The term $B$ describes the presence of an additional linewidth reduction mechanism based on gain-index coupling as expressed by Henry's factor. However, we note that, due the factor $\alpha_H$ in Eq. \ref{eq4}, the $B$-term based linewidth reduction can only be present, if $\alpha_H$ is nonzero, i.e., if the laser linewidth is already broadened by gain-index coupling [term $(1+\alpha_H^2)$ in the numerator of Eq.~\ref{eq1}]. $B$ is biggest at the rising edge of the Vernier filter’s reflection peak, where $d\ln{| r_0(\omega) |}/d\omega$ is positive. The effect can be described as a negative optical feedback mechanism, where making the resonator loss steeply frequency dependent compensates for spontaneous emission-induced index and frequency changes~\cite{vahala_1984APL, komljenovic_2017AS}. Similarly, also the intensity noise can be reduced with frequency dependent loss~\cite{newkirk_1991JQE}.

To make an optimum choice of parameters when considering the effects that determine the laser linewidth, there are two main routes to reduce the linewidth. The first and most effective one is to increase the photon lifetime and thus the phase memory time of the resonator. However, as the intrinsic loss in diode laser amplifiers is high, often higher than 90\% in double pass due to the typically very large values of $\gamma_g \approx 10^3/m$, the light in an extended cavity diode laser essentially performs only a single roundtrip before it is lost. Increasing the photon lifetime can thus not be achieved with increasing the reflectance of the feedback circuit, $R_0$. Instead, an optically long feedback arm, $L_f \gg L_g$, is required. Via a large value of $d \phi_0(\omega)/ d \omega$, the feedback essentially works as a double-pass optical delay line, similar to a delay line in an optoelectronic oscillator~\cite{tang_2018OE}. This approach is expected to yield an approximately quadratic reduction of linewidth vs. increasing length, provided that optical loss in the feedback (and thus also in the Vernier filter) does not dominate the laser cavity roundtrip loss. The second route to a narrower linewidth is increasing the laser intracavity power, specifically the power in the gain section, which means that $P_b$ needs to be increased (or its co-factor in the denominator of Eq.~\ref{eq1} by more power at the other laser ports, e.g., by rising $P_f$). Higher intracavity power improves the ratio of phase preserving stimulated emission over randomly phased spontaneous emission. With a given laser cavity design, the roundtrip loss is given, such that increasing the power requires stronger pumping. Via this route Eq.~\ref{eq1} predicts a linewidth narrowing inversely with increasing output power, i.e., in proportion with $X=P_p/P_{\textrm{th}}-1$, where $P_p$ is the pump power and $P_{\textrm{th}}$ is the threshold pump power.

\begin{figure}[tbp] \label{Bandwidth_extrapolation}
	\centering
	\includegraphics[width=0.6\linewidth]{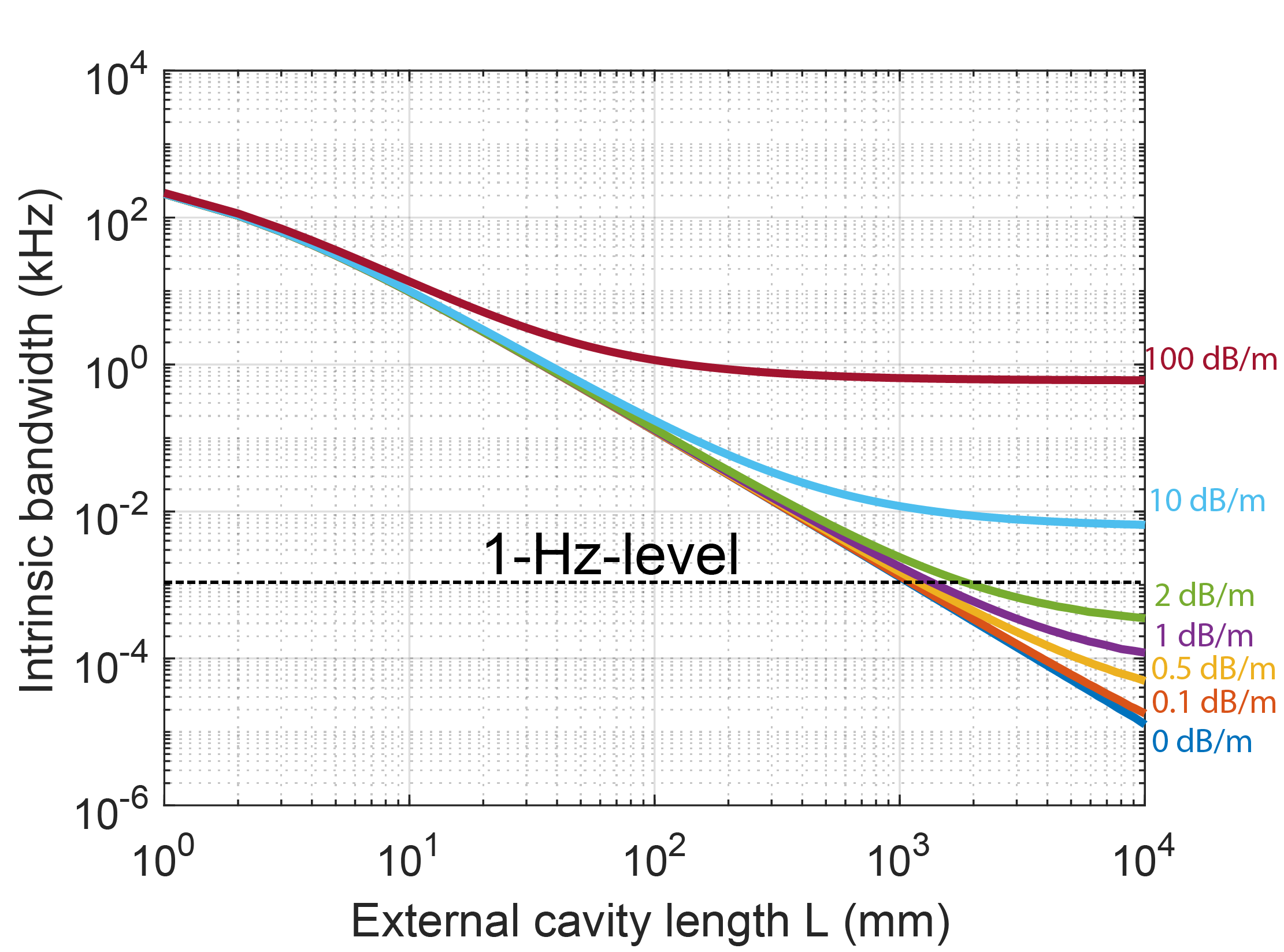}
	\caption{\label{fig:Bandwidth_extrapolation} Intrinsic laser linewidth of hybrid integrated InP-Si$_3$N$_4$ lasers, calculated as function of the single-pass optical length of the cavity extension, $L_f$, using Eq. \ref{eq1}. The length extension includes that light performs  multiple passes through the resonators of a Vernier filter circuit. See Table~\ref{tab:parameters} for the value of the parameters used in the calculation. The actual amount of feedback from the extended cavity arm to the gain section, i.e., the value of $T_f$, depends on the waveguide loss constant $\gamma_{F}$ in the feedback circuitry, which we have varied between zero and 100~dB/m. Similar values can be found in \cite{fan_2016PJ}.}
\end{figure}
\begin{table}[tbp]
	\centering
	\caption{Overview of the parameters used in calculating the intrinsic linewidth of hybrid integrated InP-Si$_3$N$_4$ lasers shown in Fig.~\ref{fig:Bandwidth_extrapolation}.}
	\begin{tabular}{c|l|c|c}
		\textbf{parameter}   & \textbf{description} & \textbf{value} & \textbf{unit} \\
		\hline
		$\lambda$            & wavelength & 1.55 & $\mu$m \\
		$P_{b}$              & output power & 1.0 & mW \\
		$\alpha_{H}$         & linewidth enhancement factor & 5 &  \\
		$\eta_{\textrm{sp}}$ & spontaneous emission factor & 2.0 & \\
		$L_{g}$              & Length gain section & 700 & $\mu$m \\ 
		$R_b$                & Power reflection back facet & 0.9 & \\
		$R_f$                & Power reflection loop mirror & 0.5 & \\
		$T_{c}$              & mode coupling loss & 0.9 & \\
		$n_{g}$              & group index gain section & 3.6 & \\
		$n_{f}$              & group index Si$_3$N$_4$ section & 1.715 \\
		$\gamma_{g}$         & loss gain section & 13 & cm$^{-1}$ \\
	\end{tabular}
	
	\label{tab:parameters}
\end{table}
To give a quantitative estimate on what intrinsic linewidth values can be expected with low-loss waveguide feedback circuits, such as with using Si$_3$N$_4$ circuits, Fig.~\ref{fig:Bandwidth_extrapolation} presents a prediction of the linewidth vs. the optical length of the cavity extension using Eq.~\ref{eq1}. The parameters used in the calculation are given in Table~\ref{tab:parameters}. To provide a conservative estimate, and in order to avoid discussing specifically designed Vernier transmission spectra for each feedback length, the calculations are performed with setting $B$ to zero. This corresponds to the laser frequency tuned to the center of the Vernier resonance, such as for maximizing the laser power. If taking $B$ into account, via proper tuning to exploit the the mentioned negative optical feedback, a factor in the order of $\alpha_H^2$ narrower linewidth may still be achieved. This would require a proper fine-tuning of the laser frequency to the low-frequency side of the Vernier resonance, for instance, with an adjustable phase section in the feedback circuit or with a pump current fine-tuning to provide a phase shift in the gain section. The calculations show that feedback circuits with less than 2~dB waveguide loss and being 1~m long promise linewidths as narrow as a few Hertz. Such loss and length requirements appear realistic, when comparing with previously demonstrated values. The lowest propagation loss observed in Si$_3$N$_4$ waveguides is below 0.1~dB/m~\cite{bauters_2011OE}. Meter-sized and highly frequency selective coupled-resonator circuits have been realized as well with the Si$_3$N$_4$ platform~\cite{taddei_2018PTL}, such that reaching a 1-Hz-linewidth seems possible with a dedicated laser design.     

In the following we present a set of recent examples of hybrid integrated InP-Si$_3$N$_4$ diode lasers that we have fabricated and characterized, in order to give an overview on current and future options for versatile on-chip light sources. 

\section{Hybrid lasers with two microring resonators and single gain section} \label{section4}

\begin{figure}[tbp] \label{packaged_laser}
	\centering
	\includegraphics[width=0.4\linewidth]{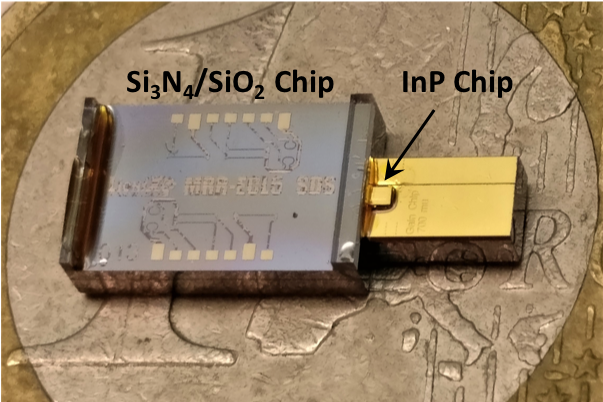}
	\includegraphics[width=0.4\linewidth]{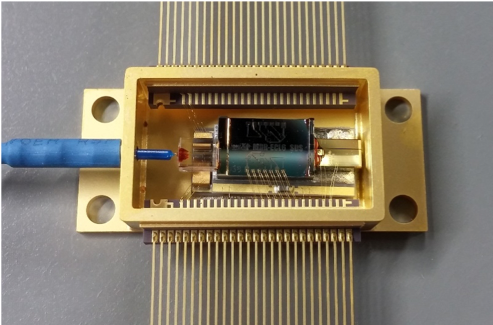}\\
	\includegraphics[width=0.55\linewidth]{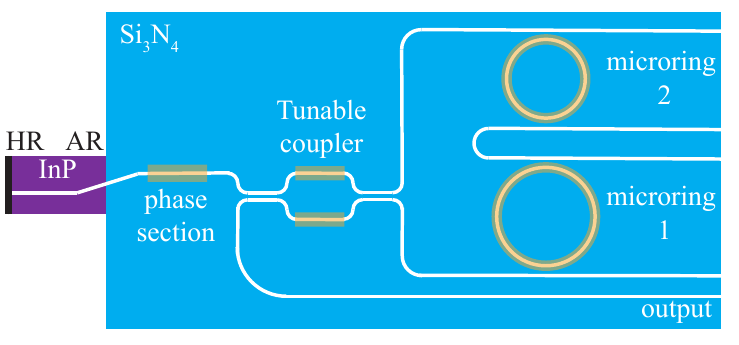}
	\caption{\label{fig:packaged_laser} (\textbf{upper left}) Photograph of the integrated Si$_3$N$_4$ and InP chips in comparison with a one-Euro-coin. (\textbf{upper right}) The hybrid integrated laser packaged into a standard butterfly housing. The generated light leaves the Si$_3$N$_4$ waveguide via a single-mode, polarization maintaining fiber. (\textbf{lower}) Schematic view of a two-ring hybrid laser. Heaters are indicated by the orange color. The output fiber is butt-coupled to the output port.} 
\end{figure}

A schematic view of a InP-Si$_3$N$_4$ hybrid laser comprising a single gain section and a Vernier filter consisting of two microring resonators is shown in Fig~\ref{fig:packaged_laser}.
For stable operation, such hybrid lasers are usually assembled in a butterfly package as shown in Fig.~\ref{fig:packaged_laser}. The package contains a Peltier element and thermistor for temperature control and stabilization of the laser chip. The bond pads on the chip are wire bonded to the butterfly pins for electrical access. Single-mode polarization maintaining fibers are attached to the output waveguides. The fiber is terminated with an angled facet FC/APC connector to prevent undesired reflections back into the laser. 
The lasers are operated after mounting on printed circuit boards that provide  multi-channel USB-controlled voltages and currents to the laser. LabVIEW or Python interfaces simplify retrieving measurement data and enable a systematic and reproducible characterization of the lasers' properties. If required, software feedback loops can be programmed that automatically optimize the laser output during parameter sweeps.

\begin{figure}[tbp] \label{24mW_turnon_Albert}
	\centering
	\includegraphics[width=0.5\linewidth]{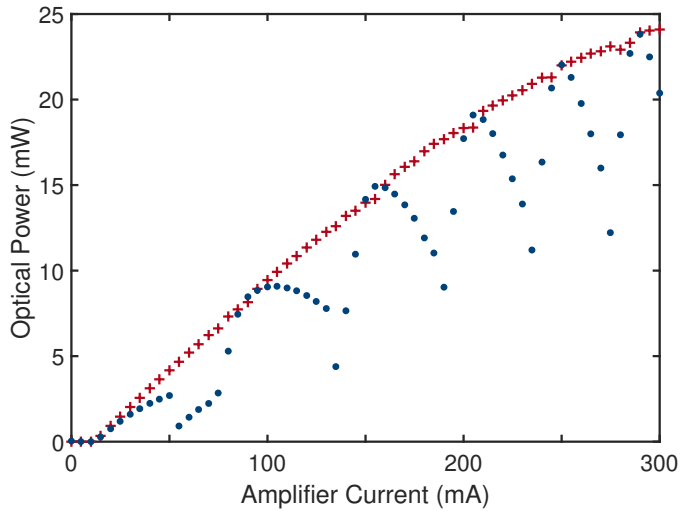}
	\caption{\label{fig:24mW_turnon_Albert} Fiber-coupled output power of a hybrid integrated InP-Si$_3$N$_4$ laser measured as function of the amplifier current (blue circles). The laser wavelength is set to 1576~nm via a Vernier filter formed by two tunable microring resonators. The tunable output coupler is set to 80\% outcoupling. With automatically maximizing the output via a phase shifter between gain section and Vernier filter (auto tuning), the output power is steadily increasing with pump current (red crosses).}
\end{figure}

In Fig.~\ref{fig:24mW_turnon_Albert} the fiber-coupled output power of a laser with two microring resonators is shown as function of the amplifier current. The Vernier filter, having a free spectral range (FSR) of 50 nm, was set to a wavelength of 1576~nm, which is near the optimum settings for this laser.  This particular laser, which is shown schematically in Fig.~\ref{fig:packaged_laser}, possesses a tunable output coupling between the gain section and Vernier filter, realized as a tunable Mach-Zehnder interferometer. In addition, the cavity length can be adjusted with a $2\pi$-phase shifter located between gain section and Vernier filter. When only increasing the amplifier current, while keeping all other laser parameters constant, the output power shows an overall increase which is, however, interrupted by power drops (blue dots). These power drops are likely initiated by a rise of temperature in the gain waveguide with increasing pump current~\cite{buus_2005BOOK}, leading to a change in refractive index. This tunes the overall laser cavity length and eventually brings the oscillating cavity mode out of resonance with the Vernier filter, seen as a power drop. With further increasing the pump current, a next cavity mode comes into Vernier resonance (longitudinal mode hop) which increases the output power again. The described mechanism involves a hysteresis because the index of the gain section is intensity dependent due to gain-index coupling, and because changing the optical power levels changes also the thermal conditions. 

To obtain a continuously increasing output power, we use an automatic readjustment of the optical cavity length by adjusting the phase section for maximize output power. With the automatic phase tuning turned on, the laser output is seen to increase approximately continuously with pump current (red crosses). With the investigated laser we measure a maximum fiber-coupled optical power of 24 mW, which is more than the previous reported values of 1.7~mW~\cite{fan_2016PJ}, 7.4~mW~\cite{oldenbeuving_2013LPL}, and 10~mW~\cite{lin_2018PJ} obtained with similar lasers. When increasing the output coupling from zero to 100\%, the threshold current increases from 8 to 19~mA, and the slope efficiency increases from zero to 0.13~mW/mA.

\begin{figure}[tbp] \label{120nm_tuning_Albert}
	\centering
	\includegraphics[width=0.8\linewidth]{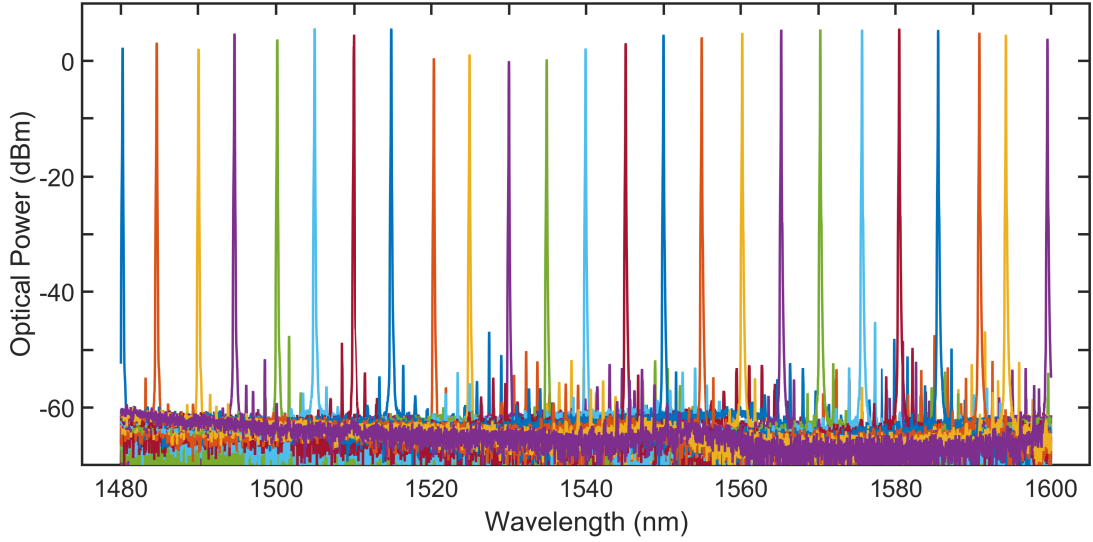}
	\caption{\label{fig:120nm_tuning_Albert} Spectral coverage (tuning range) of the hybrid integrated laser obtained with stepwise tuning the Vernier filter, followed by adjustment of the phase section for maximum power. The output coupling was set to about 80\% and the pump current to its maximum value of 300~mA. The individual spectra are recorded with 0.1~nm resolution bandwidth as measured with the OSA. The measurements show a spectral coverage of 120~nm, which is the widest range achieved for hybrid or heterogeneously integrated InP lasers.}
\end{figure}

To demonstrate that the laser can cover a broad spectral bandwidth with single-frequency output, Fig.~\ref{fig:120nm_tuning_Albert} shows a series of superimposed laser output spectra recorded with an optical spectrum analyzer. The individual, single-frequency spectra are obtained by tuning both resonators in the Vernier filter. In the example shown here the wavelength steps are approximately 5~nm. The side mode suppression ratio is as high as 63~dB, measured with 0.01~nm resolution near 1550~nm wavelength. The broadest tuning range is observed with the amplifier set to its specified maximum current of 300~mA. We note that the Vernier FSR of 50~nm would normally limit laser operation to a 50~nm wide interval as well, after which the output wavelength would hop back to the beginning of the interval. However, we note that also the output coupler is spectrally dependent, and that this dependence can be tuned. We made use of this extra tunability to extend the spectral coverage by more than a factor of two, to a range of 120~nm. This exceeds the so far widest range of 75~nm obtained with a monolithically integrated InP laser~\cite{latkowski_2015PJ} and also the 110-nm range obtained with a heterogeneously integrated InP laser~\cite{tran_2020JSTQE}, while also providing an order of magnitude more power at the edges of the tuning range.

For determining the intrinsic linewidth, we measured the power spectral density (PSD) of frequency noise with a high-finesse resonator that is slowly locked to the average laser wavelength (HighFinesse LWA-1k-1550). Frequency noise spectra display the squared and averaged frequency excursions with regard to the average frequency versus     the radio frequency, $f$, at which they occur. Slow frequency excursions are usually largest, and become smaller with increasing frequency, often with approximately a $1/f$-law~\cite{stephan_2005PRA}, also named flicker noise or technical noise. At high noise frequencies, the spectrum flattens off to a certain white noise level~\cite{didomenico_2010AO}. The height of the white-noise level is proportional to the intrinsic laser linewidth with a factor of $\pi$ if the spectrum is measured single-sided, and with a factor of $2\pi$ for double-sided spectra~\cite{stephan_2005PRA, llopis_2011OL}. To obtain the lowest linewidth, we used a low-noise current source (ILX Lightwave LDX-3620). For the PSD measurement, the laser output was set to 10 mw at 1550 nm, with the phase section set to maximize the output. 

\begin{figure}[tbp] \label{PSD_2ring_Albert}
	\centering
	\includegraphics[width=0.5\linewidth]{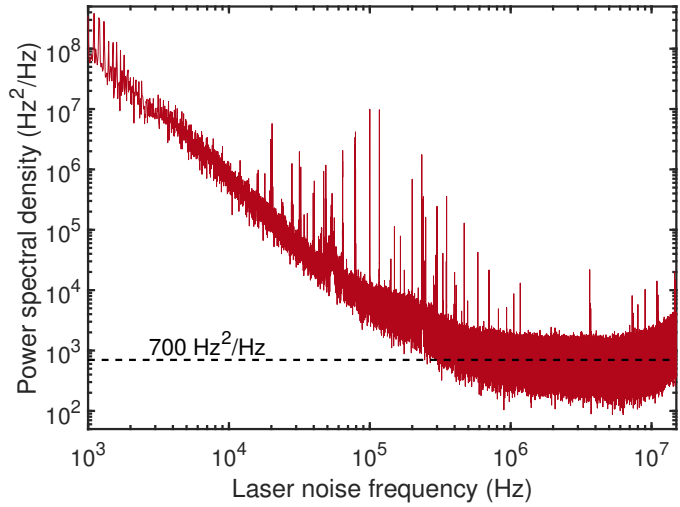}
	\caption{\label{fig:PSD_2ring_Albert} Single-sided frequency noise power spectral density of a hybrid integrated laser with two Si$_3$N$_4$ microring resonators as Vernier filter. The spectrum is measured at a wavelength of 1550~nm with a pump current of 300~mA. Taking the average noise between 1.3 and 3.5~MHz as the upper limit for the white noise level of the laser, we obtain a white-noise level of  700~Hz$^2$/Hz (dashed line). This value, via multiplying with $\pi$, corresponds to an intrinsic laser linewidth of 2.2~kHz.}
\end{figure}
Figure~\ref{fig:PSD_2ring_Albert} shows the measured frequency noise spectrum, displaying $1/f$-noise and levelling off at 700$\pm$200~Hz$^2$/Hz beyond 1~MHz noise frequency. Spurious narrowband peaks can be observed, which we address to RF-pickup. The intrinsic linewidth determined from the upper limit of the white-noise part in the spectrum is 2.2$\pm$0.7~kHz. This value is clearly smaller than the 24-kHz linewidth reported before for a similar hybrid laser with 2 ring resonators~\cite{fan_2014SPIE}. We address this mainly to the higher laser power (24~mW vs. 4.7~mW). We note that, meanwhile, we achieve as the fiber coupled output from the described type of laser reliably a power above 40~mW, and sometimes above 50~mW.

\section{High power hybrid integrated lasers with two gain sections}

Basically all applications of integrated lasers would benefit from increasing the available output power. An obvious advantage lies in easier overcoming the pump threshold of integrated nonlinear oscillators, e.g., parametric oscillators such as Kerr comb generators, or Brillouin lasers. Another advantage of higher laser power is that the signal-to-noise ratio in the after detection increases proportional to the optical power, because the RF signal power increases quadratically with the optical power whereas the RF shot noise power increases linearly. Therefore, higher output power enables, for instance in sensing, to increase the fundamental sensitivity or speed of detection. Similarly in fiber communications and microwave photonics, the ultimate (quantum limited) signal-to-noise power ratio of RF signal transmission through analog photonic links increases in proportion with the optical power~\cite{yariv_1990OL}.

In addition to the named fundamental noise, of which the influence can be reduced via increased power, lasers often show excess noise, i.e., power fluctuations caused by technical perturbations. A standard measure to quantify the total noise is the so-called relative intensity noise, RIN, which entails measuring the average power fluctuation divided by the average power. The importance of reducing RIN is given by the circumstance that all optical measurements, e.g. also of wavelength or linewidth, are finally based on photodetection where RIN forms a limiting factor~\cite{morton_2018JLT}. Because the RF powers belonging to RIN and to a signal both grow quadratically with the optical power, whereas the shot noise power grows only proportionally, the effect of RIN becomes ultimately domninant. In this case, the noise can only be reduced with reducing the RIN of the laser, which underlines the importance of lasers with low RIN. Only if RIN is not dominant, the signal to noise ratio can be increased with increasing the power, and the transition between RIN and shot noise determines the maximum useful power. Optimum is thus to realize RIN as low as shot noise at maximum power.  

Hybrid and heterogeneous integrated diode laser are very attractive for integration in photonic circuits. However, even if offering ultra-narrow linewidth, such lasers have so far been limited to an output in the order of 25~mW, and also the RIN-levels should be reduced. Here, we present, a hybrid integrated diode laser with so far the highest output power, and with a RIN-level close to the shot-noise (quantum) limit.

\begin{figure}[tbp] \label{dual_gain_circuit_Joern}
	\centering
	\includegraphics[width=0.75\linewidth]{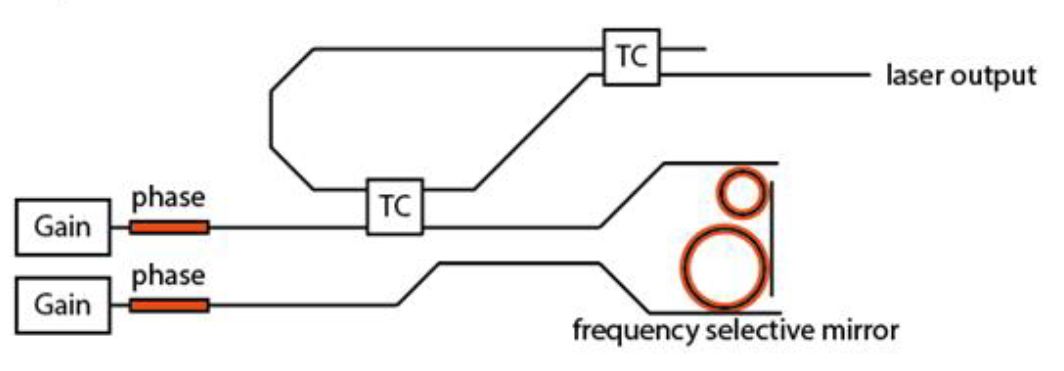}
	\includegraphics[width=0.6\linewidth]{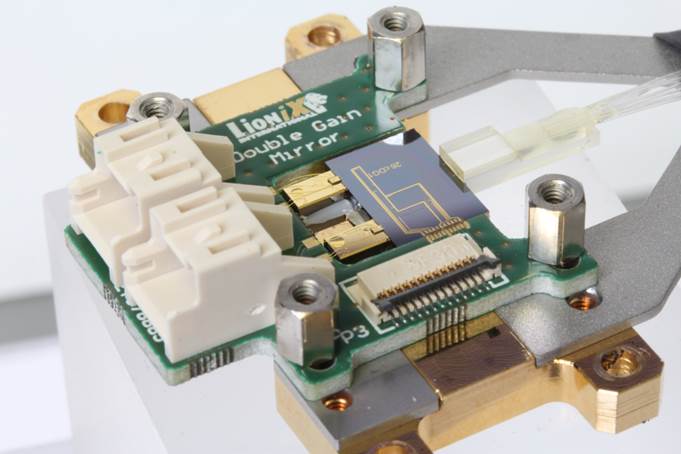}
	\caption{\label{fig:dual_gain_circuit_Joern} (\textbf{upper}) Functional design of the dual-gain laser waveguide circuitry. Two gain sections with HR coated back facets form the two ends of the laser cavity. The intracavity Vernier filter (frequency selective mirror) is passed twice per cavity roundtrip. Two tunable couplers (TC) divert the laser output to a single output port. (\textbf{lower}) Photograph of a dual-gain laser.}
\end{figure}
The functional design of the waveguide circuit of the laser is shown in Fig.~\ref{fig:dual_gain_circuit_Joern}. To increase the output power, two 700~$\mu$m long prototype semiconductor amplifiers are used, one at each end of the laser cavity. The HR coated back facets of the gain elements form the two cavity end mirrors. A Si$_3$N$_4$ waveguide circuit is used for low-loss extension of the cavity length by multiple roundtrips through two micro-resonators in Vernier configuration (FSR 208~GHz and 215~GHz). The Vernier filter, used as intracavity frequency selective mirror, is passed twice per cavity roundtrip which yields a longer cavity length and sharper spectral filtering in comparison to using a Vernier filter inside a loop mirror. The bi-directional output from a tunable Mach-Zehnder output coupler is superimposed into a single output waveguide with a second tunable coupler. The pump current to the gain sections as well as the thermo-optically controlled tuning of the ring resonators and couplers can be individually adjusted. The output is coupled to a standard polarization maintaining fiber with a coupling loss of 0.5~dB.

\begin{figure}[tbp] \label{dual_gain_output_power}
	\centering
	\includegraphics[width=0.5\linewidth]{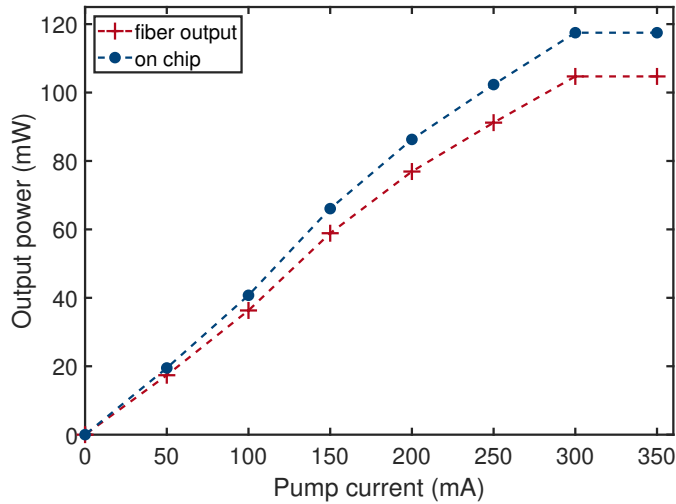}
	\caption{\label{fig:dual_gain_output_power} Output power of hybrid integrated diode laser with two gain sections (dual-gain laser) measured versus the pump current. Data points are connected with dashed lines in order to guide the eye. The maximum fiber coupled output is 105 mW, which corresponds to 117 mW at the on-chip output port.}
\end{figure}
Fig.~\ref{fig:dual_gain_output_power} shows the output power of this novel type of laser measured versus the total pump current. When applying a pump current of 300~mA to both gain sections, we achieve a maximum fiber coupled output power of 105~mW. This corresponds to an on-chip power of 117~mW. To our knowledge these values are the highest power ever achieved with a hybrid or heterogeneously integrated diode laser~\cite{Epping_2019LC}. Comparing with the output from a single gain section integrated with a similar Si$_3$N$_4$ feedback waveguide circuit shows that using two gain sections doubles the output power. Measuring the output versus tuning of the Vernier filter we observe more than 70~mW of fiber coupled output across a 100~nm wide range (from 1470 to 1570~nm), with a side mode suppression ratio of more than 50~dB.

For characterizing the noise properties of the laser we measured frequency noise and intensity noise. Amplitude spectra of frequency noise are measured at frequencies of up to 30~MHz (HighFinesse LWA-1k-1550). We note that beyond 10 to 20~MHz, the spectra are dominated by electronic noise and thus cannot be reliably addressed to laser noise. Relative intensity noise spectra are recorded by sending the laser output power to a fast photodiode and recording the signal with a 25-GHz-RF spectrum analyzer.

\begin{figure}[tbp] \label{dual_gain_noise}
	\centering
	\includegraphics[width=0.45\linewidth]{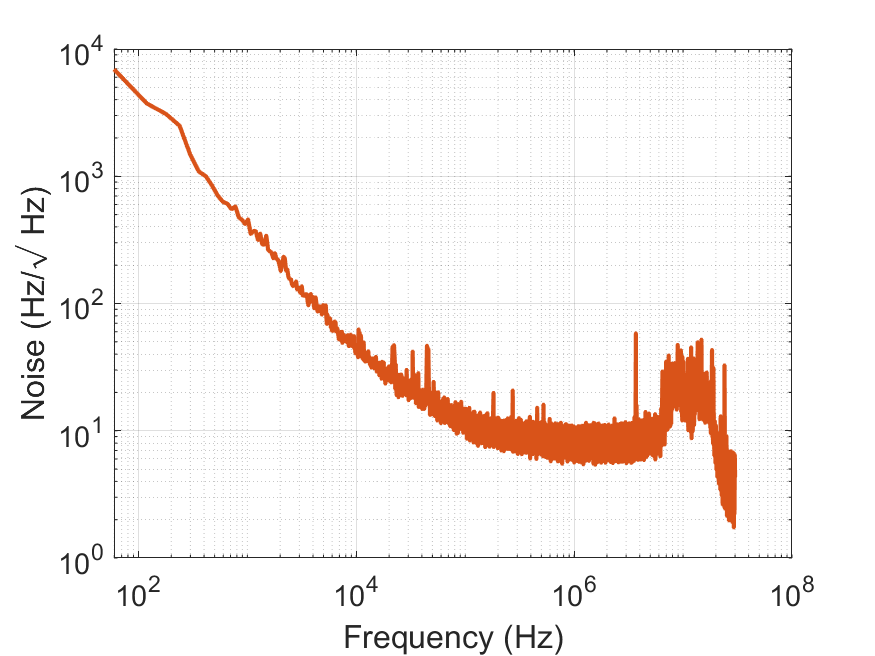}
	\includegraphics[width=0.45\linewidth]{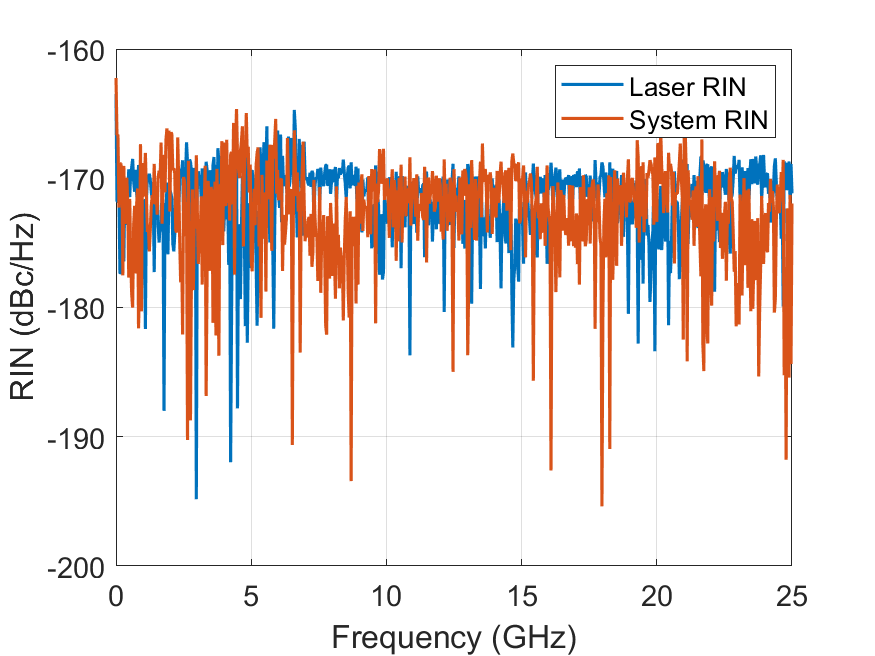}
	\caption{\label{fig:dual_gain_noise} (\textbf{left}) Single-sided frequency noise amplitude spectrum of a dual-gain hybrid integrated laser. The upper limit for the white noise level between 1 and 2 MHz corresponds to an intrinsic linewidth of 320~Hz. (\textbf{right}) The lowest relative intensity noise (RIN) is about -170~dBc/Hz, which is close to the electronic background of detection and also close to the calculated quantum (shot noise) limit of -172~dBc/Hz.}
\end{figure}

Fig.~\ref{fig:dual_gain_noise} (left panel) shows a frequency noise spectrum recorded with 100~mA pump current to both gain sections (approximately 40~mW output power). This specific current was chosen because here the laser shows single-frequency oscillation near the gain maximum without the need to apply additional tuning voltages across the waveguide heaters. Turning off the heater voltages was found to reduce pickup noise from the heater drivers. The intrinsic linewidth corresponding to the upper limit of white noise in the spectrum is about 320 Hz.

Compared to the linewidth of a laser with a single gain section described above, we address the 7-times lower linewidth to two main differences. The first is that the laser power is about 4-times higher, which should yield an 4-times narrower linewidth. The second difference is that the Vernier filter is not used as end mirror but as intracavity filter. In this case, with each laser cavity roundtrip, the light has to pass twice through the  Vernier filter, which doubles the effect of cavity length extension. From the resonator-based part of length extension, given that the resonators have the same lengths and coupling coefficients (10\%) as in the single-gain laser, we estimate that the dual-gain laser possesses a factor of 1.3 longer optical roundtrip length. In Eq.~\ref{eq3} this corresponds to a factor 1.3 larger frequency dependence of the phase shift, which yields a factor of 1.8 linewidth reduction in Eq.~\ref{eq2}, giving a total linewidth reduction by a factor of 6.4 in Eq.~\ref{eq1}, which is in reasonable agreement with the experimental linewidth ratio. A dependence of the linewidth upon fine-tuning via the $B$-term in Eq.~\ref{eq4}, as seen in si-InP lasers~\cite{tran_2020JSTQE} is currently being investigated.

The measured RIN spectrum is shown as the blue trace in the right panel of Fig.~\ref{fig:dual_gain_noise}. It can be seen that the noise is very low, near the electronic background noise (orange trace). The optical noise is generally at the level of -170~dBc/Hz, except for noise values near -165~dBc/Hz in smaller intervals below 7 GHz. For comparison we calculate the shot-noise limited RIN from $S_{I,\textrm{sn}}(f)=(2h\nu)/P_0$ for a laser power of $P_0$=40~mW  at a light frequency of $\nu$=193 THz, from which we obtain a value of -172~dBc/Hz. The comparison shows that the laser intensity noise is within a few dB of the shot-noise level, i.e., the intensity noise is approximately as low as the fundamental quantum limit and almost free of technical noise.

Summarizing these experimental data, hybrid integrated lasers with two gain sections appear very promising for delivering a so far unmatched combination of highest optical power, ultra-low linewidth and lowest intensity noise near the quantum limit. Such type of hybrid integrated lasers therefore look very promising for on-chip optical carrier generation in integrated microwave photonics. A thorough investigation of detailed properties and implementation in photonic circuits is underway.

\section{Hybrid integrated laser with record-low linewidth} \label{section6}

In the previous sections, intrinsic linewidths below 2.2~kHz and 320~Hz are reported with single and dual-gain lasers, respectively. Further line narrowing should be possible with further increasing the laser intracavity power. However, this would require to integrate even more or stronger gain sections while, according to Eq.~\ref{eq1}, one expects line narrowing only inversely proportional with power. A somewhat more attractive option is to extend the cavity length because of two reasons. First, making use of a given power, it can solely be based on extending the passive part of the laser cavity, i.e., the effective length of the Si$_3$N$_4$ circuitry. Second, the linewidth narrowing follows a steeper law, i.e., a quadratic decrease of linewidth with increasing cavity length, via the $F^2$-factor in the denominator in Eq.~\ref{eq1}. Obtaining a quadratic decrease requires, however, that the active and passive roundtrip loss, expressed by $\gamma_m$ and $\gamma_{\textrm{tot}}$, do not increase too much with increasing cavity length. The regime of quadratic reduction of linewidth with cavity length can also be noticed in Fig. \ref{fig:Bandwidth_extrapolation} as a negative slope of magnitude 2, until the overall loss in the cavity extension (expressed by $1-R_0$) becomes relevant compared to the intrinsic loss as expressed by $1-R_i$.

In pursuing this strategy we follow up an earlier version with a prototype gain element and 290~Hz linewidth~\cite{fan_2017CLEO}. The improved version presented here~\cite{fan_2019ARXIV} uses a slightly more powerful gain element, however, the main difference to the lasers discussed in the previous sections is an about 10-times longer optical cavity roundtrip length of $\approx$~0.5 m on the feedback chip. The 1000~$\mu$m long diode amplifier carries a 90\% reflective coating at its back facet and is optically coupled with a low-loss Si$_3$N$_4$ circuit. The circuit comprises three cascaded microring resonators, each equipped with  10\% power couplers. The microring resonators possess radii of $R =$ 99, 102~$\mu$m (average FSR 278~GHz, quality factor $Q~\approx 2,000$), and $R=1485$~$\mu$m (FSR 18.6~GHz, $Q~\approx 290,000$). The waveguide end mirror is formed by a Sagnac loop mirror, such that the three micro resonators are passed twice per laser cavity roundtrip. For low-noise pumping of the gain section we use a battery-driven power supply (ILX Lightwave, LDX-3620).

Regarding the coarse operation parameters, the laser shows similar properties as the lasers with two microring resonators. The threshold pump current is about 42~mA and a maximum fiber coupled output power from the Sagnac output port is 23~mW at a pump current of 320~mA. The spectral coverage of the laser with more than 1~mW single-frequency output is wider than 70~nm, with a side mode suppression higher than 60~dB. Thermo-optic tuning by acting on the heater of the ring resonators can be done via longitudinal mode hops in steps of 2~nm and 0.15~nm, which are the FSRs of the small and big microring resonators. Fine tuning can be achieved either with small changes of the diode pump current, by acting on the heaters of the Mach-Zehnder coupler for the Sagnac loop mirror, or with a phase section between the microring resonators and the gain section. 

\begin{figure}[tbp] \label{three_ring_linewidth}
	\centering
	\includegraphics[width=0.6\linewidth]{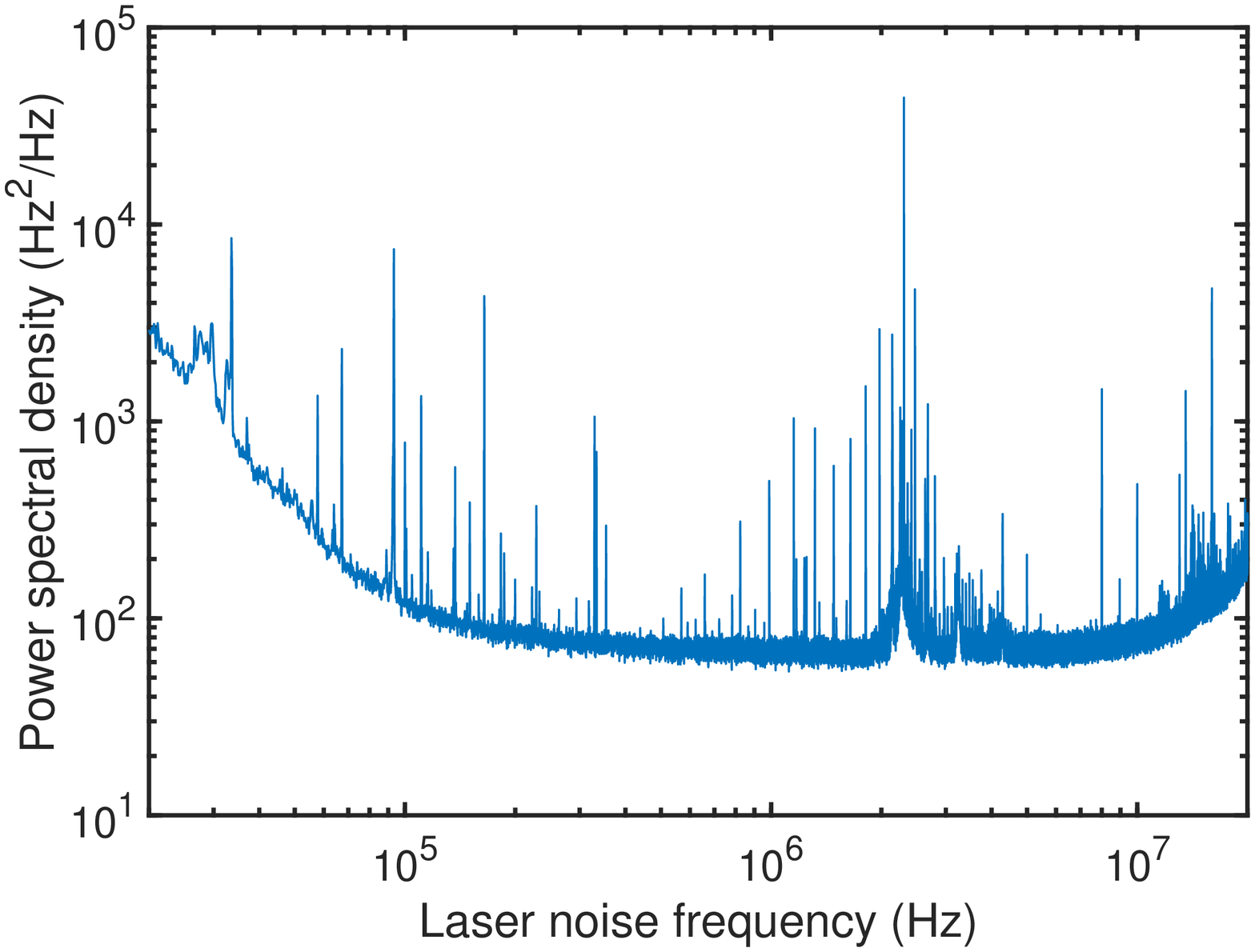}
	\includegraphics[width=0.6\linewidth]{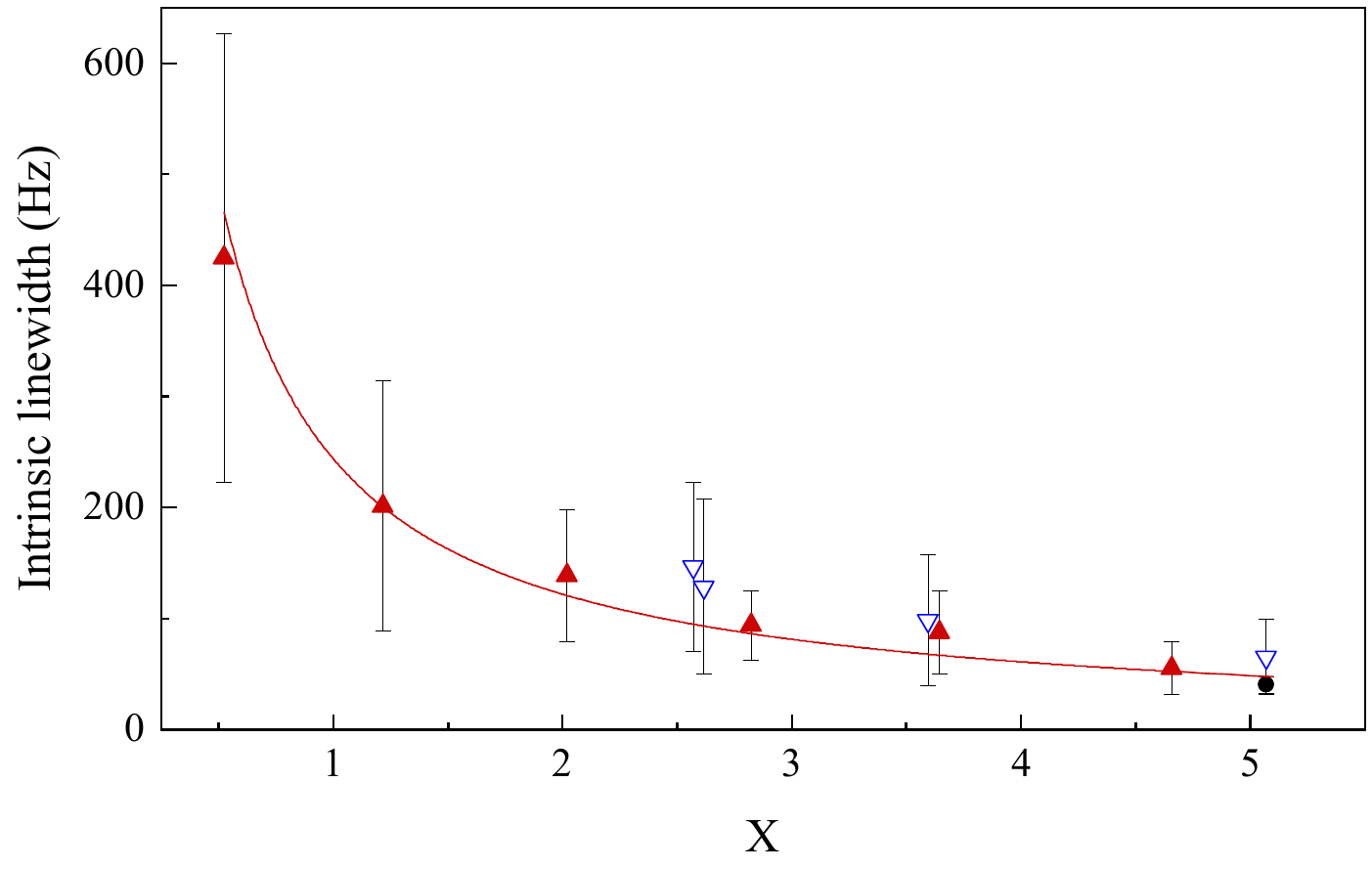}
	\caption{\label{fig:three_ring_linewidth} (\textbf{upper}) Example of a single-sided power spectrum of frequency noise recorded with a hybrid integrated laser with three microring resonators, operating at an output power of 3~mW. The maximum white noise level of 65~Hz$^2$/Hz between 1 and 2~MHz corresponds to an intrinsic linewidth of about 210 Hz. (\textbf{lower}) Lorentzian linewidth component measured as function the laser output power which is proportional to the pump threshold factor, $X$. The solid curve shows the theoretically expected inverse dependence on the output power, according to Eq.~\ref{eq1}. The lowest linewidth achieved with this laser is 40~Hz, measured with a pump current of 255~mA.}
\end{figure}
Figure~\ref{fig:three_ring_linewidth} summarizes spectral linewidth measurements using delayed self-heterodyne detection performed with two independent setups. The first uses a Mach-Zehnder interferometer with 5.4~m optical arm length difference, a 40-MHz acousto-optic modulator, and two photodiodes for balanced detection. The beat signal is recorded versus time and analyzed with a computer to obtain the power spectral density of frequency noise. The second uses an arm length difference of 20~km and an 80-MHz modulator. Here the time-averaged power spectrum of the beat signal is recorded with an RF spectrum analyzer. The beat spectra resemble Voigt profiles, where the Lorentzian linewidth component is given by the intrinsic white-noise component in the spectral power density~\cite{didomenico_2010AO}. We obtain the intrinsic component with Lorentzian fits to the wings of the measured RF line where the Lorentzian shape is minimally obstructed, i.e., avoiding the low-frequency noise regime near the line center, as well as the range close to the electronic noise floor. Linewidth measurements are carried out at various different pump currents at a wavelength near the center of the gain spectrum.

The upper panel of Fig.~\ref{fig:three_ring_linewidth} displays an example of a frequency noise spectrum, recorded at relatively low output power of 3 mW (85 mW pump current). The spectrum shows a number of narrow peaks due to RF pickup and, after levelling off at about 65 Hz$^2$/Hz, shows a slight rise, likely due to electronic amplifier noise. In spite of the relatively low laser power, the corresponding intrinsic linewidth is rather narrow, 210~Hz, which is thus clearly the effect of a long resonator providing a long photon lifetime.
The lower panel of the figure shows how the laser linewidth decreases with increasing pump power. The latter is expressed as increasing threshold factor, which specifies the normalized pump power above threshold, $X$, which is proportional to the laser output power. The various symbols indicate measurements during up and down scans of the pump current (filled and unfilled symbols, respectively). The smallest linewidth obtained from the frequency noise spectrum recorded at the highest pump current of 255~mA~\cite{fan_2019ARXIV}) is about 40~Hz, indicated by the round filled symbol at $X=5.17$. The power dependent linewidth is in good agreement with the theoretically expected decrease with laser power, following a $1/X$-power dependence (red fit curve).

An important observation in Fig.~\ref{fig:three_ring_linewidth} is that the measured linewidth narrowing does not show any saturation with output power, while other lasers often display a lowest linewidth-value, or even an linewidth increase vs. power~\cite{vilenchik_2015PROC, santis_2018PNAS, melnik_2006OE, redlich_2017JSTQE, andreou_2019}. A possible explanation for down-scalability of the linewidth with power, here and even more so with the dual-gain laser, is the absence of noticeable nonlinear effects in the laser cavity, specifically, in the Si$_3$N$_4$ feedback resonators where the power is highest. For instance, we estimate~\cite{fan_2019ARXIV} that, with output powers of a few tens of mW, several Watt of power can be present in the microring resonators, which corresponds to intensities of several hundred MW/cm$^2$. Such intensities would lead to significant nonlinear loss in other waveguide materials, specifically in semiconductors due to a much smaller bandgap. In the Si$_3$N$_4$ platform, such losses are many orders of magnitude lower~\cite{kruckel_2017OE}. Stimulated Brillouin scattering (SBS) is another effect that can manifest as a nonlinear loss in the laser feedback circuit. This process is mediated by optoacoustic interactions in a medium, and has been observed in various waveguide platforms including silica~\cite{gundavarapu_2019NPhot}, silicon~\cite{vanlaer_2015NP}, and silicon nitride~\cite{gyger_2019ARXIV}. The strength (or the intrinsic gain) of SBS is mainly dictated by the material properties, including refractive index, acoustic damping, and photoeleastic constant, as well as the optoacoustic overlap of the waveguiding structure~\cite{eggleton_2019NP}. This intrinsic gain is very small in the silicon nitride waveguide geometry used in the feedback circuit. When compared to silicon the SBS intrinsic gain of silicon nitride is approximately 130 times lower, so the SBS effect is negligible in the feedback circuit.

The absence of noticeable nonlinear loss opens the interesting potential for power based linewidth narrowing to an extent that is not possible with hybrid or heterogeneous integration of silicon-photonic circuits, or with fully monolithic semiconductor lasers. Summarizing this section, the intrinsic linewidth of 40~Hz, as is also plotted in Fig.~\ref{fig:overview_linewidth_ref} as uppermost and latest data point, is the smallest value ever measured with any hybrid or heterogeneously integrated diode laser. We conclude that further upscaling of the resonator length to the order of meters on a chip~\cite{taddei_2018PTL} with simultaneously increased power appears very promising for approaching the 1-Hz linewidth level.

\section{Dual-wavelength, multi-wavelength and visible wavelength lasers} \label{section DMVlasers}

So far we have described work on hybrid integrated lasers that provide a continuous-wave output in the form of a spectrally narrowband, single optical frequency with constant power. However, there is highest interest also in multi-frequency sources, so-called optical frequency combs or mode-locked light sources, specifically, for dual-comb sensing~\cite{coddington_2016O}, metrology~\cite{udem_2002N}, coherent optical communications \cite{pfeifle_2014NPhot}, and microwave photonics~\cite{marpaung_2019NP}. Similarly, dual-wavelength lasers are of great importance for optical generation and distribution of high-purity microwave and THz radiation, for communication, sensing and metrology~\cite{pillet_2008JLT}. Finally, hybrid lasers with high coherence and wavelength tunable output will find numerous applications when realizing them in various different wavelength ranges. For instance, improved time-keeping on board of satellites requires narrow linewidth integrated lasers at a larger variety of wavelengths in the infrared and visible. For instance, operating a Sr lattice clock~\cite{takamoto_2005N} requires a narrow linewidth at 698 nm and also at further transitions to provide excitation, re-pumping or trapping. Other applications for narrow linewidth visible lasers on a chip will be found in quantum technology and sensing~\cite{wicht_2017SPIE, maze_2008N, noelleke_2018SPIE, reimer_2018FoO}. Classical sensing benefits as well from visible narrow linewidth sources, such as cavity-enhanced Raman detection~\cite{wang_2019ASR} of gases. In the following we report some of our experimental progress and preparations on hybrid integrated diode comb lasers, dual-frequency diode lasers and hybrid integrated lasers for the visible range.

\subsection{Diode comb lasers}
For exploiting the full potential in applications, there are two central requirements regarding the coherence of comb sources. The first is a highly equidistant spacing of the comb lines with fixed mutual phasing. This is usually fulfilled without additional effort, because mutual phase locking via injection locking through nonlinear sideband generation is what underlies all mode-locking mechanisms. The second requirement is that the spectral linewidth of the individual comb lines
has to be extremely narrow, preferably in the kHz range or below. This corresponds to a low jitter in time-resolved detection, and is also what enables coherent multi-heterodyne (e.g., dual-comb) measurements with phase sensitivity and maximum signal-to-noise ratio~\cite{coddington_2008PRL}.

Most attractive candidates for applications are chip-based diode laser frequency combs~\cite{wang_2017LSA}, due to their direct excitation with an electric current. However, diode laser combs usually fail to meet the requirement for narrowband comb lines. Just as with single-frequency lasers, the reason is a short photon lifetime due to a short cavity length, high optical roundtrip loss, and strong gain-index coupling. With monolithically integrated diode lasers, even with an extended cavity length, the linewidths typically remain in the MHz-range~\cite{cheung_2010PTL}.

In terms of cavity lifetime and thus the intrinsic linewidth, Kerr combs form a highly promising alternative to diode laser combs, especially since their recent hybrid integration with diode pump lasers~\cite{stern_2018N, raja_2019NatCom} within the same integrated photonic circuit. The reason for a long cavity lifetime is low roundtrip loss in Kerr resonators, due to fabrication with a dielectric (large electronic bandgap) waveguide platform. This provides a narrow linewidth of the individual comb frequencies and also a much wider spectral coverage than with lasers. On the other hand, in Kerr comb oscillators the photon lifetime cannot be extended much with a longer cavity length, because the oscillation threshold goes up with the mode volume. The main disadvantage of Kerr comb oscillators, compared to diode laser combs, is essentially the introduction of an additional pump threshold and a generally higher complexity. In order to bypass the latter, chip-based frequency comb lasers with an extended cavity have already been investigated in the form of heterogeneously integrated mode-locked lasers~\cite{srinivasan_2014FO,davenport_2018PR}. The narrowest intrinsic linewidth reported so far for a passively mode-locked and heterogeneously integrated InP-Si laser is 250~kHz~\cite{wang_2017LSA}.

\begin{figure}[tbp] \label{comb_laser_general}
	\centering
	\includegraphics[width=0.45\linewidth]{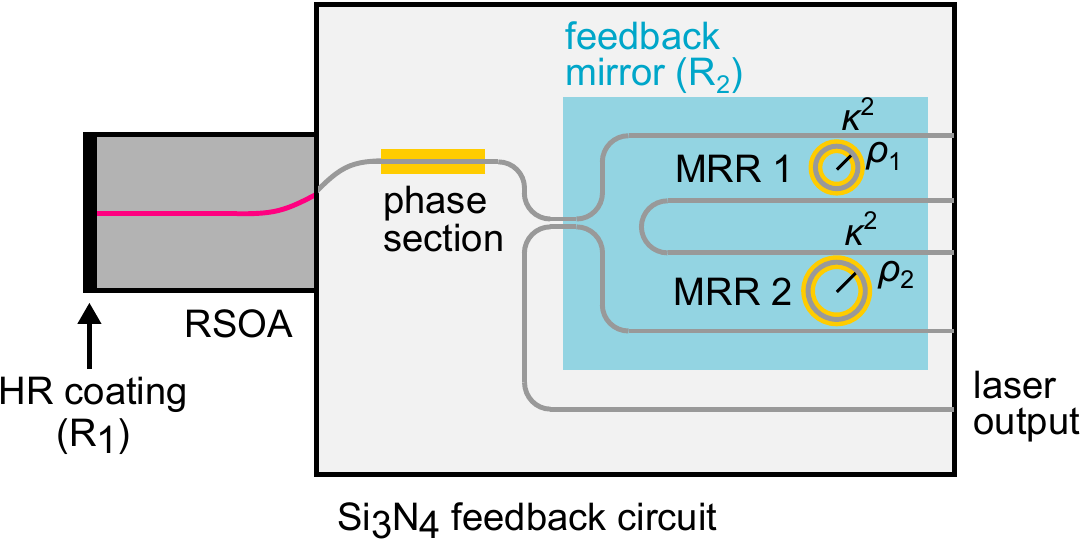}
	\includegraphics[width=0.45\linewidth]{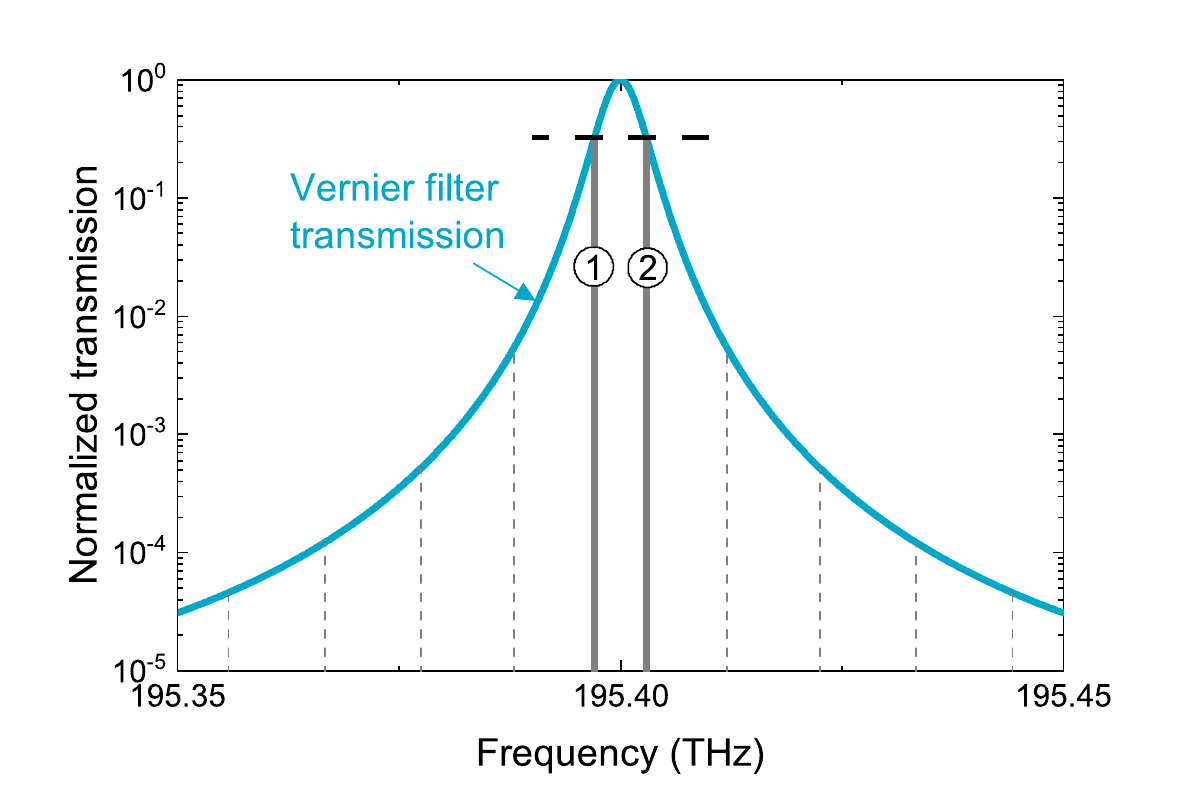}
	\caption{\label{fig:comb_laser_general} (\textbf{left}) Schematic diagram of the hybrid waveguide laser. The back facet of the gain element (RSOA) is HR coated for a reflectance $R_1$=90\%. The Si$_3$N$_4$ circuit contains a phase section, and a Vernier feedback circuit (blue shaded area) with an effective power reflectance $R_0$ based on microring resonators (MRRs) with radii $\rho_1$=136.5 $\mu$m and $\rho_2$=140.9 $\mu$m and power coupling coefficient $\kappa^2$=10\%. The phase section and the MRRs can be tuned using resistive electric heaters. (\textbf{right}) Calculated transmission spectrum of the Vernier filter (solid blue line) and center frequencies of the longitudinal laser cavity modes (gray solid and dashed lines). The  mode frequencies are tuned via the phase section to establish equal transmission through the Vernier filter for two neighboring modes (1) and (2). Nonlinear generation of optical beat frequencies is then found to generate a frequency comb.}
\end{figure}
In order to provide narrower linewidth in diode laser combs we have investigated comb generation with a hybrid integrated InP-Si$_3$N$_4$ laser as shown in Fig.~\ref{fig:comb_laser_general}~\cite{mak_2019OE}. We use a standard low loss Si$_3$N$_4$ feedback circuit to increase the photon lifetime and thereby decrease the intrinsic linewidth of the individual comb frequencies. With essentially the same basic circuitry as was presented in Sect.~\ref{section4} for single-frequency generation, we extend the optical cavity roundtrip length to approximately 6~cm. However, to generate a frequency comb with mutually phase-locked phases, we adjust the phase section for achieving equal transmission through the Vernier filter for two neighboring modes as depicted in the right panel of Fig.~\ref{fig:comb_laser_general}. Once the laser oscillates at these two modes simultaneously through well-balanced roundtrip losses, nonlinear mixing in the semiconductor gain section generates further optical sidebands. The newly generated sidebands are amplified in the laser gain, establishing a frequency comb.

It is important to note in Fig.~\ref{fig:comb_laser_general} that the calculated center frequencies of the cold cavity modes are not exactly equidistant. This is due to the dispersion of the transmission resonance of the microrings. In Eq.~\ref{eq3} this is expressed as non-linear frequency dependence of the roundtrip phase through the Vernier filter, which corresponds to a frequency dependent cavity length. If the laser would just display multi-mode oscillation with random mutual phasing, one would observe a distribution of different beat frequencies in the laser output due to the non-equidistant cold-cavity modes. On the contrary, if the laser is mode locked, i.e., having mutually phase-locked modes, this would be seen as a single beat frequency, due to a uniform (equidistant) spacing of the light frequencies. 

\begin{figure}[tbp] \label{comb_spectra}
	\centering
	\includegraphics[width=0.45\linewidth]{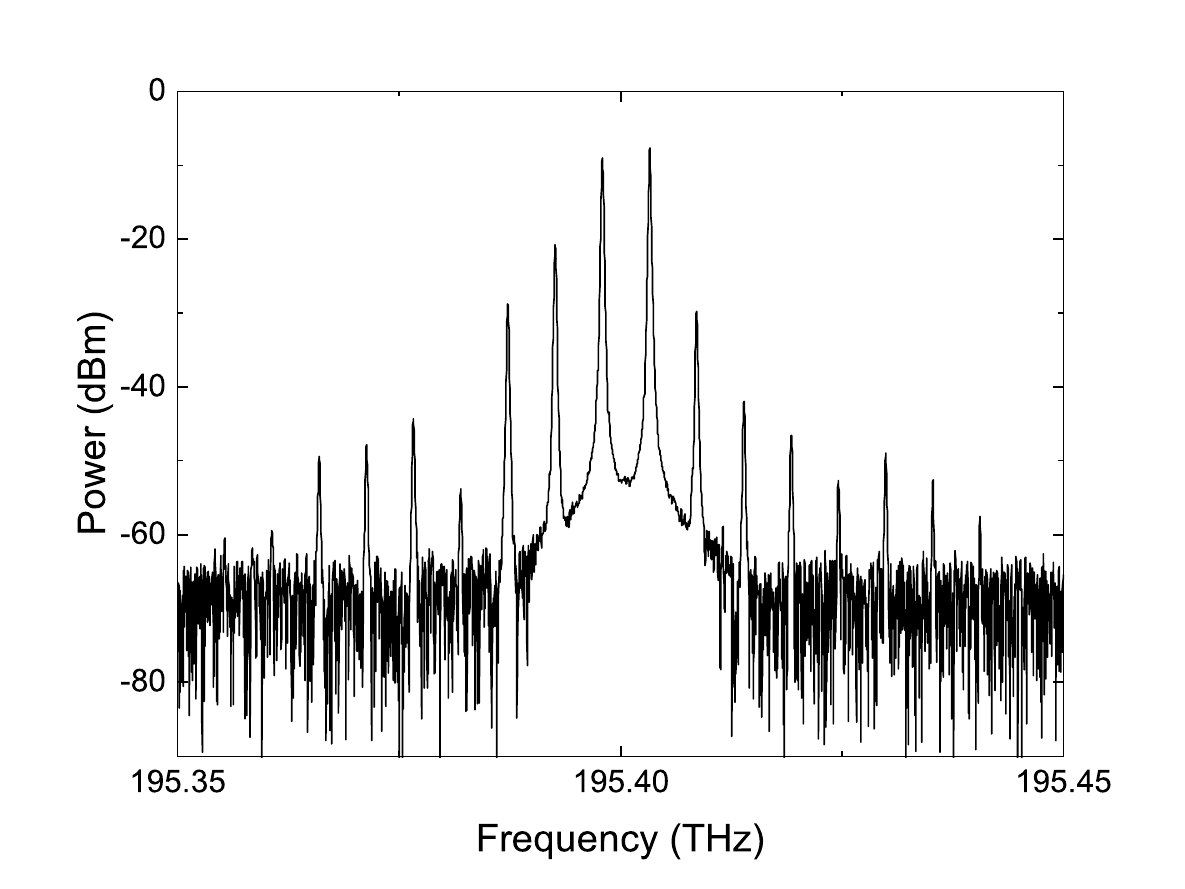}
	\includegraphics[width=0.45\linewidth]{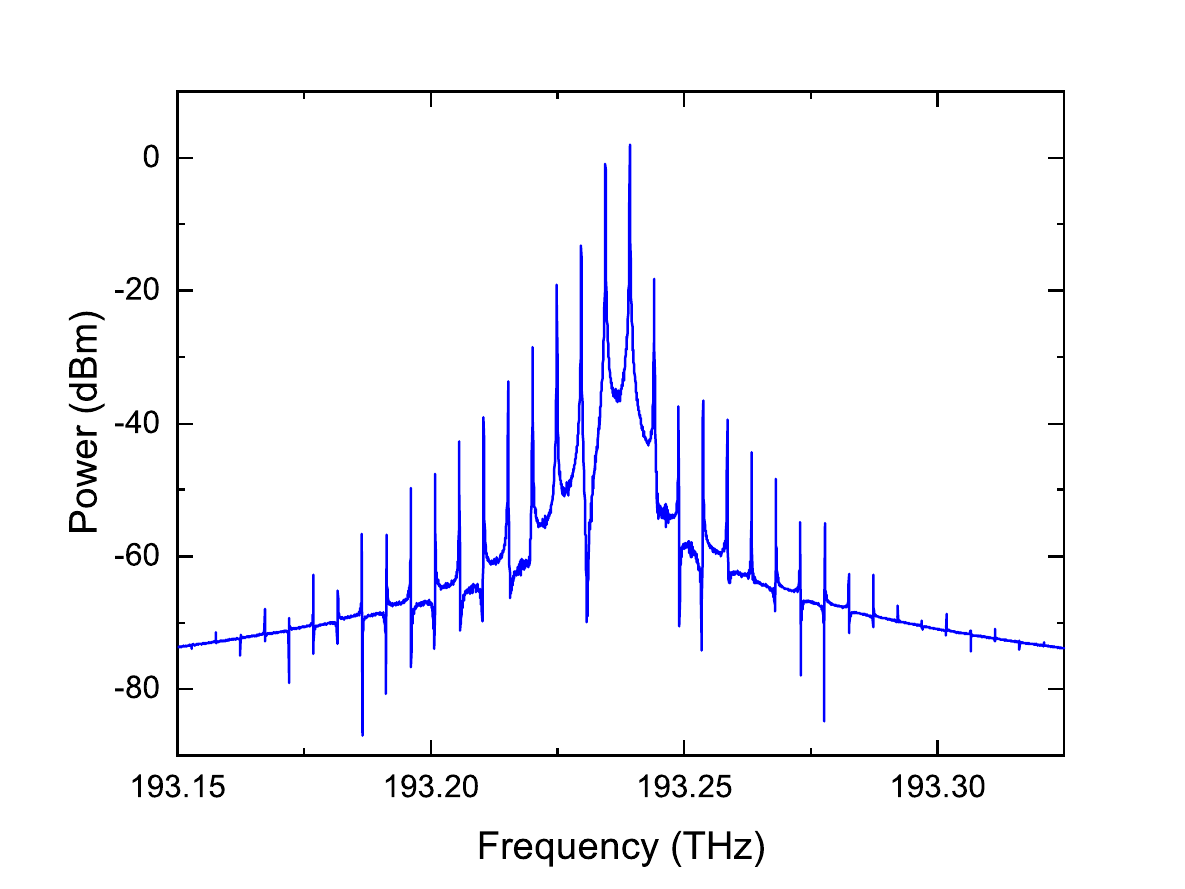}
	\caption{\label{fig:comb_spectra} (\textbf{left}) Frequency comb spectrum generated with a hybrid-integrated InP-Si$_3$N$_4$ laser. The total fiber-coupled output power is 2~mW. (\textbf{right}) Calculated frequency comb spectrum using the experimental parameters as listed in the appendix of \cite{mak_2019OE}. The calculation is based on a transmission line model to represent the spatial, spectral and temporal distribution of light and charge carriers in the gain section, while using an analytically calculated, complex-valued field amplitude reflectivity spectrum to model the feedback circuit \cite{fan_2017OE}.}
\end{figure}
Fig.~\ref{fig:comb_spectra} shows a measured comb spectrum (left panel) comprising 17~lines. The lines are strictly equidistant with a spacing of 5.5~GHz within the optical resolution of the spectrum. The right panel displays a calculated spectrum, obtained with a tranmission line model~\cite{vpi_2019} as described in~\cite{fan_2017OE}. It can be seen that there is good agreement with the experimental data. To verify the equidistance of the experimental comb lines more precisely, we recorded the RF mode beating with a fast photodiode and a RF spectrum analyzer. The measurements show a single and narrowband RF frequency at around 5.5~GHz which corresponds to the beating of directly neighboring modes, and shows narrowband harmonics of the beat due to beating of modes with non-direct neighbors. The single fundamental beat frequency shows a narrow intrinsic linewidth of approximately 18 kHz. We recall as described with Fig.~\ref{fig:comb_laser_general} that the absence of mode-locking would generate multiple fundamental beat frequencies due to the non-equidistant spacing of the cold cavity. Having observed a single, narrow linewidth fundamental beat frequency proves an equal spacing of the generated light frequencies with high precision, i.e., it confirms that the generated frequency comb is mode-locked (phase-locked).

At this point one may wonder why laser oscillation off the center of the cold cavity modes is possible here. The reason is the high laser cavity roundtrip loss, which imposes a wide band width on the cold cavity mode. As described in Sect.~\ref{section3}, and to more detail in Ref.~\cite{fan_2019ARXIV}, the laser roundtrip losses are high. This is mainly due to the high intrinsic waveguide loss in semiconductor amplifiers, and due to loss caused by integration with a different waveguide platform, as was expressed as a low effective reflectance $R_i$ in Fig.~\ref{fig:hybrid_laser_general}. Our estimates show that $R_i$ is rather small, in the order of a few percent, with approximately 98\% of the light lost per roundtrip. Calculating the according FWHM spectral bandwidth of the cavity then yields values larger than the mode spacing. This is what enables mode-locking even far off the mode center frequencies.

Having demonstrated frequency comb generation, the central point and main motivation of the investigation was to narrow the intrinsic linewidth of single comb lines via an extended photon lifetime gained by hybrid integration with a low-loss Si$_3$N$_4$ circuit. To measure the linewidth of single comb lines, we performed beat measurements between the hybrid laser and an independent reference laser. As reference laser we used an extended cavity laser (SANTEC TSL-210) with an intrinsic linewidth of 6 kHz. This value was determined with delayed self-heterodyne measurements as described in Sect.~\ref{section6}. The beat measurements with the hybrid laser yielded very narrow intrinsic linewidths of the individual comb lines, with an average value as small as 34 kHz. This values is a factor 7 lower than the previously smallest linewidth for any chip integrated frequency comb diode laser~\cite{wang_2017LSA}.

We note that the observed linewidth already approaches similar values as Kerr combs pumped by narrow-linewidth diode lasers~\cite{stern_2018N}. In that sense, the approach to control frequency comb diode lasers with low-loss hybrid integrated circuits bears much promise, because one can hope to reach ultra-narrowband linewidth of the comb lines similar to the 40-Hz-level described in Sect.~\ref{section6} and further progressing towards the 1-Hz-level as extrapolated in Fig.~\ref{fig:Bandwidth_extrapolation}. On the other hand, hard challenges are to be faced. A first challenge is that cavity extension via microring resonators does not allow to generate wide combs due to sharp frequency selection. A possible path towards broader comb spectra is using modified feedback circuits with a spectrally flattened transmission. Another challenge lies in the circumstance that extending the cavity for linewidth narrowing reduces the cavity mode spacing and thus lowers the generated RF beat frequencies. While this is convenient for detection with low-speed electronics equipment, certain applications have much stronger interest in increasing the mode beating frequency into the GHz and THz range. This might require to restrict oscillation to only a few modes at large spectral distance, in spite of dense cavity mode spacing.

\subsection{Dual-wavelength lasers}
In order to investigate such scenarios we are currently investigating hybrid lasers with two separately adjustable Vernier filters to provide, in a first step, dual-wavelength sources for generation of microwave and THz signals. Dual-wavelength sources have been investigated extensively using a large variety of different approaches. This includes rare-earth-doped bulk solid state lasers~\cite{pillet_2008JLT} and fiber lasers, the latter yielding linewidths of the microwave beat frequency in the order of 80~kHz~\cite{chen_2006TMTT}. Aiming on applications where size, weight and power consumption are highly important, rare-earth-doped waveguide lasers have widely been explored as well~\cite{grivas_2016PQE}. However, these lasers require optical pumping which introduces additional complexity, whereas semiconductor lasers operate with direct electronic pumping.

In order to synchronize the frequency fluctuations at the two output wavelengths for providing a narrow linewidth of the beat, there was early work on DBR lasers with gratings containing two spatial periods, operated with a single gain section. There was, however, no report on the linewidth of the individual laser wavelengths or the beat frequency output~\cite{iio_1995PTL}. A dual wavelength laser at 1.35 $\mu$m was reported based on two transverse gratings and where the modes spatially overlap to a degree that increases with power\cite{pozzi_2006PTL}. Indeed, with FWHM linewidths of 60~MHz for the individual lasers the beat showed common-mode noise rejection seen as the beat linewidth reducing from 140 to 40~MHz with increasing power. High power output of more than 70~mW was demonstrated with two separate Bragg lasers, i.e., with two gratings surrounding two gain sections and the output combined with a Y-junction and being subsequently amplified~\cite{price_2007PTL}. However, the intrinsic linewidth of the individual lasers was not narrower than 900~kHz. A monolithically integrated dual-wavelength DBR laser in the 1.3~$\mu$m range was realized for THz~generation~\cite{kim_2013LPL}. Individual laser linewidths between 6 and 9~MHz were measured, depending on where the laser operates within its few-nanometer tuning range. The linewidth of the THz radiation was not given, likely, because it is very difficult to measure electronically at high THz frequencies. It remained thus open whether operation in the same cavity, with the same gain section and a dual-period gratings had synchronized the optical frequency fluctuations for a line narrowing of the THz~signal below the individual optical linewidths. At lower beat frequencies around 100~GHz, the lowest FWHM linewidth of beat frequencies were generated with monolithically integrated lasers based on arrayed waveguide gratings (AWG) and reached  250~kHz~\cite{carpintero_2012OL} and 56~kHz~\cite{guzman_2014URSI}. The work that comes closest to our own investigations is the recent realization of two hybrid integrated diode lasers with the same Si$_3$N$_4$ waveguide chip, although with separate feedback circuits~\cite{zhu_2019OE}. The lasers described here are based on two separate semiconductor gain sections, InP and GaAs, to obtain simultaneous operation at two largely different wavelengths, near 1.5 and 1~$\mu$m. A promising application would be driving difference-frequency generation at 3 to 5~$\mu$m wavelength in a compact format, such as for mid-IR molecular fingerprint detection. The intrinsic spectral linewidths of the individual lasers were measured as 18 and 70~kHz, respectively. 

\begin{figure}[tbp] \label{dual_wavelength_circuit}
	\centering
	\includegraphics[width=0.6\linewidth]{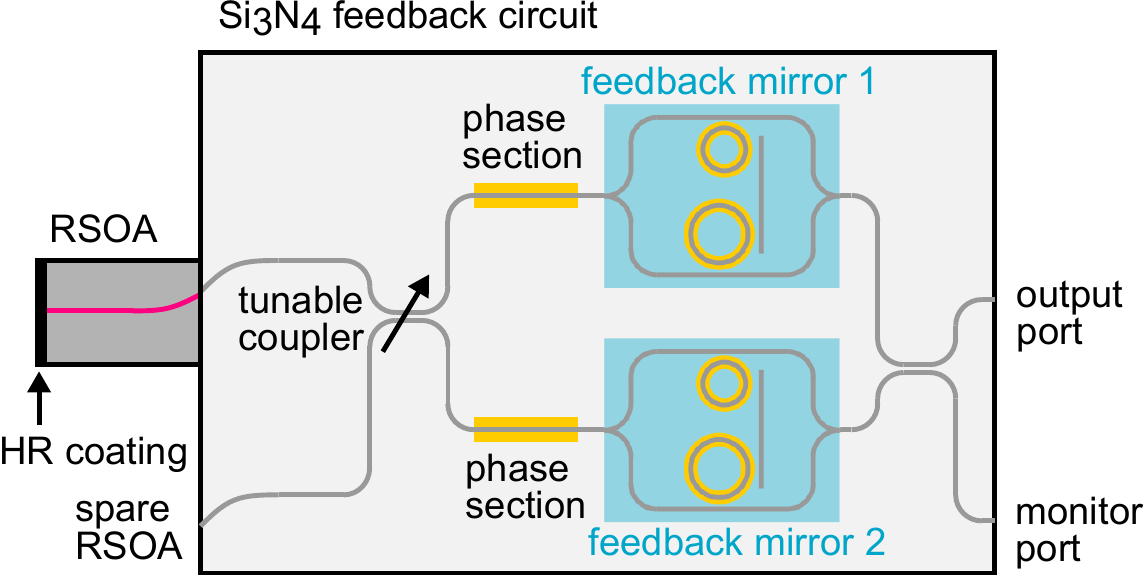}
	\caption{\label{fig:dual_wavelength_circuit} Schematic waveguide design of the hybrid integrated dual-wavelength laser. Two separately tunable Vernier mirrors are used for feedback to the same gain section. A tunable coupler allows to adjust the relative strength of feedback from the two Vernier mirrors, in order to maintain dual-wavelength oscillation in spite of gain competition.}
\end{figure}
In our work in progress, we realized a dual-wavelength laser based on dual Vernier feedback with a single gain section as shown in Fig.~\ref{fig:dual_wavelength_circuit}. Two equally dimensioned Vernier feedback circuits, each equipped with two tunable microring resonators are used to initiate laser oscillation at two widely and independently tunable wavelengths. Such tuning may also involve modulation of one or both of the wavelengths. The same gain chip is used for amplification at both wavelengths. This aims on synchronizing the influence of index fluctuations on the respective cavity lengths, i.e., to increase the common-mode noise rejection that reduces the linewidth of the microwave beat frequency. In order to counteract spectral condensation to a single wavelength via gain competition, the relative strengths of feedback from the Vernier circuits can be adjusted with a tunable Mach-Zehnder coupler. The superimposed output can be monitored at two exit ports.

\begin{figure}[tbp] \label{dual_spectra}
	\centering
	\includegraphics[width=0.45\linewidth]{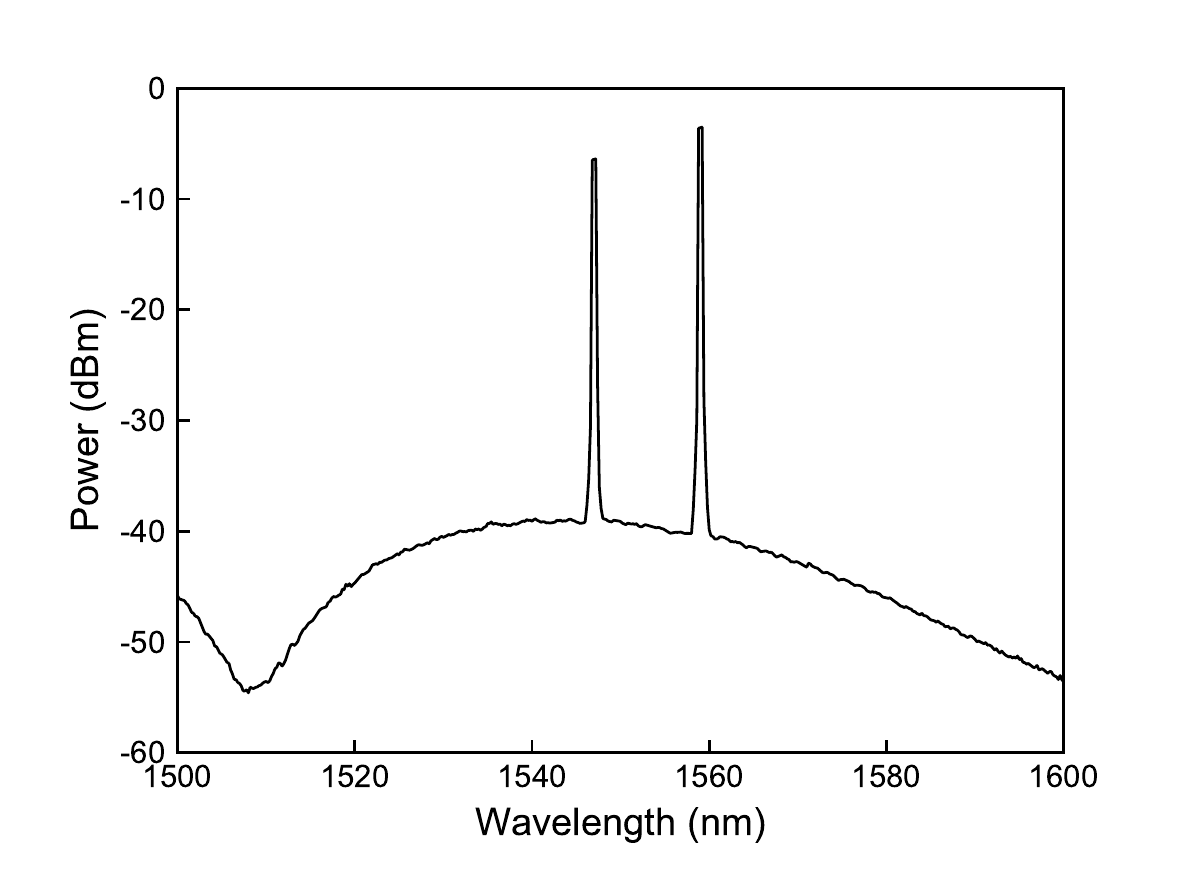} \\
	\includegraphics[width=0.45\linewidth]{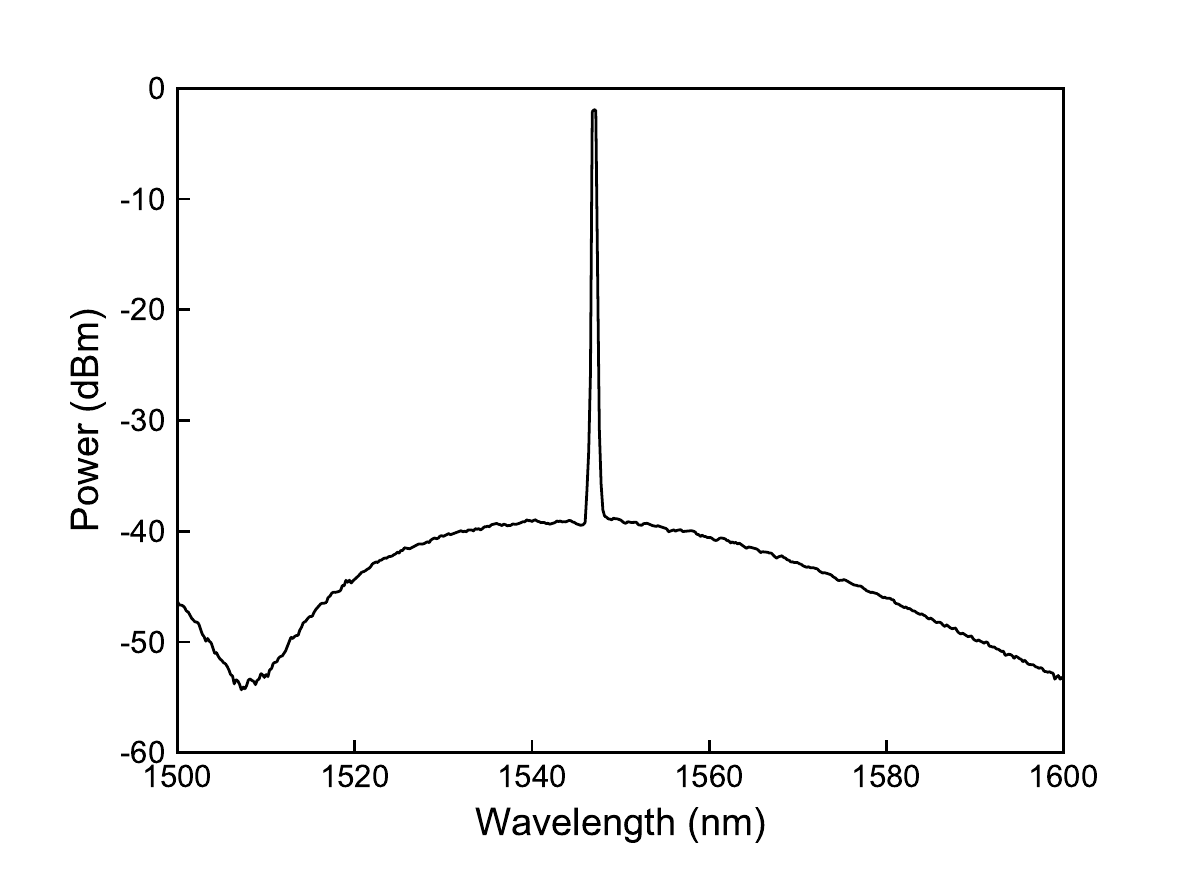}
	\includegraphics[width=0.45\linewidth]{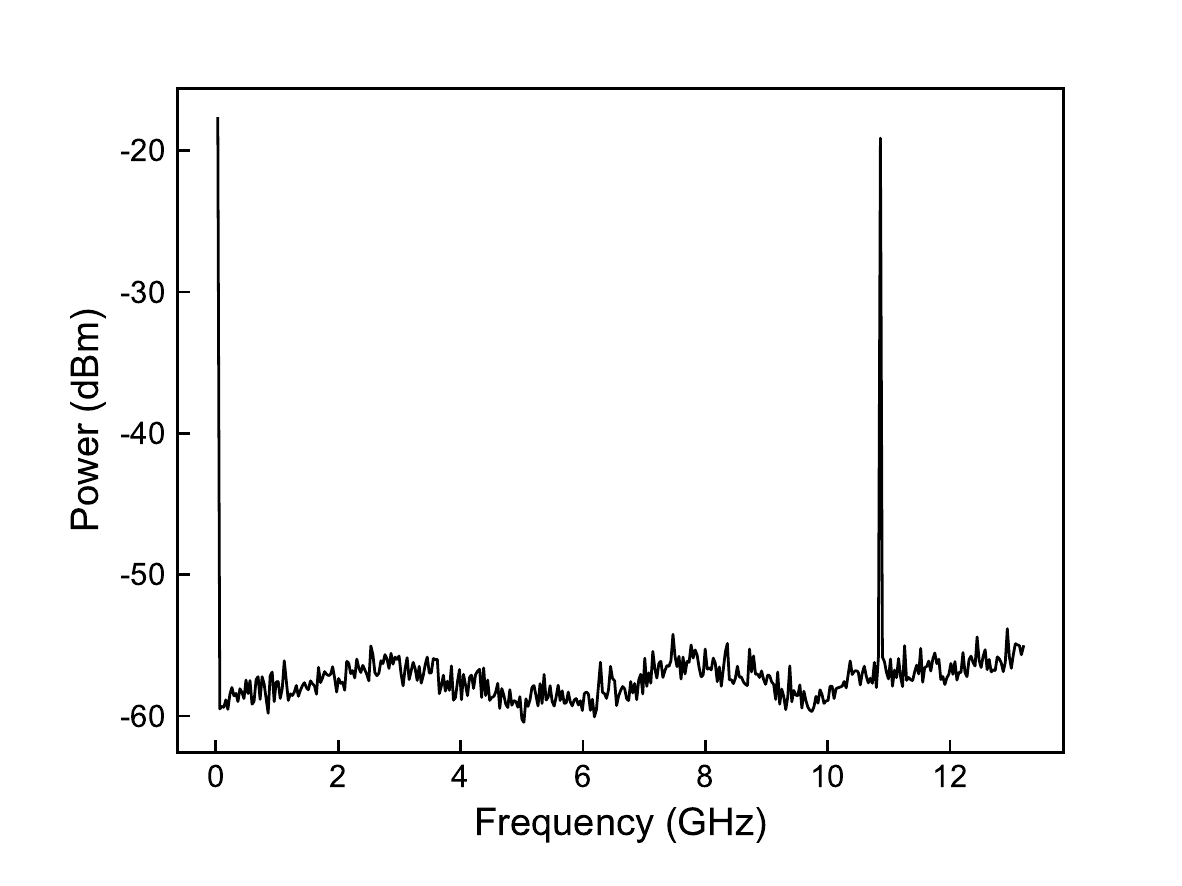}
	\caption{\label{fig:dual_spectra} (\textbf{upper}) Dual-wavelength optical output spectrum showing operation with a 12~nm spacing in wavelength. (\textbf{bottom left}) Output wavelength difference tuned to $\approx$~0.1 nm (beat frequency $\approx$~11GHz), which is below the resolution of the optical spectrum analyzer. (\textbf{bottom right}) The according beat frequency detected near 11~GHz with a radio frequency analyzer.}
\end{figure}
Fig.~\ref{fig:dual_spectra} displays two output spectra obtained with an optical spectrum analyzer set to 1 nm resolution, and one spectrum measured with a RF spectrum analyzer behind a fast photodiode. The upper panel shows the spectrum after tuning the two wavelengths to a separation of 12~nm (1.5~THz) measured with the optical spectrum analyzer. The side mode suppression with regard to the spontaneous emission background is between 40 and 50~dB. The specific spectral shape of the background, with a minimum at around 1510~nm, is caused by a small path length difference of the Mach-Zehnder arms of the tunable coupler. This can be concluded because the wavelength of the minimum is adjustable with the heaters on top of the coupler. The bottom left spectrum shows the two wavelengths tuned to almost the same value (0.09~nm difference) which is not resolved by the optical spectrum analyzer. To increase the resolution we sent the laser output to a fast photodiode and recorded the signal with an RF spectrum analyzer (bottom right panel). The recording shows that the two wavelengths are tuned to a difference frequency of about 11~GHz.

Using this laser, ongoing and future measurements aim to measure also the linewidth of the beat frequency and compare it with the intrinsic linewidth of the two individual laser output frequencies. We expect to observe widely and arbitrary tunable microwave and THz-generation with linewidths in the tens of kHz range or below. Such experiments might provide one of the lowest RF linewidths generated by integrated diode lasers in chip-sized format.

\subsection{Visible wavelength hybrid integrated lasers}

One of the great promises of Si$_3$N$_4$ waveguide circuits is their excellent transparency also in the near-infrared and visible range. The transmission window coarsely spans from 400~nm to 2.3~$\mu$m, while the Si$_3$N$_4$ core on its own provides transparency even up to 8~$\mu$m. In particular, its transparency in the visible, where silicon is strongly absorbing, is expected to secure a central function for Si$_3$N$_4$ circuits as well as for hybrid lasers in the visible based on Si$_3$N$_4$ feedback circuits. For applications such as named at the beginning of Sect.~\ref{section DMVlasers} visible hybrid Si$_3$N$_4$ diode lasers are of great potential.

Outside the 1.5~$\mu$m range, to our knowledge, Si$_3$N$_4$-based hybrid integrated diode lasers have only been demonstrated in the near-infrared, near 1~$\mu$m wavelength \cite{numata_2012IOPConf, bovington_2014OL, zhuang_2011OE}, which is of interest to compete with highly coherent monolithic Nd:YAG bulk ring lasers~\cite{kane_1985OL}. So far, there is no demonstration of operation in the visible. One of the reasons is that hitherto Si$_3$N$_4$ has primarily been employed in single-pass applications where loss is less critical~\cite{hosseini_2015OE, epping_2015OE}, with the exception of  a resonator-based visible spectrometer~\cite{fan_2018OL}.

We aim on realizing a visible hybrid laser with narrow linewidth and tunable near 690~nm, and here we report on the preparation and characterization of appropriate Si$_3$N$_4$ feedback circuits. Currently under investigation is realizing appropriate waveguide and circuit design parameters, i.e., the waveguide cross section, resonator radii and coupling constants. Obtaining appropriate parameters is much more challenging than near 1.5~$\mu$m in the infrared, mainly due to the much smaller wavelength. For instance, a proper mode field needs to be designed that counteracts potentially increased Rayleigh scattering while allowing curvatures that extends the Vernier free spectral range for matching a typical gain bandwidth of about 15~nm. Furthermore, proper waveguide tapers have to be designed for efficient coupling to an anti-reflection coated optical gain chip that operates in the visible.

\begin{figure}[tbp] \label{visible_Vernier}
	\centering
	\includegraphics[width=0.4\linewidth]{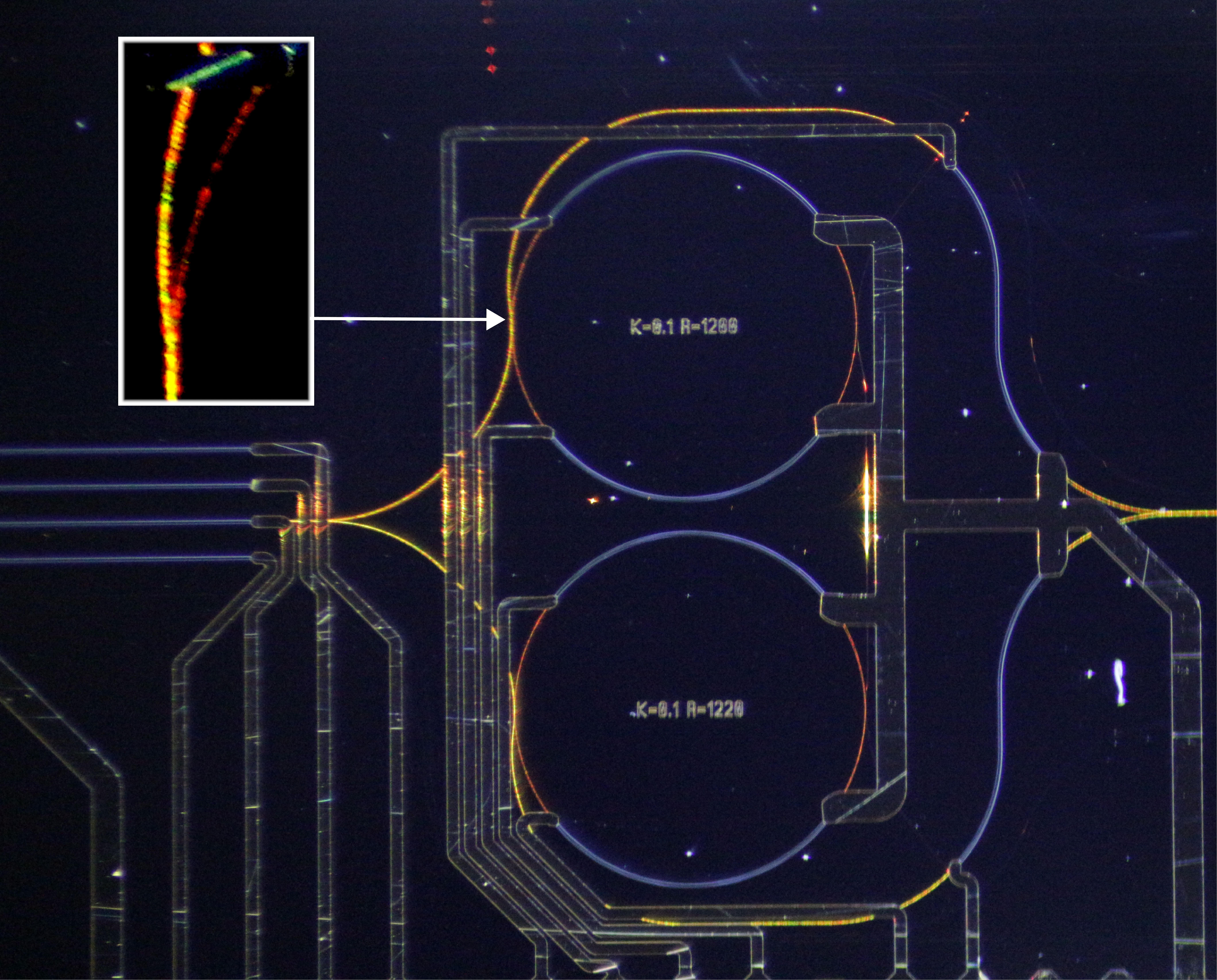}
	\hspace{0.5cm}
	\includegraphics[width=0.4\linewidth]{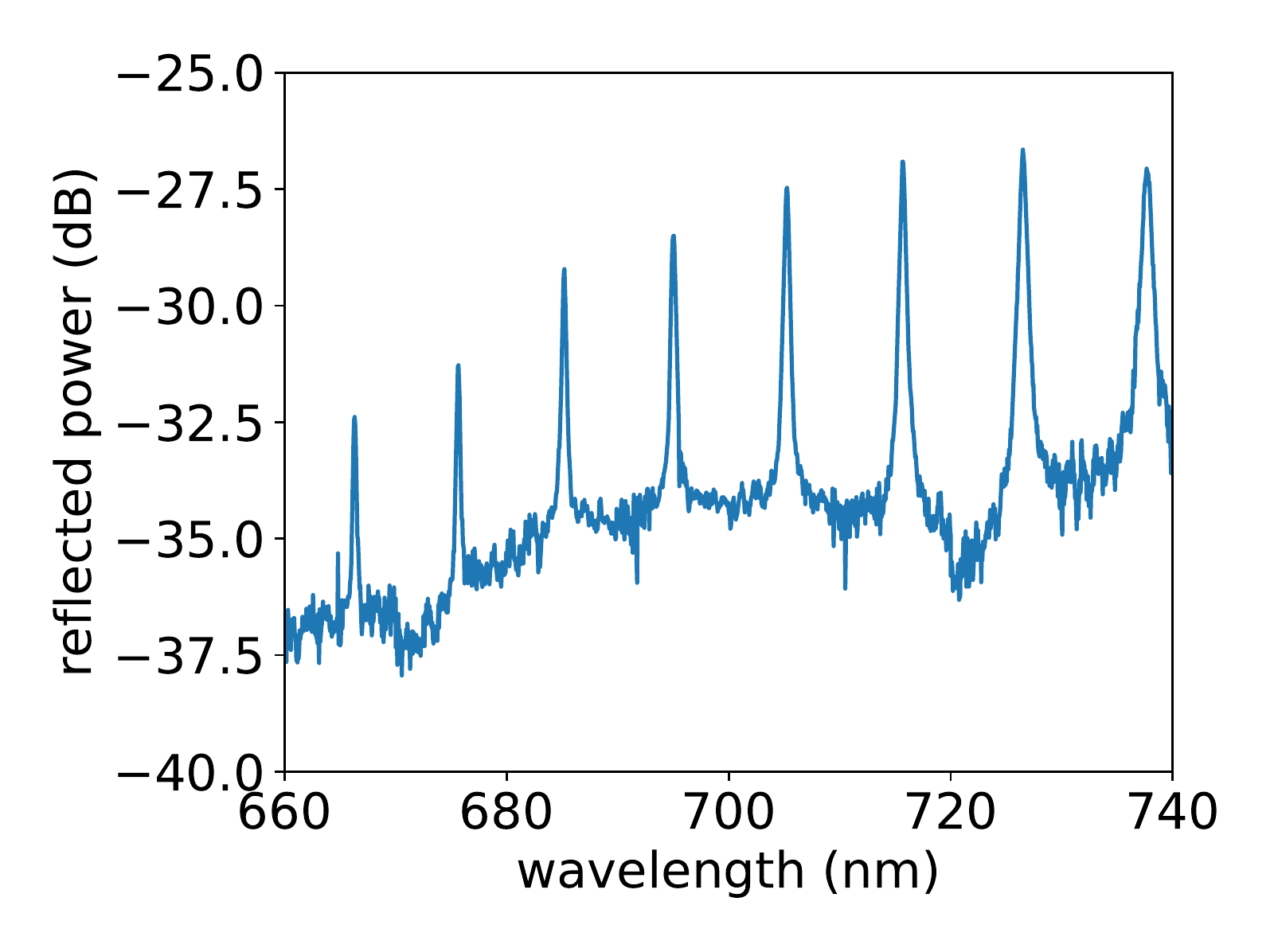} 
	\caption{\label{fig:visible_Vernier} (\textbf{left}) Top view of a dual microring resonator Vernier filter circuit when injecting TE-polarized, white light from the left. The two resonators are located on the left-hand side of the chip, but most light bypasses the resonators. Insert shows a zoom-in of the coupler region with enhanced contrast, showing red light circulating within the ring. (\textbf{right}) Transmission spectrum of a Vernier filter recorded with 0.2~nm resolution. The radii of the microring resonators are $\rho_1$~=~1200 and $\rho_2$~=~1205~$\mu$m, with a specified power coupling of 5\% for the add and drop ports.}
\end{figure}
Fig.~\ref{fig:visible_Vernier} gives a coarse overview over current activities. The left panel shows scattered light from a dual microring resonator Vernier filter designed for TE-polarized red light, recorded with a top view camera when injected with TE-polarized white light from a supercontinuum source. The insert depicts an enlarged section of the coupler region with enhanced contrast, and clearly shows that red light is circulating inside the ring. The right panel displays an example of a measured feedback spectrum, showing Vernier reflection peaks with a free spectral range of $\approx$~10~nm). We note that the optical spectrum analyzer used does not resolve the much narrower bandwidth of the Vernier reflection peaks estimated to be around 1.5~pm (1~GHz). Figure~\ref{fig:visible_Vernier} confirms for the first time the design and operation of a Vernier filter for TE-polarized red light.

The next set of experiments will concentrate on characterization of losses in the circuit and losses caused by coupling to the circuit. Thereafter, first feedback experiments aim on demonstrating laser oscillation.

\section{Conclusions}

To summarize, we have investigated a variety of hybrid integrated diode lasers in the 1.55 $\mu$m wavelength range based on InP semiconductor optical amplifiers, using low-loss dielectric feedback circuits fabricated with the Si$_3$N$_4$ waveguide platform. The fundamental key properties of the latter are lowest propagation loss, including lowest nonlinear loss due to a wide bandgap, transparency that reaches also across the visible range, and a high index contrast with the SiO$_2$ cladding. The importance of these properties is that they are central to introducing a long photon lifetime into otherwise lossy laser resonators, that well-defined and tunable spectral properties can be implemented in laser resonators, such as high-Q filters and interferometers, and that these functionalities can be carried over from their main current use in the infrared to other spectral ranges, specifically also the visible. The overall impact is a record increase of coherence properties, i.e., of spectral quality, of spectral controllability, of spectral coverage, and of output power with low intensity noise with on-chip light generation using diode lasers.

With this approach, the investigated hybrid lasers make optimum use of the best of two integrated photonic platforms: i) semiconductor amplifiers provide all the active optical functions, specifically, light amplification with wide spectral coverage, highest speed and electrical-to-optical efficiency, and nonlinear mixing to generate sidebands and optical comb spectra; ii) Si$_3$N$_4$ provides maximally passive optical functionalities that enable to propagate, interfere, spectrally shape and store light without losing it.

The investigated lasers, based on amplification in InP semiconductor gain sections, were selected to cover and optimize a range of different operational modes of lasers.  Specifically, these are single-frequency operation with ultranarrow intrinsic linewidth, wide spectral coverage, high-power output, low intensity noise, dual-wavelength and frequency comb operation. State-of-the-art output properties were presented, such as a record-low intrinsic linewidth of 40 Hz, a record-high output power above 100 mW, and a record-wide spectral coverage of more than 120 nm. A lowest level of relative intensity noise (RIN) of -170 dBc/Hz was demonstrated, which is close to the fundamental shot noise (quantum) limit.

A great benefit of the Si$_3$N$_4$ platform is its compatibility with CMOS fabrication equipment, which has led to an impressive maturity enabling to reproducible fabricate complex and thus highly functional circuits. Examples are coherent optical receivers and transmitters~\cite{wang_2014OL}, optical beamforming networks~\cite{liu_2018JSTQE,visscher_2019EuCAP}, and circuits may be expanded to operate entire arrays of lasers~\cite{oldenbeuving_2010OE}, to provide redundancy or to coherently add their outputs via mutual locking~\cite{fan_2014OL}. There is also compatibility with microfluidics~\cite{kuswandi_2007ACA,artundo_2017OP} and thus significant potential for lab-on-the chip and bio sensing applications~\cite{ymeti_2007NL,porcel_2019OLT}. 

The excellent compatibility with seamless integration in complex photonic circuits, paired with highest performance point to a great potential of hybrid integrated lasers in applications. 

\vspace{6pt} 

\section*{funding}
This research was funded in parts by the IOP Photonic Devices program of RVO (Rijksdienst voor Ondernemend Nederland), a division of the Ministry for Economic Affairs, the Netherlands, by the European Union’s Horizon 2020 research and innovation programme under grant agreements 780502 (3PEAT), 688519 (PIX4life), and 762055 (BlueSpace), and by the Dutch Research Council (NWO) projects "Light and sound-based signal processing in silicon nitride" (Vidi program, project 15702), "Functional hybrid technologies" (Memphis II program, project 13537) and "On-chip photonic control of gigahertz phonons" (Start Up program, project STU.018-2.002).

\section*{acknowledgments}
We thank H.M.J. Bastiaens and Y. Klaver for helpful discussions and support.

\section{Conflicts of interest}
The authors declare no conflict of interest.


\begin{thebibliography}{100}
	\newcommand{\enquote}[1]{``#1''}
	
	\bibitem{biesheuvel_2016NCom}
	J.~Biesheuvel, J.~P. Karr, L.~Hilico, K.~S.~E. Eikema, W.~Ubachs, and J.~C.~J.
	Koelemeij, \enquote{Probing {QED} and fundamental constants through laser
		spectroscopy of vibrational transitions in {HD+},}
	{\protect\JournalTitle{Nature Communications}} \textbf{7}, 10385 (2016).
	
	\bibitem{gabrielse_2002PRL}
	G.~Gabrielse, N.~S. Bowden, P.~Oxley, A.~Speck, C.~H. Storry, J.~N. Tan,
	M.~Wessels, D.~Grzonka, W.~Oelert, G.~Schepers, T.~Sefzick, J.~Walz,
	H.~Pittner, T.~W. H{\"a}nsch, and E.~A. Hessels, \enquote{Background-free
		observation of cold antihydrogen with field-ionization analysis of its
		states,} {\protect\JournalTitle{Physical Review Letters}} \textbf{89}, 213401
	(2002).
	
	\bibitem{hansch_1972NPS}
	T.~W. H{\"a}nsch, I.~S. Shahin, and A.~L. Schawlow, \enquote{Optical resolution
		of the {L}amb shift in atomic hydrogen by laser saturation spectroscopy,}
	{\protect\JournalTitle{Nature Physical Science}} \textbf{235}, 63--65 (1972).
	
	\bibitem{bagdonaite_2014AJ}
	J.~Bagdonaite, W.~Ubachs, M.~T. Murphy, and J.~B. Whitmore, \enquote{Analysis
		of molecular hydrogen absorption toward {QSO} {B}0642-5038 for a varying
		proton-to-electron mass ratio,} {\protect\JournalTitle{The Astrophysical
			Journal}} \textbf{782}, 10 (2014).
	
	\bibitem{abbott_2016PRL}
	B.~P. Abbott and et~al., \enquote{Observation of gravitational waves from a
		binary black hole merger,} {\protect\JournalTitle{Physical Review Letters}}
	\textbf{116}, {061102} (2016).
	
	\bibitem{crump_2006CLEO}
	P.~{Crump}, M.~{Grimshaw}, {Jun Wang}, {Weimin Dong}, {Shiguo Zhang}, {Suhit
		Das}, J.~{Farmer}, M.~{DeVito}, L.~S. {Meng}, and J.~K. {Brasseur},
	\enquote{85{\%} power conversion efficiency 975-nm broad area diode lasers at
		-50 $^{o}${C}, 76{\%} at 10 $^{o}${C},} in \emph{2006 Conference on Lasers
		and Electro-Optics and 2006 Quantum Electronics and Laser Science
		Conference,}  (2006), p. JWB24.
	
	\bibitem{leong_2005SMS}
	W.~H. Leong, W.~J. Staszewski, B.~C. Lee, and F.~Scarpa, \enquote{Structural
		health monitoring using scanning laser vibrometry: {III}. {L}amb waves for
		fatigue crack detection,} {\protect\JournalTitle{Smart Materials and
			Structures}} \textbf{14}, 1387--1395 (2005).
	
	\bibitem{Jac14}
	\enquote{See, e.g., distributed fiber sensing of pipelines, oil and gas wells,
		borders, railways, roadways, power plants or utilities,}
	\url{http://www.optasense.com/wp-cont7nt/uploads/2015/01/FiberOpticAug14.pdf}.
	Accessed: 2019-10-06.
	
	\bibitem{technobis_2017}
	\enquote{See, e.g., fiber sensing in aerospace, automotive, high-tech, civil,
		lifescience, or pharmaceutical applications,} \url{http://www.technobis.com}.
	Accessed: 2019-10-06.
	
	\bibitem{he_2011NT}
	L.~He, {\c S}.~{\"O}zdemir, J.~Zhu, W.~Kim, and L.~Yang, \enquote{Detecting
		single viruses and nanoparticles using whispering gallery microlasers,}
	{\protect\JournalTitle{Nature Nanotechnology}} \textbf{6}, 428--432 (2011).
	
	\bibitem{alalusi_2009SPIE}
	M.~Alalusi, P.~Brasil, S.~Lee, P.~Mols, L.~Stolpner, A.~Mehnert, and S.~Li,
	\enquote{{Low noise planar external cavity laser for interferometric fiber
			optic sensors},} in \emph{Proceedings of SPIE,} , vol. 7316 E.~Udd, H.~H. Du,
	and A.~Wang, eds., International Society for Optics and Photonics (SPIE,
	2009), pp. 235 -- 247.
	
	\bibitem{rothberg_2017OLE}
	S.~Rothberg, M.~Allen, P.~Castellini, D.~D. Maio, J.~Dirckx, D.~Ewins,
	B.~Halkon, P.~Muyshondt, N.~Paone, T.~Ryan, H.~Steger, E.~Tomasini,
	S.~Vanlanduit, and J.~Vignola, \enquote{An international review of laser
		{D}oppler vibrometry: {M}aking light work of vibration measurement,}
	{\protect\JournalTitle{Optics and Lasers in Engineering}} \textbf{99}, 11--22
	(2017).
	
	\bibitem{hecht_2018OPN}
	J.~Hecht, \enquote{Lidar for self-driving cars,} {\protect\JournalTitle{Optics
			\& Photonics News}} \textbf{29}, 26--33 (2018).
	
	\bibitem{tran_2017OE}
	M.~A. Tran, T.~Komljenovic, J.~C. Hulme, M.~Kennedy, D.~J. Blumenthal, and
	J.~E. Bowers, \enquote{Integrated optical driver for interferometric optical
		gyroscopes,} {\protect\JournalTitle{Optics Express}} \textbf{25}, 3826--3840
	(2017).
	
	\bibitem{srinivasan_2014OE}
	S.~Srinivasan, R.~Moreira, D.~Blumenthal, and J.~E. Bowers, \enquote{Design of
		integrated hybrid silicon waveguide optical gyroscope,}
	{\protect\JournalTitle{Optics Express}} \textbf{22}, 24988--24993 (2014).
	
	\bibitem{gundavarapu_2019NPhot}
	S.~Gundavarapu, G.~M. Brodnik, M.~Puckett, T.~Huffman, D.~Bose, R.~Behunin,
	J.~Wu, T.~Qiu, C.~Pinho, N.~Chauhan, J.~Nohava, P.~T. Rakich, K.~D. Nelson,
	M.~Salit, and D.~J. Blumenthal, \enquote{Sub-{H}ertz fundamental linewidth
		photonic integrated {B}rillouin laser,} {\protect\JournalTitle{Nature
			Photonics}} \textbf{13}, 60--67 (2019).
	
	\bibitem{lezius_2016O}
	M.~Lezius, T.~Wilken, C.~Deutsch, M.~Giunta, O.~Mandel, A.~Thaller,
	V.~Schkolnik, M.~Schiemangk, A.~Dinkelaker, A.~Kohfeldt, A.~Wicht,
	M.~Krutzik, A.~Peters, O.~Hellmig, H.~Duncker, K.~Sengstock,
	P.~Windpassinger, K.~Lampmann, T.~H{\"u}lsing, T.~W. H{\"a}nsch, and
	R.~Holzwarth, \enquote{Space-borne frequency comb metrology,}
	{\protect\JournalTitle{Optica}} \textbf{3}, 1381--1387 (2016).
	
	\bibitem{newman_2019O}
	Z.~L. Newman, V.~Maurice, T.~Drake, J.~R. Stone, T.~C. Briles, D.~T. Spencer,
	C.~Fredrick, Q.~Li, D.~Westly, B.~R. Ilic, B.~Shen, M.-G. Suh, K.~Y. Yang,
	C.~Johnson, D.~M.~S. Johnson, L.~Hollberg, K.~J. Vahala, K.~Srinivasan, S.~A.
	Diddams, J.~Kitching, S.~B. Papp, and M.~T. Hummon, \enquote{Architecture for
		the photonic integration of an optical atomic clock,}
	{\protect\JournalTitle{Optica}} \textbf{6}, 680--685 (2019).
	
	\bibitem{jiang_2011NP}
	Y.~Jiang, A.~Ludlow, N.~Lemke, R.~Fox, J.~Sherman, L.-S. Ma, and C.~Oates,
	\enquote{Making optical atomic clocks more stable with $10^{-16}$-level laser
		stabilization,} {\protect\JournalTitle{Nature Photonics}} \textbf{5},
	158--161 (2011).
	
	\bibitem{ESA16}
	\enquote{{GALILEO} begins serving the globe,} \url{https://www.esa.int}.
	Accessed: 2019-10-06.
	
	\bibitem{gill_2005Met}
	P.~Gill, \enquote{Optical frequency standards,}
	{\protect\JournalTitle{Metrologia}} \textbf{42}, S125--S137 (2005).
	
	\bibitem{spillane_2002N}
	S.~M. Spillane, T.~J. Kippenberg, and K.~J. Vahala, \enquote{Ultralow-threshold
		{R}aman laser using a spherical dielectric microcavity,}
	{\protect\JournalTitle{Nature}} \textbf{415}, 621--623 (2002).
	
	\bibitem{li_2017O}
	J.~Li, M.-G. Suh, and K.~Vahala, \enquote{Microresonator {B}rillouin
		gyroscope,} {\protect\JournalTitle{Optica}} \textbf{4}, 346--348 (2017).
	
	\bibitem{stern_2018N}
	B.~Stern, X.~Ji, Y.~Okawachi, A.~L. Gaeta, and M.~Lipson,
	\enquote{Battery-operated integrated frequency comb generator,}
	{\protect\JournalTitle{Nature}} \textbf{562}, 401--405 (2018).
	
	\bibitem{raja_2019NatCom}
	A.~S. Raja, A.~S. Voloshin, H.~Guo, S.~E. Agafonova, J.~Liu, A.~S.
	Gorodnitskiy, M.~Karpov, N.~G. Pavlov, E.~Lucas, R.~R. Galiev, A.~E.
	Shitikov, J.~D. Jost, M.~L. Gorodetsky, and T.~J. Kippenberg,
	\enquote{Electrically pumped photonic integrated soliton microcomb,}
	{\protect\JournalTitle{Nature Communications}} \textbf{10}, 680 (2019).
	
	\bibitem{pavlov_2018NPhot}
	N.~G. Pavlov, S.~Koptyaev, G.~V. Lihachev, A.~S. Voloshin, A.~S. Gorodnitskiy,
	M.~V. Ryabko, S.~V. Polonsky, and M.~L. Gorodetsky, \enquote{Narrow-linewidth
		lasing and soliton {K}err microcombs with ordinary laser diodes,}
	{\protect\JournalTitle{Nature Photonics}} \textbf{12}, 694--698 (2018).
	
	\bibitem{wang_2017LSA}
	Z.~Wang, K.~Van~Gasse, V.~Moskalenko, S.~Latkowski, E.~Bente, B.~Kuyken, and
	G.~Roelkens, \enquote{A {III}-{V}-on-{Si} ultra-dense comb laser,}
	{\protect\JournalTitle{Light: Science \& Applications}} \textbf{6},
	e16260--e16260 (2017).
	
	\bibitem{coddington_2009NPhot}
	I.~Coddington, W.~C. Swann, L.~Nenadovic, and N.~R. Newbury, \enquote{Rapid and
		precise absolute distance measurements at long range,}
	{\protect\JournalTitle{Nature Photonics}} \textbf{3}, 351--356 (2009).
	
	\bibitem{suh_2018Science}
	M.-G. Suh and K.~J. Vahala, \enquote{Soliton microcomb range measurement,}
	{\protect\JournalTitle{Science}} \textbf{359}, 884--887 (2018).
	
	\bibitem{rieker_2014O}
	G.~B. Rieker, F.~R. Giorgetta, W.~C. Swann, J.~Kofler, A.~M. Zolot, L.~C.
	Sinclair, E.~Baumann, C.~Cromer, G.~Petron, C.~Sweeney, P.~P. Tans,
	I.~Coddington, and N.~R. Newbury, \enquote{Frequency-comb-based remote
		sensing of greenhouse gases over kilometer air paths,}
	{\protect\JournalTitle{Optica}} \textbf{1}, 290--298 (2014).
	
	\bibitem{coddington_2016O}
	I.~Coddington, N.~Newbury, and W.~Swann, \enquote{Dual-comb spectroscopy,}
	{\protect\JournalTitle{Optica}} \textbf{3}, 414--426 (2016).
	
	\bibitem{bao_2019O}
	C.~Bao, M.-G. Suh, and K.~Vahala, \enquote{Microresonator soliton dual-comb
		imaging,} {\protect\JournalTitle{Optica}} \textbf{6}, 1110--1116 (2019).
	
	\bibitem{spencer_2018N}
	D.~T. Spencer, T.~Drake, T.~C. Briles, J.~Stone, L.~C. Sinclair, C.~Fredrick,
	Q.~Li, D.~Westly, B.~R. Ilic, A.~Bluestone, N.~Volet, T.~Komljenovic,
	L.~Chang, S.~H. Lee, D.~Y. Oh, M.-G. Suh, K.~Y. Yang, M.~H.~P. Pfeiffer,
	T.~J. Kippenberg, E.~Norberg, L.~Theogarajan, K.~Vahala, N.~R. Newbury,
	K.~Srinivasan, J.~E. Bowers, S.~A. Diddams, and S.~B. Papp, \enquote{An
		optical-frequency synthesizer using integrated photonics,}
	{\protect\JournalTitle{Nature}} \textbf{557}, 81--85 (2018).
	
	\bibitem{reimer_2018FoO}
	C.~Reimer, Y.~Zhang, P.~Roztocki, S.~Sciara, L.~R. Cort{\'e}s, M.~Islam,
	B.~Fischer, B.~Wetzel, A.~C. Cino, S.~T. Chu, B.~Little, D.~Moss, L.~Caspani,
	J.~Aza{\~{n}}a, M.~Kues, and R.~Morandotti, \enquote{On-chip frequency combs
		and telecommunications signal processing meet quantum optics,}
	{\protect\JournalTitle{Frontiers of Optoelectronics}} \textbf{11}, 134--147
	(2018).
	
	\bibitem{kikuchi_2016JLT}
	K.~Kikuchi, \enquote{Fundamentals of coherent optical fiber communications,}
	{\protect\JournalTitle{Journal of Lightwave Technology}} \textbf{34},
	157--179 (2016).
	
	\bibitem{yue_2019PJ}
	Y.~Yue, Q.~Wang, and J.~Anderson, \enquote{Experimental investigation of 400
		{Gb/s} data center interconnect using unamplified high-baud-rate and
		high-order {QAM} single-carrier signal,} {\protect\JournalTitle{Applied
			Sciences}} \textbf{9}, 2455 (2019).
	
	\bibitem{winzer_2006ProcIEEE}
	P.~J. {Winzer} and R.~J. {Essiambre}, \enquote{Advanced optical modulation
		formats,} {\protect\JournalTitle{Proceedings of the {IEEE}}} \textbf{94},
	952--985 (2006).
	
	\bibitem{zhang_2009PTL}
	S.~Zhang, P.~Y. Kam, C.~Yu, and J.~Chen, \enquote{Laser linewidth tolerance of
		decision-aided maximum likelihood phase estimation in coherent optical
		{M}-ary {PSK} and {QAM} systems,} {\protect\JournalTitle{IEEE Photonics
			Technology Letters}} \textbf{21}, 1075--1077 (2009).
	
	\bibitem{beppu_2014OFC}
	S.~{Beppu}, K.~{Kasai}, M.~{Yoshida}, and M.~{Nakazawa}, \enquote{2048 {QAM}
		(66 {Gbit}/s) single-carrier coherent optical transmission over 150 km with a
		potential {SE} of 15.3 bit/s/{H}z,} in \emph{OFC 2014,}  (2014), p. W1A.6.
	
	\bibitem{pfeifle_2014NPhot}
	J.~Pfeifle, V.~Brasch, M.~Lauermann, Y.~Yu, D.~Wegner, T.~Herr, K.~Hartinger,
	P.~Schindler, J.~Li, D.~Hillerkuss, R.~Schmogrow, C.~Weimann, R.~Holzwarth,
	W.~Freude, J.~Leuthold, T.~J. Kippenberg, and C.~Koos, \enquote{Coherent
		terabit communications with microresonator {K}err frequency combs,}
	{\protect\JournalTitle{Nature Photonics}} \textbf{8}, 375--380 (2014).
	
	\bibitem{seeds_2006JLT}
	A.~J. Seeds and K.~J. Williams, \enquote{Microwave photonics,}
	{\protect\JournalTitle{Journal of Lightwave Technology}} \textbf{24},
	4628--4641 (2006).
	
	\bibitem{capmany_2007NPhot}
	J.~Capmany and D.~Novak, \enquote{Microwave photonics combines two worlds,}
	{\protect\JournalTitle{Nature Photonics}} \textbf{1}, 319--330 (2007).
	
	\bibitem{zhuang_2008Leos}
	L.~{Zhuang}, A.~{Meijerink}, C.~G.~H. {Roeloffzen}, D.~A.~I. {Marpaung}, R.~G.
	{Heideman}, M.~{Hoekman}, A.~{Leinse}, and W.~{van Etten}, \enquote{Novel
		ring resonator-based optical beamformer for broadband phased array receive
		antennas,} in \emph{LEOS 2008 - 21st Annual Meeting of the IEEE Lasers and
		Electro-Optics Society,}  (2008), pp. 20--21.
	
	\bibitem{zhuang_2011OE}
	L.~Zhuang, D.~Marpaung, M.~Burla, W.~Beeker, A.~Leinse, and C.~Roeloffzen,
	\enquote{Low-loss, high-index-contrast {Si}$_3${N}$_4$/sio$_2$ optical
		waveguides for optical delay lines in microwave photonics signal processing,}
	{\protect\JournalTitle{Optics Express}} \textbf{19}, 23162--23170 (2011).
	
	\bibitem{burla_2013OE}
	M.~Burla, L.~R. Cort\'{e}s, M.~Li, X.~Wang, L.~Chrostowski, and J.~A. {n}a,
	\enquote{Integrated waveguide {B}ragg gratings for microwave photonics signal
		processing,} {\protect\JournalTitle{Optics Express}} \textbf{21},
	25120--25147 (2013).
	
	\bibitem{marpaung_2015O}
	D.~Marpaung, B.~Morrison, M.~Pagani, R.~Pant, D.-Y. Choi, B.~Luther-Davies,
	S.~J. Madden, and B.~J. Eggleton, \enquote{Low-power, chip-based stimulated
		brillouin scattering microwave photonic filter with ultrahigh selectivity,}
	{\protect\JournalTitle{Optica}} \textbf{2}, 76--83 (2015).
	
	\bibitem{marpaung_2019NP}
	D.~Marpaung and J.~Yao, \enquote{Integrated microwave photonics,}
	{\protect\JournalTitle{Nature Photonics}} \textbf{13}, 80--90 (2019).
	
	\bibitem{khilo_2012OE}
	A.~Khilo, S.~J. Spector, M.~E. Grein, A.~H. Nejadmalayeri, C.~W. Holzwarth,
	M.~Y. Sander, M.~S. Dahlem, M.~Y. Peng, M.~W. Geis, N.~A. DiLello, J.~U.
	Yoon, A.~Motamedi, J.~S. Orcutt, J.~P. Wang, C.~M. Sorace-Agaskar, M.~v{s}
	A.~Popovi{\'c}, J.~Sun, G.-R. Zhou, H.~Byun, J.~Chen, J.~L. Hoyt, H.~I.
	Smith, R.~J. Ram, M.~Perrott, T.~M. Lyszczarz, E.~P. Ippen, and F.~X.
	K{\"a}rtner, \enquote{Photonic {ADC}: overcoming the bottleneck of electronic
		jitter,} {\protect\JournalTitle{Optics Express}} \textbf{20}, 4454--4469
	(2012).
	
	\bibitem{carpintero_2012OL}
	G.~Carpintero, E.~Rouvalis, K.~{\L}awniczuk, M.~Fice, C.~C. Renaud, X.~J.~M.
	Leijtens, E.~A. J.~M. Bente, M.~Chitoui, F.~V. Dijk, and A.~J. Seeds,
	\enquote{95 {GHz} millimeter wave signal generation using an arrayed
		waveguide grating dual wavelength semiconductor laser,}
	{\protect\JournalTitle{Optics Letters}} \textbf{37}, 3657--3659 (2012).
	
	\bibitem{seeds_2015JLT}
	A.~J. {Seeds}, H.~{Shams}, M.~J. {Fice}, and C.~C. {Renaud},
	\enquote{Tera{H}ertz photonics for wireless communications,}
	{\protect\JournalTitle{Journal of Lightwave Technology}} \textbf{33},
	579--587 (2015).
	
	\bibitem{lang_1985JQE}
	R.~{Lang}, K.~{Vahala}, and A.~{Yariv}, \enquote{The effect of spatially
		dependent temperature and carrier fluctuations on noise in semiconductor
		lasers,} {\protect\JournalTitle{IEEE Journal of Quantum Electronics}}
	\textbf{21}, 443--451 (1985).
	
	\bibitem{ruthman_1978ProcIEEE}
	J.~{Rutman}, \enquote{Characterization of phase and frequency instabilities in
		precision frequency sources: {F}ifteen years of progress,}
	{\protect\JournalTitle{Proceedings of the IEEE}} \textbf{66}, 1048--1075
	(1978).
	
	\bibitem{thomson_1982ProcIEEE}
	D.~J. {Thomson}, \enquote{Spectrum estimation and harmonic analysis,}
	{\protect\JournalTitle{Proceedings of the IEEE}} \textbf{70}, 1055--1096
	(1982).
	
	\bibitem{didomenico_2010AO}
	G.~D. Domenico, S.~Schilt, and P.~Thomann, \enquote{Simple approach to the
		relation between laser frequency noise and laser line shape,}
	{\protect\JournalTitle{Applied Optics}} \textbf{49}, 4801--4807 (2010).
	
	\bibitem{barnes_1971IEEETransIM}
	J.~A. {Barnes}, A.~R. {Chi}, L.~S. {Cutler}, D.~J. {Healey}, D.~B. {Leeson},
	T.~E. {McGunigal}, J.~A. {Mullen}, W.~L. {Smith}, R.~L. {Sydnor}, R.~F.~C.
	{Vessot}, and G.~M.~R. {Winkler}, \enquote{Characterization of frequency
		stability,} {\protect\JournalTitle{IEEE Transactions on Instrumentation and
			Measurement}} \textbf{IM-20}, 105--120 (1971).
	
	\bibitem{schawlow_1958PR}
	A.~L. Schawlow and C.~H. Townes, \enquote{Infrared and optical masers,}
	{\protect\JournalTitle{Physical Review}} \textbf{112}, 1940--1949 (1958).
	
	\bibitem{fleming_1981APL}
	M.~W. Fleming and A.~Mooradian, \enquote{Fundamental line broadening of
		single‐mode {(GaAl)As} diode lasers,} {\protect\JournalTitle{Applied
			Physics Letters}} \textbf{38}, 511--513 (1981).
	
	\bibitem{henry_1982JQE}
	C.~Henry, \enquote{Theory of the linewidth of semiconductor lasers,}
	{\protect\JournalTitle{IEEE Journal of Quantum Electronics}} \textbf{18},
	259--264 (1982).
	
	\bibitem{wiseman_1999PRA}
	H.~M. Wiseman, \enquote{Light amplification without stimulated emission:
		{B}eyond the standard quantum limit to the laser linewidth,}
	{\protect\JournalTitle{Physical Review A}} \textbf{60}, 4083--4093 (1999).
	
	\bibitem{drever_1983APB}
	R.~W.~P. Drever, J.~L. Hall, F.~V. Kowalski, J.~Hough, G.~M. Ford, A.~J.
	Munley, and H.~Ward, \enquote{Laser phase and frequency stabilization using
		an optical resonator,} {\protect\JournalTitle{Applied Physics B}}
	\textbf{31}, 97--105 (1983).
	
	\bibitem{day_1992JQE}
	T.~{Day}, E.~K. {Gustafson}, and R.~L. {Byer}, \enquote{Sub-hertz relative
		frequency stabilization of two-diode laser-pumped {Nd}:{YAG} lasers locked to
		a {F}abry-{P}\'erot interferometer,} {\protect\JournalTitle{IEEE Journal of
			Quantum Electronics}} \textbf{28}, 1106--1117 (1992).
	
	\bibitem{lin_2012OL}
	Q.~Lin, M.~A.~V. Camp, H.~Zhang, B.~Jelenkovi{\'c}, and V.~Vuleti{\'c},
	\enquote{Long-external-cavity distributed {Bragg} reflector laser with
		subkilohertz intrinsic linewidth,} {\protect\JournalTitle{Optics Letters}}
	\textbf{37}, 1989--1991 (2012).
	
	\bibitem{ward_2005JSTQE}
	A.~J. {Ward}, D.~J. {Robbins}, G.~{Busico}, E.~{Barton}, L.~{Ponnampalam},
	J.~P. {Duck}, N.~D. {Whitbread}, P.~J. {Williams}, D.~C.~J. {Reid}, A.~C.
	{Carter}, and M.~J. {Wale}, \enquote{Widely tunable {DS-DBR} laser with
		monolithically integrated {SOA}: design and performance,}
	{\protect\JournalTitle{IEEE Journal of Selected Topics in Quantum
			Electronics}} \textbf{11}, 149--156 (2005).
	
	\bibitem{lavery_2013JLT}
	D.~{Lavery}, R.~{Maher}, D.~S. {Millar}, B.~C. {Thomsen}, P.~{Bayvel}, and
	S.~J. {Savory}, \enquote{Digital coherent receivers for long-reach optical
		access networks,} {\protect\JournalTitle{Journal of Lightwave Technology}}
	\textbf{31}, 609--620 (2013).
	
	\bibitem{akulova_2002JSTQE}
	Y.~A. Akulova, G.~A. Fish, P.-C. Koh, C.~L. Schow, P.~Kozodoy, A.~P. Dahl,
	S.~Nakagawa, M.~C. Larson, M.~P. Mack, T.~A. Strand, C.~W. Coldren,
	E.~Hegblom, S.~K. Penniman, T.~Wipiejewski, and L.~A. Coldren,
	\enquote{Widely tunable electroabsorption-modulated sampled-grating {DBR}
		laser transmitter,} {\protect\JournalTitle{IEEE Journal on Selected Topics in
			Quantum Electronics}} \textbf{8}, 1349--1357 (2002).
	
	\bibitem{okai_1990PTL}
	M.~{Okai}, T.~{Tsuchiya}, K.~{Uomi}, N.~{Chinone}, and T.~{Harada},
	\enquote{Corrugation-pitch-modulated {MQW-DFB} laser with narrow spectral
		linewidth (170 {kHz}),} {\protect\JournalTitle{IEEE Photonics Technology
			Letters}} \textbf{2}, 529--530 (1990).
	
	\bibitem{price_2006PTL}
	R.~K. {Price}, J.~J. {Borchardt}, V.~C. {Elarde}, R.~B. {Swint}, and J.~J.
	{Coleman}, \enquote{Narrow-linewidth asymmetric cladding distributed {B}ragg
		reflector semiconductor lasers at 850 nm,} {\protect\JournalTitle{IEEE
			Photonics Technology Letters}} \textbf{18}, 97--99 (2006).
	
	\bibitem{spiessberger_2011APB}
	S.~Spie{\ss}berger, M.~Schiemangk, A.~Wicht, H.~Wenzel, G.~Erbert, and
	G.~Tr{\"a}nkle, \enquote{{DBR} laser diodes emitting near 1064 nm with a
		narrow intrinsic linewidth of 2 {kHz},} {\protect\JournalTitle{Applied
			Physics B}} \textbf{104}, 813 (2011).
	
	\bibitem{toptica_2019}
	\enquote{Tunable diode lasers, toptica photonics application notes,}
	\url{https://www.toptica.com/fileadmin/Editors_English/11\_brochures\_datasheets/01\_brochures/toptica\_BR_Scientific\_Lasers.pdf}.
	Accessed: 2019-10-18.
	
	\bibitem{luvsandamdin_2014OE}
	E.~Luvsandamdin, C.~K\"{u}rbis, M.~Schiemangk, A.~Sahm, A.~Wicht, A.~Peters,
	G.~Erbert, and G.~Tr{\"a}nkle, \enquote{Micro-integrated extended cavity
		diode lasers for precision potassium spectroscopy in space,}
	{\protect\JournalTitle{Optics Express}} \textbf{22}, 7790--7798 (2014).
	
	\bibitem{liang_2015NC}
	W.~Liang, V.~S. Ilchenko, D.~Eliyahu, A.~A. Savchenkov, A.~B. Matsko,
	D.~Seidel, and L.~Maleki, \enquote{Ultralow noise miniature external cavity
		semiconductor laser,} {\protect\JournalTitle{Nature Communications}}
	\textbf{6}, 7371 (2015).
	
	\bibitem{wei_2016OE}
	F.~Wei, F.~Yang, X.~Zhang, D.~Xu, M.~Ding, L.~Zhang, D.~Chen, H.~Cai, Z.~Fang,
	and G.~Xijia, \enquote{Subkilohertz linewidth reduction of a {DFB} diode
		laser using self-injection locking with a fiber {B}ragg grating
		{F}abry-{P}{\'e}rot cavity,} {\protect\JournalTitle{Optics Express}}
	\textbf{24}, 17406--17415 (2016).
	
	\bibitem{morton_2018JLT}
	P.~A. {Morton} and M.~J. {Morton}, \enquote{High-power, ultra-low noise hybrid
		lasers for microwave photonics and optical sensing,}
	{\protect\JournalTitle{Journal of Lightwave Technology}} \textbf{36},
	5048--5057 (2018).
	
	\bibitem{zhao_2010OC}
	Y.~Zhao, J.~Zhang, J.~Stuhler, G.~Schuricht, F.~Lison, Z.~Lu, and L.~Wang,
	\enquote{Sub-hertz frequency stabilization of a commercial diode laser,}
	{\protect\JournalTitle{Optics Communications}} \textbf{283}, 4696--4700
	(2010).
	
	\bibitem{alnis_2008PRA}
	J.~Alnis, A.~Matveev, N.~Kolachevsky, T.~Udem, and T.~W. H{\"a}nsch,
	\enquote{Subhertz linewidth diode lasers by stabilization to vibrationally
		and thermally compensated ultralow-expansion glass {F}abry-{P}{\'e}rot
		cavities,} {\protect\JournalTitle{Physical Review A}} \textbf{77}, {053809}
	(2008).
	
	\bibitem{stoehr_2006OL}
	H.~Stoehr, F.~Mensing, J.~Helmcke, and U.~Sterr, \enquote{Diode laser with 1
		{Hz} linewidth,} {\protect\JournalTitle{Optics Letters}} \textbf{31},
	736--738 (2006).
	
	\bibitem{roeloffzen_2013OE}
	C.~G.~H. Roeloffzen, L.~Zhuang, C.~Taddei, A.~Leinse, R.~G. Heideman, P.~W.~L.
	van Dijk, R.~M. Oldenbeuving, D.~A.~I. Marpaung, M.~Burla, and K.~J. Boller,
	\enquote{Silicon nitride microwave photonic circuits,}
	{\protect\JournalTitle{Optics Express}} \textbf{21}, 22937--22961 (2013).
	
	\bibitem{marpaung_2013LPR}
	D.~Marpaung, C.~Roeloffzen, R.~Heideman, A.~Leinse, S.~Sales, and J.~Capmany,
	\enquote{Integrated microwave photonics,} {\protect\JournalTitle{Laser \&
			Photonics Reviews}} \textbf{7}, 506--538 (2013).
	
	\bibitem{doylend_2012OL}
	J.~K. Doylend, M.~J.~R. Heck, J.~T. Bovington, J.~D. Peters, M.~L. Davenport,
	L.~A. Coldren, and J.~E. Bowers, \enquote{Hybrid {III-V} silicon photonic
		source with integrated {1D} free-space beam steering,}
	{\protect\JournalTitle{Optics Letters}} \textbf{37}, 4257--4259 (2012).
	
	\bibitem{matsumoto_2010OFCC}
	T.~Matsumoto, A.~Suzuki, M.~Takahashi, S.~Watanabe, S.~Ishii, K.~Suzuki,
	T.~Kaneko, H.~Yamazaki, and N.~Sakuma, \enquote{Narrow spectral linewidth
		full band tunable laser based on waveguide ring resonators with low power
		consumption,} in \emph{Optical Fiber Communication Conference,}  (Optical
	Society of America, 2010), p. OThQ5.
	
	\bibitem{nemoto_2012APE}
	K.~Nemoto, T.~Kita, and H.~Yamada, \enquote{Narrow-spectral-linewidth
		wavelength-tunable laser diode with {Si} wire waveguide ring resonators,}
	{\protect\JournalTitle{Applied Physics Express}} \textbf{5}, {082701} (2012).
	
	\bibitem{oldenbeuving_2013LPL}
	R.~M. Oldenbeuving, E.~J. Klein, H.~L. Offerhaus, C.~J. Lee, H.~Song, and K.~J.
	Boller, \enquote{25 {kHz} narrow spectral bandwidth of a wavelength tunable
		diode laser with a short waveguide-based external cavity,}
	{\protect\JournalTitle{Laser Physics Letters}} \textbf{10}, {015804} (2013).
	
	\bibitem{keyvaninia_2013OE}
	S.~Keyvaninia, G.~Roelkens, D.~V. Thourhout, C.~Jany, M.~Lamponi, A.~L.
	Liepvre, F.~Lelarge, D.~Make, G.-H. Duan, D.~Bordel, and J.-M. Fedeli,
	\enquote{Demonstration of a heterogeneously integrated {III-V/SOI} single
		wavelength tunable laser,} {\protect\JournalTitle{Optics Express}}
	\textbf{21}, 3784--3792 (2013).
	
	\bibitem{hulme_2013OE}
	J.~C. Hulme, J.~K. Doylend, and J.~E. Bowers, \enquote{Widely tunable {Vernier}
		ring laser on hybrid silicon,} {\protect\JournalTitle{Optics Express}}
	\textbf{21}, 19718--19722 (2013).
	
	\bibitem{yang_2014OE}
	S.~Yang, Y.~Zhang, D.~W. Grund, G.~A. Ejzak, Y.~Liu, A.~Novack, D.~Prather,
	A.~E.-J. Lim, G.-Q. Lo, T.~Baehr-Jones, and M.~Hochberg, \enquote{A single
		adiabatic microring-based laser in 220 nm silicon-on-insulator,}
	{\protect\JournalTitle{Optics Express}} \textbf{22}, 1172--1180 (2014).
	
	\bibitem{santis_2014PNAS}
	C.~T. Santis, S.~T. Steger, Y.~Vilenchik, A.~Vasilyev, and A.~Yariv,
	\enquote{High-coherence semiconductor lasers based on integral high-{Q}
		resonators in hybrid {Si/III-V} platforms,}
	{\protect\JournalTitle{Proceedings of the National Academy of Sciences}}
	\textbf{111}, 2879--2884 (2014).
	
	\bibitem{fan_2014SPIE}
	Y.~Fan, R.~M. Oldenbeuving, E.~J. Klein, C.~J. Lee, H.~Song, M.~R.~H. Khan,
	H.~L. Offerhaus, P.~J.~M. van~der Slot, and K.-J. Boller, \enquote{A hybrid
		semiconductor-glass waveguide laser,} in \emph{Proceedings of SPIE,} , vol.
	9135 (2014), p. 91351B.
	
	\bibitem{duan_2014JSTQE}
	G.~{Duan}, C.~{Jany}, A.~L. {Liepvre}, A.~{Accard}, M.~{Lamponi}, D.~{Make},
	P.~{Kaspar}, G.~{Levaufre}, N.~{Girard}, F.~{Lelarge}, J.~{Fedeli},
	A.~{Descos}, B.~{Ben Bakir}, S.~{Messaoudene}, D.~{Bordel}, S.~{Menezo},
	G.~{de Valicourt}, S.~{Keyvaninia}, G.~{Roelkens}, D.~{Van Thourhout}, D.~J.
	{Thomson}, F.~Y. {Gardes}, and G.~T. {Reed}, \enquote{Hybrid {III-V} on
		silicon lasers for photonic integrated circuits on silicon,}
	{\protect\JournalTitle{IEEE Journal of Selected Topics in Quantum
			Electronics}} \textbf{20}, 158--170 (2014).
	
	\bibitem{defelipe_2014PTL}
	D.~{de Felipe}, Z.~{Zhang}, W.~{Brinker}, M.~{Kleinert}, A.~M. {Novo},
	C.~{Zawadzki}, M.~{Moehrle}, and N.~{Keil}, \enquote{Polymer-based external
		cavity lasers: tuning efficiency, reliability, and polarization diversity,}
	{\protect\JournalTitle{IEEE Photonics Technology Letters}} \textbf{26},
	1391--1394 (2014).
	
	\bibitem{kita_2014STQE}
	T.~{Kita}, K.~{Nemoto}, and H.~{Yamada}, \enquote{Silicon photonic
		wavelength-tunable laser diode with asymmetric {Mach-Zehnder}
		interferometer,} {\protect\JournalTitle{IEEE Journal of Selected Topics in
			Quantum Electronics}} \textbf{20}, 344--349 (2014).
	
	\bibitem{dong_2014OE}
	P.~Dong, T.-C. Hu, T.-Y. Liow, Y.-K. Chen, C.~Xie, X.~Luo, G.-Q. Lo, R.~Kopf,
	and A.~Tate, \enquote{Novel integration technique for silicon/{III-V} hybrid
		laser,} {\protect\JournalTitle{Optics Express}} \textbf{22}, 26854--26861
	(2014).
	
	\bibitem{debregeas_2014ISLC}
	H.~Debregeas, C.~Ferrari, M.~A. Cappuzzo, F.~Klemens, R.~Keller, F.~Pardo,
	C.~Bolle, C.~Xie, and M.~P. Earnshaw, \enquote{2~{kHz} linewidth {C}-band
		tunable laser by hybrid integration of reflective {SOA} and {SiO}$_2$ {PLC}
		external cavity,} in \emph{2014 International Semiconductor Laser
		Conference,}  (2014), pp. 50--51.
	
	\bibitem{kobayashi_2015JLT}
	N.~Kobayashi, K.~Sato, M.~Namiwaka, K.~Yamamoto, S.~Watanabe, T.~Kita,
	H.~Yamada, and H.~Yamazaki, \enquote{Silicon photonic hybrid ring-filter
		external cavity wavelength tunable lasers,} {\protect\JournalTitle{Journal of
			Lightwave Technology}} \textbf{33}, 1241--1246 (2015).
	
	\bibitem{tang_2015OL}
	R.~Tang, T.~Kita, and H.~Yamada, \enquote{Narrow-spectral-linewidth silicon
		photonic wavelength-tunable laser with highly asymmetric {Mach-Zehnder}
		interferometer,} {\protect\JournalTitle{Optics Letters}} \textbf{40},
	1504--1507 (2015).
	
	\bibitem{srinivasan_2015PJ}
	S.~{Srinivasan}, M.~{Davenport}, T.~{Komljenovic}, J.~{Hulme}, D.~T. {Spencer},
	and J.~E. {Bowers}, \enquote{Coupled-ring-resonator-mirror-based
		heterogeneous {III/V} silicon tunable laser,} {\protect\JournalTitle{IEEE
			Photonics Journal}} \textbf{7}, 1--8 (2015).
	
	\bibitem{santis_2015CLEO}
	C.~T. Santis, Y.~Vilenchik, A.~Yariv, N.~Satyan, and G.~Rakuljic,
	\enquote{Sub-{kHz} quantum linewidth semiconductor laser on silicon chip,} in
	\emph{CLEO: 2015 Postdeadline Paper Digest,}  (Optical Society of America,
	2015), p. JTh5A.7.
	
	\bibitem{komljenovic_2015JSTQE}
	T.~{Komljenovic}, S.~{Srinivasan}, E.~{Norberg}, M.~{Davenport}, G.~{Fish}, and
	J.~E. {Bowers}, \enquote{Widely tunable narrow-linewidth monolithically
		integrated external-cavity semiconductor lasers,} {\protect\JournalTitle{IEEE
			Journal of Selected Topics in Quantum Electronics}} \textbf{21}, 214--222
	(2015).
	
	\bibitem{kita_2016STQE}
	T.~{Kita}, R.~{Tang}, and H.~{Yamada}, \enquote{Narrow spectral linewidth
		silicon photonic wavelength tunable laser diode for digital coherent
		communication system,} {\protect\JournalTitle{IEEE Journal of Selected Topics
			in Quantum Electronics}} \textbf{22}, 23--34 (2016).
	
	\bibitem{fan_2017CLEO}
	Y.~Fan, R.~M. Oldenbeuving, C.~G. Roeloffzen, M.~Hoekman, D.~Geskus, R.~G.
	Heideman, and K.-J. Boller, \enquote{290~{Hz} intrinsic linewidth from an
		integrated optical chip-based widely tunable {InP}-{Si}$_3${N}$_4$ hybrid
		laser,} in \emph{Conference on Lasers and Electro-Optics,}  (Optical Society
	of America, 2017), p. JTh5C.9.
	
	\bibitem{komljenovic_2017JLT}
	T.~{Komljenovic}, S.~{Liu}, E.~{Norberg}, G.~A. {Fish}, and J.~E. {Bowers},
	\enquote{Control of widely tunable lasers with high-{Q} resonator as an
		integral part of the cavity,} {\protect\JournalTitle{Journal of Lightwave
			Technology}} \textbf{35}, 3934--3939 (2017).
	
	\bibitem{stern_2017OL}
	B.~Stern, X.~Ji, A.~Dutt, and M.~Lipson, \enquote{Compact narrow-linewidth
		integrated laser based on a low-loss silicon nitride ring resonator,}
	{\protect\JournalTitle{Optics Letters}} \textbf{42}, 4541--4544 (2017).
	
	\bibitem{verdier_2018JLT}
	A.~{Verdier}, G.~{de Valicourt}, R.~{Brenot}, H.~{Debregeas}, P.~{Dong},
	M.~{Earnshaw}, H.~{Carr 'e8re}, and Y.~{Chen}, \enquote{Ultrawideband
		wavelength-tunable hybrid external-cavity lasers,}
	{\protect\JournalTitle{Journal of Lightwave Technology}} \textbf{36}, 37--43
	(2018).
	
	\bibitem{li_2018JLT}
	Y.~{Li}, Y.~{Zhang}, H.~{Chen}, S.~{Yang}, and M.~{Chen}, \enquote{Tunable
		self-injected {F}abry-{P}{\'e}rot laser diode coupled to an external high-{Q}
		{Si}$_3${N}$_4$/{SiO}$_2$ microring resonator,}
	{\protect\JournalTitle{Journal of Lightwave Technology}} \textbf{36},
	3269--3274 (2018).
	
	\bibitem{tran_2018ECOC}
	M.~A. {Tran}, D.~{Huang}, T.~{Komljenovic}, S.~{Liu}, L.~{Liang}, M.~{Kennedy},
	and J.~E. {Bowers}, \enquote{Multi-ring mirror-based narrow-linewidth
		widely-tunable lasers in heterogeneous silicon photonics,} in \emph{2018
		European Conference on Optical Communication (ECOC),}  (2018), pp. 1--3.
	
	\bibitem{zhu_2018CLEO}
	Y.~Zhu, S.~Zeng, X.~Zhao, Y.~Zhao, and L.~Zhu, \enquote{Narrow-linewidth,
		tunable external cavity diode lasers through hybrid integration of
		quantum-well/quantum-dot {SOA}s with {Si}$_3${N}$_4$ microresonators,} in
	\emph{Conference on Lasers and Electro-Optics,}  (Optical Society of America,
	2018), p. SW4B.2.
	
	\bibitem{huang_2019O}
	D.~Huang, M.~A. Tran, J.~Guo, J.~Peters, T.~Komljenovic, A.~Malik, P.~A.
	Morton, and J.~E. Bowers, \enquote{High-power sub-{kHz} linewidth lasers
		fully integrated on silicon,} {\protect\JournalTitle{Optica}} \textbf{6},
	745--752 (2019).
	
	\bibitem{xiang_2019OL}
	C.~Xiang, P.~A. Morton, and J.~E. Bowers, \enquote{Ultra-narrow linewidth laser
		based on a semiconductor gain chip and extended {Si}$_3${N}$_4$ {Bragg}
		grating,} {\protect\JournalTitle{Optics Letters}} \textbf{44}, 3825--3828
	(2019).
	
	\bibitem{fan_2019ARXIV}
	Y.{Fan}, \enquote{Ultra-narrow linewidth hybrid integrated semiconductor
		laser,}  (2019).
	
	\bibitem{santis_2018PNAS}
	C.~T. Santis, Y.~Vilenchik, N.~Satyan, G.~Rakuljic, and A.~Yariv,
	\enquote{Quantum control of phase fluctuations in semiconductor lasers,}
	{\protect\JournalTitle{Proceedings of the National Academy of Sciences}}
	\textbf{115}, E7896--E7904 (2018).
	
	\bibitem{klein_2012IEEECInPRM}
	H.~{Klein}, C.~{Wagner}, W.~{Brinker}, F.~{Soares}, D.~{de Felipe}, Z.~{Zhang},
	C.~{Zawadzki}, N.~{Keil}, and M.~{Moehrle}, \enquote{Hybrid {InP}-polymer 30
		nm tunable {DBR} laser for 10 {Gbit}/s direct modulation in the {C}-band,} in
	\emph{2012 International Conference on Indium Phosphide and Related
		Materials,}  (2012), pp. 20--21.
	
	\bibitem{numata_2010OE}
	K.~Numata, J.~Camp, M.~A. Krainak, and L.~Stolpner, \enquote{Performance of
		planar-waveguide external cavity laser for precision measurements,}
	{\protect\JournalTitle{Optics Express}} \textbf{18}, 22781--22788 (2010).
	
	\bibitem{numata_2012IOPConf}
	K.~Numata and J.~Camp, \enquote{Precision laser development for interferometric
		space missions {NGO}, {SGO}, and {GRACE} follow-on,}
	{\protect\JournalTitle{Journal of Physics: Conference Series}} \textbf{363},
	{012054} (2012).
	
	\bibitem{kita_2015APL}
	T.~Kita, R.~Tang, and H.~Yamada, \enquote{Compact silicon photonic
		wavelength-tunable laser diode with ultra-wide wavelength tuning range,}
	{\protect\JournalTitle{Applied Physics Letters}} \textbf{106}, 111104 (2015).
	
	\bibitem{tran_2020JSTQE}
	M.~A. {Tran}, D.~{Huang}, J.~{Guo}, T.~{Komljenovic}, P.~A. {Morton}, and J.~E.
	{Bowers}, \enquote{Ring-resonator based widely-tunable narrow-linewidth
		{Si}/{InP} integrated lasers,} {\protect\JournalTitle{IEEE Journal of
			Selected Topics in Quantum Electronics}} \textbf{26}, 1500514 (2020).
	
	\bibitem{roelkens_2010LPR}
	G.~Roelkens, L.~Liu, D.~Liang, R.~Jones, A.~Fang, B.~Koch, and J.~Bowers,
	\enquote{{III-V}/silicon photonics for on-chip and intra-chip optical
		interconnects,} {\protect\JournalTitle{Laser \& Photonics Reviews}}
	\textbf{4}, 751--779 (2010).
	
	\bibitem{yariv_2007OE}
	A.~Yariv and X.~Sun, \enquote{Supermode {Si/III-V} hybrid lasers, optical
		amplifiers and modulators: {A} proposal and analysis,}
	{\protect\JournalTitle{Optics Express}} \textbf{15}, 9147--9151 (2007).
	
	\bibitem{vilenchik_2015PROC}
	Y.~Vilenchik, C.~T. Santis, S.~T. Steger, N.~Satyan, and A.~Yariv,
	\enquote{Theory and observation on non-linear effects limiting the coherence
		properties of high-{Q} hybrid {Si/III-V} lasers,} in \emph{Novel in-plane
		semiconductor lasers XIV,} , vol. 9382 (International Society for Optics and
	Photonics, 2015), p. 93820N.
	
	\bibitem{kuyken_2017NANOPHOT}
	B.~Kuyken, F.~Leo, S.~Clemmen, U.~Dave, R.~Van~Laer, T.~Ideguchi, H.~Zhao,
	X.~Liu, J.~Safioui, S.~Coen, S.~Gorza, S.~K. Selvaraja, S.~Massar, R.~M.
	Osgood, P.~Verheyen, J.~Van~Campenhout, R.~Baets, W.~M.~J. Green, and
	G.~Roelkens, \enquote{Nonlinear optical interactions in silicon waveguides,}
	{\protect\JournalTitle{Nanophotonics}} \textbf{6}, 377--392 (2017).
	
	\bibitem{taballione_2019OE}
	C.~Taballione, T.~A.~W. Wolterink, J.~Lugani, A.~Eckstein, B.~A. Bell,
	R.~Grootjans, I.~Visscher, D.~Geskus, C.~G.~H. Roeloffzen, J.~J. Renema,
	I.~A. Walmsley, P.~W.~H. Pinkse, and K.-J. Boller, \enquote{8$\times$8
		reconfigurable quantum photonic processor based on silicon nitride
		waveguides,} {\protect\JournalTitle{Optics Express}} \textbf{27},
	26842--26857 (2019).
	
	\bibitem{poberaj_2012LPR}
	G.~Poberaj, H.~Hu, W.~Sohler, and P.~Günter, \enquote{Lithium niobate on
		insulator ({LNOI}) for micro-photonic devices,} {\protect\JournalTitle{Laser
			\& Photonics Reviews}} \textbf{6}, 488--503 (2012).
	
	\bibitem{chang_2017OL}
	L.~Chang, M.~H.~P. Pfeiffer, N.~Volet, M.~Zervas, J.~D. Peters, C.~L.
	Manganelli, E.~J. Stanton, Y.~Li, T.~J. Kippenberg, and J.~E. Bowers,
	\enquote{Heterogeneous integration of lithium niobate and silicon nitride
		waveguides for wafer-scale photonic integrated circuits on silicon,}
	{\protect\JournalTitle{Optics Letters}} \textbf{42}, 803--806 (2017).
	
	\bibitem{belt_2017O}
	M.~Belt, M.~L. Davenport, J.~E. Bowers, and D.~J. Blumenthal,
	\enquote{Ultra-low-loss {Ta}$_2${O}$_5$-core/{SiO}$_2$-clad planar waveguides
		on {Si} substrates,} {\protect\JournalTitle{Optica}} \textbf{4}, 532--536
	(2017).
	
	\bibitem{jung_2013OL}
	H.~Jung, C.~Xiong, K.~Y. Fong, X.~Zhang, and H.~X. Tang, \enquote{Optical
		frequency comb generation from aluminum nitride microring resonator,}
	{\protect\JournalTitle{Optics Letters}} \textbf{38}, 2810--2813 (2013).
	
	\bibitem{roeloffzen_2018JSTQE}
	C.~G.~H. {Roeloffzen}, M.~{Hoekman}, E.~J. {Klein}, L.~S. {Wevers}, R.~B.
	{Timens}, D.~{Marchenko}, D.~{Geskus}, R.~{Dekker}, A.~{Alippi},
	R.~{Grootjans}, A.~{van Rees}, R.~M. {Oldenbeuving}, J.~P. {Epping}, R.~G.
	{Heideman}, K.~{W 'f6rhoff}, A.~{Leinse}, D.~{Geuzebroek}, E.~{Schreuder},
	P.~W.~L. {van Dijk}, I.~{Visscher}, C.~{Taddei}, Y.~{Fan}, C.~{Taballione},
	Y.~{Liu}, D.~{Marpaung}, L.~{Zhuang}, M.~{Benelajla}, and K.~{Boller},
	\enquote{Low-loss {Si}$_3${N}$_4$ {TriPleX} optical waveguides: technology
		and applications overview,} {\protect\JournalTitle{IEEE Journal of Selected
			Topics in Quantum Electronics}} \textbf{24}, 1--21 (2018).
	
	\bibitem{mooradian_1981JQE}
	M.~Fleming and A.~Mooradian, \enquote{Spectral characteristics of
		external-cavity controlled semiconductor lasers,} {\protect\JournalTitle{IEEE
			Journal of Quantum Electrontrics}} \textbf{17}, 44--59 (1981).
	
	\bibitem{komljenovic_2017AS}
	T.~Komljenovic, L.~Liang, R.-L. Chao, J.~Hulme, S.~Srinivasan, M.~Davenport,
	and J.~E.~Bowers, \enquote{Widely-tunable ring-resonator semiconductor
		lasers,} {\protect\JournalTitle{Applied Sciences}} \textbf{7}, 732 (2017).
	
	\bibitem{javaloyes_2019}
	J.~Javaloyes and S.~Balle, \enquote{{Freetwm}: a simulation tool for
		semiconductor lasers,} \url{https://onl.uib.eu/Softwares/Download/}.
	Accessed: 2019-10-20.
	
	\bibitem{vpi_2019}
	\enquote{Vpi component maker photonics,}
	\url{https://www.vpiphotonics.com/Tools/PhotonicCircuits/}. Accessed:
	2019-10-20.
	
	\bibitem{fan_2017OE}
	Y.~Fan, R.~E.~M. Lammerink, J.~Mak, R.~M. Oldenbeuving, P.~J.~M. van~der Slot,
	and K.-J. Boller, \enquote{Spectral linewidth analysis of semiconductor
		hybrid lasers with feedback from an external waveguide resonator circuit,}
	{\protect\JournalTitle{Optics Express}} \textbf{25}, 32767--32782 (2017).
	
	\bibitem{henry_1986JLT}
	C.~Henry, \enquote{Theory of spontaneous emission noise in open resonators and
		its application to lasers and optical amplifiers,}
	{\protect\JournalTitle{Journal of Lightwave Technology}} \textbf{4}, 288--297
	(1986).
	
	\bibitem{patzak_1983EL}
	E.~{Patzak}, A.~{Sugimura}, S.~{Saito}, T.~{Mukai}, and H.~{Olesen},
	\enquote{Semiconductor laser linewidth in optical feedback configurations,}
	{\protect\JournalTitle{Electronics Letters}} \textbf{19}, 1026--1027 (1983).
	
	\bibitem{kazarinov_1987JQE}
	R.~Kazarinov and C.~Henry, \enquote{The relation of line narrowing and chirp
		reduction resulting from the coupling of a semiconductor laser to passive
		resonator,} {\protect\JournalTitle{IEEE Journal of Quantum Electronics}}
	\textbf{23}, 1401--1409 (1987).
	
	\bibitem{koch_1990JLT}
	T.~L. {Koch} and U.~{Koren}, \enquote{Semiconductor lasers for coherent optical
		fiber communications,} {\protect\JournalTitle{Journal of Lightwave
			Technology}} \textbf{8}, 274--293 (1990).
	
	\bibitem{ujihara_1984JQE}
	K.~{Ujihara}, \enquote{Phase noise in a laser with output coupling,}
	{\protect\JournalTitle{IEEE Journal of Quantum Electronics}} \textbf{20},
	814--818 (1984).
	
	\bibitem{bjork_1987JQE}
	G.~{Bjork} and O.~{Nilsson}, \enquote{A tool to calculate the linewidth of
		complicated semiconductor lasers,} {\protect\JournalTitle{IEEE Journal of
			Quantum Electronics}} \textbf{23}, 1303--1313 (1987).
	
	\bibitem{vahala_1983APL}
	K.~Vahala, L.~C. Chiu, S.~Margalit, and A.~Yariv, \enquote{On the linewidth
		enhancement factor $\alpha$ in semiconductor injection lasers,}
	{\protect\JournalTitle{Applied Physics Letters}} \textbf{42}, 631--633
	(1983).
	
	\bibitem{vahala_1984APL}
	K.~Vahala and A.~Yariv, \enquote{Detuned loading in coupled cavity
		semiconductor lasers—effect on quantum noise and dynamics,}
	{\protect\JournalTitle{Applied Physics Letters}} \textbf{45}, 501--503
	(1984).
	
	\bibitem{newkirk_1991JQE}
	M.~A. {Newkirk} and K.~J. {Vahala}, \enquote{Amplitude-phase decorrelation: a
		method for reducing intensity noise in semiconductor lasers,}
	{\protect\JournalTitle{IEEE Journal of Quantum Electronics}} \textbf{27},
	13--22 (1991).
	
	\bibitem{tang_2018OE}
	J.~Tang, T.~Hao, W.~Li, D.~Domenech, R.~Ba{\~n}os, P.~Mu{\~n}oz, N.~Zhu,
	J.~Capmany, and M.~Li, \enquote{Integrated optoelectronic oscillator,}
	{\protect\JournalTitle{Optics Express}} \textbf{26}, 12257--12265 (2018).
	
	\bibitem{fan_2016PJ}
	Y.~Fan, J.~P. Epping, R.~M. Oldenbeuving, C.~G.~H. Roeloffzen, M.~Hoekman,
	R.~Dekker, R.~G. Heideman, P.~J.~M. van~der Slot, and K.-J. Boller,
	\enquote{Optically integrated {InP-Si}$_3${N}$_4$ hybrid laser,}
	{\protect\JournalTitle{IEEE Photonics Journal}} \textbf{8}, 1--11 (2016).
	
	\bibitem{bauters_2011OE}
	J.~F. Bauters, M.~J.~R. Heck, D.~D. John, J.~S. Barton, C.~M. Bruinink,
	A.~Leinse, R.~e.~G. Heideman, D.~J. Blumenthal, and J.~E. Bowers,
	\enquote{Planar waveguides with less than 0.1 {dB}/m propagation loss
		fabricated with wafer bonding,} {\protect\JournalTitle{Optics Express}}
	\textbf{19}, 24090--24101 (2011).
	
	\bibitem{taddei_2018PTL}
	C.~{Taddei}, L.~{Zhuang}, C.~G.~H. {Roeloffzen}, M.~{Hoekman}, and K.~{Boller},
	\enquote{High-selectivity on-chip optical bandpass filter with sub-100-{MHz}
		flat-top and under-2 shape factor,} {\protect\JournalTitle{IEEE Photonics
			Technology Letters}} \textbf{31}, 455--458 (2019).
	
	\bibitem{buus_2005BOOK}
	J.~Buus, M.-C. Amann, and D.~J. Blumenthal, \emph{Tunable diode lasers and
		related optical sources} (Wiley, 2005).
	
	\bibitem{lin_2018PJ}
	Y.~{Lin}, C.~{Browning}, R.~B. {Timens}, D.~H. {Geuzebroek}, C.~G.~H.
	{Roeloffzen}, M.~{Hoekman}, D.~{Geskus}, R.~M. {Oldenbeuving}, R.~G.
	{Heideman}, Y.~{Fan}, K.~J. {Boller}, and L.~P. {Barry},
	\enquote{Characterization of hybrid {InP}-{TriPleX} photonic integrated
		tunable lasers based on silicon nitride ({Si}$_3${N}$_4$/{SiO}$_2$) microring
		resonators for optical coherent system,} {\protect\JournalTitle{IEEE
			Photonics Journal}} \textbf{10}, 1--8 (2018).
	
	\bibitem{latkowski_2015PJ}
	S.~{Latkowski}, A.~{H 'e4nsel}, N.~{Bhattacharya}, T.~{de Vries},
	L.~{Augustin}, K.~{Williams}, M.~{Smit}, and E.~{Bente}, \enquote{Novel
		widely tunable monolithically integrated laser source,}
	{\protect\JournalTitle{IEEE Photonics Journal}} \textbf{7}, 1--9 (2015).
	
	\bibitem{stephan_2005PRA}
	G.~M. St\'ephan, T.~T. Tam, S.~Blin, P.~Besnard, and M.~T\^etu, \enquote{Laser
		line shape and spectral density of frequency noise,}
	{\protect\JournalTitle{Physical Review A}} \textbf{71}, {043809} (2005).
	
	\bibitem{llopis_2011OL}
	O.~Llopis, P.~H. Merrer, H.~Brahimi, K.~Saleh, and P.~Lacroix, \enquote{Phase
		noise measurement of a narrow linewidth {CW} laser using delay line
		approaches,} {\protect\JournalTitle{Optics Letters}} \textbf{36}, 2713--2715
	(2011).
	
	\bibitem{yariv_1990OL}
	A.~Yariv, \enquote{Signal-to-noise considerations in fiber links with periodic
		or distributed optical amplification,} {\protect\JournalTitle{Optics
			Letters}} \textbf{15}, 1064--1066 (1990).
	
	\bibitem{Epping_2019LC}
	J.~P. Epping, R.~M. Oldenbeuving, D.~Geskus, I.~Visscher, R.~Grootjans, C.~G.
	Roeloffzen, and R.~G. Heideman, \enquote{High power, tunable, narrow
		linewidth dual gain hybrid laser,} in \emph{Laser Congress 2019 (ASSL, LAC,
		LS\&C),}  (Optical Society of America, 2019), p. ATu1A.4.
	
	\bibitem{melnik_2006OE}
	S.~Melnik, G.~Huyet, and A.~V. Uskov, \enquote{The linewidth enhancement factor
		$\alpha$ of quantum dot semiconductor lasers,} {\protect\JournalTitle{Optics
			Express}} \textbf{14}, 2950--2955 (2006).
	
	\bibitem{redlich_2017JSTQE}
	C.~{Redlich}, B.~{Lingnau}, H.~{Huang}, R.~{Raghunathan}, K.~{Schires},
	P.~{Poole}, F.~{Grillot}, and K.~{Lüdge}, \enquote{Linewidth rebroadening in
		quantum dot semiconductor lasers,} {\protect\JournalTitle{IEEE Journal of
			Selected Topics in Quantum Electronics}} \textbf{23}, 1--10 (2017).
	
	\bibitem{andreou_2019}
	S.~Andreou, K.~A. Williams, and E.~A. J.~M. Bente, \enquote{Monolithically
		integrated {InP}-based {DBR} lasers with an intra-cavity ring resonator,}
	{\protect\JournalTitle{Optics Express}} \textbf{27}, 26281--26294 (2019).
	
	\bibitem{kruckel_2017OE}
	C.~J. Kr{\"u}ckel, A.~F{\"u}l{\"o}p, Z.~Ye, P.~A. Andrekson, and
	V.~Torres-Company, \enquote{Optical bandgap engineering in nonlinear silicon
		nitride waveguides,} {\protect\JournalTitle{Optics Express}} \textbf{25},
	15370--15380 (2017).
	
	\bibitem{vanlaer_2015NP}
	R.~Van~Laer, B.~Kuyken, D.~Van~Thourhout, and R.~Baets, \enquote{Interaction
		between light and highly confined hypersound in a silicon photonic nanowire,}
	{\protect\JournalTitle{Nature Photonics}} \textbf{9}, 199--203 (2015).
	
	\bibitem{gyger_2019ARXIV}
	F.~Gyger, J.~Liu, F.~Yang, J.~He, A.~S. Raja, R.~N. Wang, S.~A. Bhave, T.~J.
	Kippenberg, and L.~Thévenaz, \enquote{Observation of stimulated {B}rillouin
		scattering in silicon nitride integrated waveguides,}  (2019).
	
	\bibitem{eggleton_2019NP}
	B.~J. Eggleton, C.~G. Poulton, P.~T. Rakich, M.~J. Steel, and G.~Bahl,
	\enquote{Brillouin integrated photonics,} {\protect\JournalTitle{Nature
			Photonics}} \textbf{13}, 664--677 (2019).
	
	\bibitem{udem_2002N}
	T.~Udem, R.~Holzwarth, and T.~W. H{\"a}nsch, \enquote{Optical frequency
		metrology,} {\protect\JournalTitle{Nature}} \textbf{416}, 233--237 (2002).
	
	\bibitem{pillet_2008JLT}
	G.~Pillet, L.~Morvan, M.~Brunel, F.~Bretenaker, D.~Dolfi, M.~Vallet, J.-P.
	Huignard, and A.~L. Floch, \enquote{Dual-frequency laser at 1.5 {$\mu$}m for
		optical distribution and generation of high-purity microwave signals,}
	{\protect\JournalTitle{Journal of Lightwave Technology}} \textbf{26},
	2764--2773 (2008).
	
	\bibitem{takamoto_2005N}
	M.~Takamoto, F.-L. Hong, R.~Higashi, and H.~Katori, \enquote{An optical lattice
		clock,} {\protect\JournalTitle{Nature}} \textbf{435}, 321--324 (2005).
	
	\bibitem{wicht_2017SPIE}
	A.~Wicht, A.~Bawamia, M.~Kr{\"u}ger, C.~K{\"u}rbis, M.~Schiemangk, R.~Smol,
	A.~Peters, and G.~Tr{\"a}nkle, \enquote{{Narrow linewidth diode laser modules
			for quantum optical sensor applications in the field and in space},} in
	\emph{Components and packaging for laser systems {III},} , vol. 10085 A.~L.
	Glebov and P.~O. Leisher, eds., International Society for Optics and
	Photonics (SPIE, 2017), pp. 103 -- 118.
	
	\bibitem{maze_2008N}
	J.~R. Maze, P.~L. Stanwix, J.~S. Hodges, S.~Hong, J.~M. Taylor, P.~Cappellaro,
	L.~Jiang, M.~V.~G. Dutt, E.~Togan, A.~S. Zibrov, A.~Yacoby, R.~L. Walsworth,
	and M.~D. Lukin, \enquote{Nanoscale magnetic sensing with an individual
		electronic spin in diamond,} {\protect\JournalTitle{Nature}} \textbf{455},
	644--647 (2008).
	
	\bibitem{noelleke_2018SPIE}
	C.~N{\"o}lleke, P.~Leisching, G.~Blume, D.~Jedrzejczyk, J.~Pohl, D.~Feise,
	A.~Sahm, and K.~Paschke, \enquote{{Frequency locking of compact laser-diode
			modules at 633 nm},} in \emph{Photonic Instrumentation Engineering V,} , vol.
	10539 Y.~G. Soskind, ed., International Society for Optics and Photonics
	(SPIE, 2018), pp. 28 -- 33.
	
	\bibitem{wang_2019ASR}
	P.~Wang, W.~Chen, F.~Wan, J.~Wang, and J.~Hu, \enquote{A review of
		cavity-enhanced raman spectroscopy as a gas sensing method,}
	{\protect\JournalTitle{Applied Spectroscopy Reviews}} pp. 1--25 (2019).
	
	\bibitem{coddington_2008PRL}
	I.~Coddington, W.~C. Swann, and N.~R. Newbury, \enquote{Coherent
		multiheterodyne spectroscopy using stabilized optical frequency combs,}
	{\protect\JournalTitle{Physical Review Letters}} \textbf{100}, {013902}
	(2008).
	
	\bibitem{cheung_2010PTL}
	S.~{Cheung}, J.~{Baek}, R.~P. {Scott}, N.~K. {Fontaine}, F.~M. {Soares},
	X.~{Zhou}, D.~M. {Baney}, and S.~J.~B. {Yoo}, \enquote{1-{GHz} monolithically
		integrated hybrid mode-locked {InP} laser,} {\protect\JournalTitle{IEEE
			Photonics Technology Letters}} \textbf{22}, 1793--1795 (2010).
	
	\bibitem{srinivasan_2014FO}
	S.~Srinivasan, M.~Davenport, M.~J.~R. Heck, J.~Hutchinson, E.~Norberg, G.~Fish,
	and J.~Bowers, \enquote{Low phase noise hybrid silicon mode-locked lasers,}
	{\protect\JournalTitle{Frontiers of Optoelectronics}} \textbf{7}, 265--276
	(2014).
	
	\bibitem{davenport_2018PR}
	M.~L. Davenport, S.~Liu, and J.~E. Bowers, \enquote{Integrated heterogeneous
		silicon/iii-v mode-locked lasers,} {\protect\JournalTitle{Photonincs
			Research}} \textbf{6}, 468--478 (2018).
	
	\bibitem{mak_2019OE}
	J.~Mak, A.~van Rees, Y.~Fan, E.~J. Klein, D.~Geskus, P.~J.~M. van~der Slot, and
	K.-J. Boller, \enquote{Linewidth narrowing via low-loss dielectric waveguide
		feedback circuits in hybrid integrated frequency comb lasers,}
	{\protect\JournalTitle{Optics Express}} \textbf{27}, 13307--13318 (2019).
	
	\bibitem{chen_2006TMTT}
	{Xiangfei Chen}, {Zhichao Deng}, and {Jianping Yao}, \enquote{Photonic
		generation of microwave signal using a dual-wavelength
		single-longitudinal-mode fiber ring laser,} {\protect\JournalTitle{IEEE
			Transactions on Microwave Theory and Techniques}} \textbf{54}, 804--809
	(2006).
	
	\bibitem{grivas_2016PQE}
	C.~Grivas, \enquote{Optically pumped planar waveguide lasers: {P}art {II}:
		{G}ain media, laser systems, and applications,}
	{\protect\JournalTitle{Progress in Quantum Electronics}} \textbf{45-46},
	3--160 (2016).
	
	\bibitem{iio_1995PTL}
	S.~{Iio}, M.~{Suehiro}, T.~{Hirata}, and T.~{Hidaka},
	\enquote{Two-longitudinal-mode laser diodes,} {\protect\JournalTitle{IEEE
			Photonics Technology Letters}} \textbf{7}, 959--961 (1995).
	
	\bibitem{pozzi_2006PTL}
	F.~{Pozzi}, R.~M. {De La Rue}, and M.~{Sorel}, \enquote{Dual-wavelength
		{InAlGaAs}–{InP} laterally coupled distributed feedback laser,}
	{\protect\JournalTitle{IEEE Photonics Technology Letters}} \textbf{18},
	2563--2565 (2006).
	
	\bibitem{price_2007PTL}
	R.~K. {Price}, V.~B. {Verma}, K.~E. {Tobin}, V.~C. {Elarde}, and J.~J.
	{Coleman}, \enquote{Y-branch surface-etched distributed {B}ragg reflector
		lasers at 850 nm for optical heterodyning,} {\protect\JournalTitle{IEEE
			Photonics Technology Letters}} \textbf{19}, 1610--1612 (2007).
	
	\bibitem{kim_2013LPL}
	N.~Kim, H.-C. Ryu, D.~Lee, S.-P. Han, H.~Ko, K.~Moon, J.-W. Park, M.~Y. Jeon,
	and K.~H. Park, \enquote{Monolithically integrated optical beat sources
		toward a single-chip broadband terahertz emitter,}
	{\protect\JournalTitle{Laser Physics Letters}} \textbf{10}, {085805} (2013).
	
	\bibitem{guzman_2014URSI}
	R.~Guzm{\'a}n, A.~Jimenez, V.~Corral, G.~Carpintero, X.~Leijtens, and
	K.~Lawniczuk, \enquote{Narrow linewidth dual-wavelength laser sources based
		on {AWG} for the generation of millimeter wave signals,} in \emph{XXIX
		Simposium Nacional de la Uni{\'o}n Cient{\'i}fica Internacional de Radio,
		Septiembre 3-5, 2014, Valencia, Spain,}  (2014).
	
	\bibitem{zhu_2019OE}
	Y.~Zhu and L.~Zhu, \enquote{Narrow-linewidth, tunable external cavity dual-band
		diode lasers through {InP}/{GaAs}-{Si}$_3${N}$_4$ hybrid integration,}
	{\protect\JournalTitle{Optics Express}} \textbf{27}, 2354--2362 (2019).
	
	\bibitem{bovington_2014OL}
	J.~T. Bovington, M.~J.~R. Heck, and J.~E. Bowers, \enquote{Heterogeneous lasers
		and coupling to {Si}$_3${N}$_4$ near 1060 nm,} {\protect\JournalTitle{Optics
			Letters}} \textbf{39}, 6017--6020 (2014).
	
	\bibitem{kane_1985OL}
	T.~J. Kane and R.~L. Byer, \enquote{Monolithic, unidirectional single-mode
		{Nd}:{YAG} ring laser,} {\protect\JournalTitle{Optics Letters}} \textbf{10},
	65--67 (1985).
	
	\bibitem{hosseini_2015OE}
	N.~Hosseini, R.~Dekker, M.~Hoekman, M.~Dekkers, J.~Bos, A.~Leinse, and
	R.~Heideman, \enquote{Stress-optic modulator in triplex platform using a
		piezoelectric lead zirconate titanate (pzt) thin film,}
	{\protect\JournalTitle{Optics Express}} \textbf{23}, 14018--14026 (2015).
	
	\bibitem{epping_2015OE}
	J.~P. Epping, T.~Hellwig, M.~Hoekman, R.~Mateman, A.~Leinse, R.~G. Heideman,
	A.~van Rees, P.~J. van~der Slot, C.~J. Lee, C.~Fallnich, and K.-J. Boller,
	\enquote{On-chip visible-to-infrared supercontinuum generation with more than
		495 {THz} spectral bandwidth,} {\protect\JournalTitle{Optics Express}}
	\textbf{23}, 19596--19604 (2015).
	
	\bibitem{fan_2018OL}
	T.~Fan, Z.~Xia, A.~Adibi, and A.~A. Eftekhar, \enquote{Highly-uniform
		resonator-based visible spectrometer on a {Si}$_3${N}$_4$ platform with
		robust and accurate post-fabrication trimming,} {\protect\JournalTitle{Optics
			Letters}} \textbf{43}, 4887--4890 (2018).
	
	\bibitem{wang_2014OL}
	J.~Wang, S.~Chen, and D.~Dai, \enquote{Silicon hybrid demultiplexer with 64
		channels for wavelength/mode-division multiplexed on-chip optical
		interconnects,} {\protect\JournalTitle{Optics Letters}} \textbf{39},
	6993--6996 (2014).
	
	\bibitem{liu_2018JSTQE}
	Y.~{Liu}, A.~R. {Wichman}, B.~{Isaac}, J.~{Kalkavage}, E.~J. {Adles}, T.~R.
	{Clark}, and J.~{Klamkin}, \enquote{Ultra-low-loss silicon nitride optical
		beamforming network for wideband wireless applications,}
	{\protect\JournalTitle{IEEE Journal of Selected Topics in Quantum
			Electronics}} \textbf{24}, 1--10 (2018).
	
	\bibitem{visscher_2019EuCAP}
	I.~{Visscher}, C.~{Roeloffzen}, C.~{Taddei}, M.~{Hoekman}, L.~{Wevers},
	R.~{Grootjans}, P.~{Kapteijn}, D.~{Geskus}, A.~{Alippi}, R.~{Dekker},
	R.~{Oldenbeuving}, J.~{Epping}, R.~B. {Timens}, E.~{Klein}, A.~{Leinse},
	P.~v.~{Dijk}, and R.~{Heideman}, \enquote{Broadband true time delay microwave
		photonic beamformer for phased array antennas,} in \emph{2019 13th European
		Conference on Antennas and Propagation (EuCAP),}  (2019), pp. 1--5.
	
	\bibitem{oldenbeuving_2010OE}
	R.~M. Oldenbeuving, C.~J. Lee, P.~D. van Voorst, H.~L. Offerhaus, and K.-J.
	Boller, \enquote{Modeling of mode locking in a laser with spatially separate
		gain media,} {\protect\JournalTitle{Optics Express}} \textbf{18},
	22996--23008 (2010).
	
	\bibitem{fan_2014OL}
	Y.~Fan, R.~M. Oldenbeuving, M.~R.~H. Khan, C.~G.~H. Roeloffzen, E.~J. Klein,
	C.~J. Lee, H.~L. Offerhaus, and K.-J. Boller, \enquote{Q-factor measurements
		through injection locking of a semiconductor-glass hybrid laser with unknown
		intracavity losses,} {\protect\JournalTitle{Optics Letters}} \textbf{39},
	1748--1751 (2014).
	
	\bibitem{kuswandi_2007ACA}
	B.~Kuswandi, Nuriman, J.~Huskens, and W.~Verboom, \enquote{Optical sensing
		systems for microfluidic devices: A review,} {\protect\JournalTitle{Analytica
			Chimica Acta}} \textbf{601}, 141--155 (2007).
	
	\bibitem{artundo_2017OP}
	I.~Artundo, \enquote{Photonic integration: new applications are visible,}
	{\protect\JournalTitle{Optik \& Photonik}} \textbf{12}, 22--25 (2017).
	
	\bibitem{ymeti_2007NL}
	A.~Ymeti, J.~Greve, P.~V. Lambeck, T.~Wink, van H{\"o}vell, Beumer, R.~R. Wijn,
	R.~G. Heideman, V.~Subramaniam, and J.~S. Kanger, \enquote{Fast,
		ultrasensitive virus detection using a young interferometer sensor,}
	{\protect\JournalTitle{Nano Letters}} \textbf{7}, 394--397 (2007).
	
	\bibitem{porcel_2019OLT}
	M.~A. Porcel, A.~Hinojosa, H.~Jans, A.~Stassen, J.~Goyvaerts, D.~Geuzebroek,
	M.~Geiselmann, C.~Dominguez, and I.~Artundo, \enquote{Silicon nitride
		photonic integration for visible light applications,}
	{\protect\JournalTitle{Optics \& Laser Technology}} \textbf{112}, 299--306
	(2019).
	
\end{thebibliography}
\end{document}